\documentclass[submission,copyright,creativecommons]{eptcs}
\providecommand{\event}{GandALF 2016} 
\usepackage{breakurl}             

\title{Cycle Detection in Computation Tree Logic}
\author{Ga\"{e}lle Fontaine
\institute{Universidad de Chile \\ Santiago de Chile, Chile}
\and
Fabio Mogavero 
\institute{University of Oxford \\ Oxford, UK}
\and
Aniello Murano 
\institute{University of Naples \\ Naples, Italy}
\and
Giuseppe Perelli 
\institute{University of Oxford \\ Oxford, UK}
\and
Loredana Sorrentino 
\institute{University of Naples \\ Naples, Italy}
}
\def\titlerunning{Cycle Detection in Computation Tree Logic}
\def\authorrunning{G. Fontaine et al.}

\usepackage{cite}

\usepackage[final]{microtype}

\hypersetup
	{
	pdfsubject	=	{},
	pdfkeywords	=	{},
	pdfproducer	=	{},
	pdfcreator	=	{}
	}

\usepackage{etoolbox}
\usepackage{amsthm}

\AtEndPreamble
	{

	\usepackage{changebar}

	\usepackage{fixltx2e}

	\usepackage{enumerate}

	}

\usepackage{xargs}
\usepackage{xspace}
\usepackage{xstring}
\usepackage{boolexpr}

\usepackage[cmex10]{mathtools}
\usepackage{amsxtra}
\usepackage{amssymb}
\usepackage{amsfonts}
\usepackage{mathrsfs}
\usepackage{mathdots}
\usepackage{latexsym}
\usepackage{textcomp}
\usepackage{pifont}
\usepackage{wasysym}

\newcommand{\argemp}[2]
	{\if&#1&\else#2\fi}

\newcommand{\argdef}[2]
	{\if&#1&#2\else#1\fi}

\newcommand{\argint}[3]
	{\if&#2&\else#1#2#3\fi}

\newcommand{\argext}[3]
	{\if&#1&#3\else#1\if&#3&\else#2#3\fi\fi}

\newcommandx{\argsubsup}[3][2=, 3=]
	{\def\argsubscript{{#2}}\def\argsuperscript{{#3}}#1}

\newcommandx{\argind}[9][2=, 3=, 4=, 5=, 6=, 7=, 8=, 9=]
	{%
	\switch[#1=]%
		\case{0}#2%
		\case{1}#3%
		\case{2}#4%
		\case{3}#5%
		\case{4}#6%
		\case{5}#7%
		\case{6}#8%
		\case{7}#9%
		\otherwise\ensuremath{\clubsuit}%
	\endswitch%
	}

\newcommand{\arga}[1]
	{#1}
\newcommand{\argb}[2]
	{\argext{\arga{#1}}{, \allowbreak}{#2}}
\newcommand{\argc}[3]
	{\argext{\argb{#1}{#2}}{, \allowbreak}{#3}}
\newcommand{\argd}[4]
	{\argext{\argc{#1}{#2}{#3}}{, \allowbreak}{#4}}
\newcommand{\arge}[5]
	{\argext{\argd{#1}{#2}{#3}{#4}}{, \allowbreak}{#5}}
\newcommand{\argf}[6]
	{\argext{\arge{#1}{#2}{#3}{#4}{#5}}{, \allowbreak}{#6}}
\newcommand{\argg}[7]
	{\argext{\argf{#1}{#2}{#3}{#4}{#5}{#6}}{, \allowbreak}{#7}}
\newcommand{\argh}[8]
	{\argext{\argg{#1}{#2}{#3}{#4}{#5}{#6}{#7}}{, \allowbreak}{#8}}
\newcommand{\argi}[9]
	{\argext{\argh{#1}{#2}{#3}{#4}{#5}{#6}{#7}{#8}}{, \allowbreak}{#9}}

\newcommand{\argj}[9]
	{%
	\def\valarga{#1}%
	\def\valargb{#2}%
	\def\valargc{#3}%
	\def\valargd{#4}%
	\def\valarge{#5}%
	\def\valargf{#6}%
	\def\valargg{#7}%
	\def\valargh{#8}%
	\def\valargi{#9}%
	\argauxj%
	}
\newcommand{\argk}[9]
	{%
	\def\valarga{#1}%
	\def\valargb{#2}%
	\def\valargc{#3}%
	\def\valargd{#4}%
	\def\valarge{#5}%
	\def\valargf{#6}%
	\def\valargg{#7}%
	\def\valargh{#8}%
	\def\valargi{#9}%
	\argauxk%
	}
\newcommand{\argl}[9]
	{%
	\def\valarga{#1}%
	\def\valargb{#2}%
	\def\valargc{#3}%
	\def\valargd{#4}%
	\def\valarge{#5}%
	\def\valargf{#6}%
	\def\valargg{#7}%
	\def\valargh{#8}%
	\def\valargi{#9}%
	\argauxl%
	}
\newcommand{\argm}[9]
	{%
	\def\valarga{#1}%
	\def\valargb{#2}%
	\def\valargc{#3}%
	\def\valargd{#4}%
	\def\valarge{#5}%
	\def\valargf{#6}%
	\def\valargg{#7}%
	\def\valargh{#8}%
	\def\valargi{#9}%
	\argauxm%
	}
\newcommand{\argn}[9]
	{%
	\def\valarga{#1}%
	\def\valargb{#2}%
	\def\valargc{#3}%
	\def\valargd{#4}%
	\def\valarge{#5}%
	\def\valargf{#6}%
	\def\valargg{#7}%
	\def\valargh{#8}%
	\def\valargi{#9}%
	\argauxn%
	}
\newcommand{\argo}[9]
	{%
	\def\valarga{#1}%
	\def\valargb{#2}%
	\def\valargc{#3}%
	\def\valargd{#4}%
	\def\valarge{#5}%
	\def\valargf{#6}%
	\def\valargg{#7}%
	\def\valargh{#8}%
	\def\valargi{#9}%
	\argauxo%
	}
\newcommand{\argp}[9]
	{%
	\def\valarga{#1}%
	\def\valargb{#2}%
	\def\valargc{#3}%
	\def\valargd{#4}%
	\def\valarge{#5}%
	\def\valargf{#6}%
	\def\valargg{#7}%
	\def\valargh{#8}%
	\def\valargi{#9}%
	\argauxp%
	}
\newcommand{\argq}[9]
	{%
	\def\valarga{#1}%
	\def\valargb{#2}%
	\def\valargc{#3}%
	\def\valargd{#4}%
	\def\valarge{#5}%
	\def\valargf{#6}%
	\def\valargg{#7}%
	\def\valargh{#8}%
	\def\valargi{#9}%
	\argauxq%
	}
\newcommand{\argr}[9]
	{%
	\def\valarga{#1}%
	\def\valargb{#2}%
	\def\valargc{#3}%
	\def\valargd{#4}%
	\def\valarge{#5}%
	\def\valargf{#6}%
	\def\valargg{#7}%
	\def\valargh{#8}%
	\def\valargi{#9}%
	\argauxr%
	}

\newcommand{\argauxj}[1]
	{%
	\argext%
		{\argi{\valarga}{\valargb}{\valargc}{\valargd}{\valarge}{\valargf}{\valargg}
			{\valargh}{\valargi}}
		{, \allowbreak}{#1}%
	}
\newcommand{\argauxk}[2]
	{\argext{\argauxj{#1}}{, \allowbreak}{#2}}
\newcommand{\argauxl}[3]
	{\argext{\argauxk{#1}{#2}}{, \allowbreak}{#3}}
\newcommand{\argauxm}[4]
	{\argext{\argauxl{#1}{#2}{#3}}{, \allowbreak}{#4}}
\newcommand{\argauxn}[5]
	{\argext{\argauxm{#1}{#2}{#3}{#4}}{, \allowbreak}{#5}}
\newcommand{\argauxo}[6]
	{\argext{\argauxn{#1}{#2}{#3}{#4}{#5}}{, \allowbreak}{#6}}
\newcommand{\argauxp}[7]
	{\argext{\argauxo{#1}{#2}{#3}{#4}{#5}{#6}}{, \allowbreak}{#7}}
\newcommand{\argauxq}[8]
	{\argext{\argauxp{#1}{#2}{#3}{#4}{#5}{#6}{#7}}{, \allowbreak}{#8}}
\newcommand{\argauxr}[9]
	{\argext{\argauxq{#1}{#2}{#3}{#4}{#5}{#6}{#7}{#8}}{, \allowbreak}{#9}}

\newcommand{\txtfnt}[2][]
	{{%
	\IfStrEq{#1}{}
		{#2}
		{%
		\StrLeft{#1}{2}[\optbgn]%
		\StrGobbleLeft{#1}{2}[\optend]%
		\IfStrEqCase{\optbgn}
			{%
			{Rm}{\rmfamily\txtfnt[\optend]{#2}}%
			{Sf}{\sffamily\txtfnt[\optend]{#2}}%
			{Tt}{\ttfamily\txtfnt[\optend]{#2}}%
			{Up}{\upshape\txtfnt[\optend]{#2}}%
			{It}{\itshape\txtfnt[\optend]{#2}}%
			{Sl}{\slshape\txtfnt[\optend]{#2}}%
			{Sc}{\scshape\txtfnt[\optend]{#2}}%
			{Md}{\mdseries\txtfnt[\optend]{#2}}%
			{Bf}{\bfseries\txtfnt[\optend]{#2}}%
			{Em}{\emph{\txtfnt[\optend]{#2}}}%
			}
			[\ensuremath{\clubsuit}]%
		}%
	}}

\newcommand{\txtsub}[2][]
	{\argemp{#2}{\ensuremath{_{\text{\txtfnt[#1]{#2}}}}}}

\newcommand{\txtsup}[2][]
	{\argemp{#2}{\ensuremath{^{\text{\txtfnt[#1]{#2}}}}}}

\newcommandx{\txt}[4][1=, 3=, 4=]
	{\text{\txtfnt[#1]{#2}\ensuremath{\txtsub[#1]{#3}\txtsup[#1]{#4}}}}

\newcommandx{\txtarg}[5][1=, 3=, 4=]
	{{\txt[#1]{#2}[#3][#4]\argint{(}{#5}{)}}}

\newcommand{\txtstyname}{RmScMd}
\newcommand{\txtname}[1][]
	{\txt[\argdef{#1}{\txtstyname}]}
\newcommand{\txtargname}[1][]
	{\txtarg[\argdef{#1}{\txtstyname}]}

\newcommand{\txtstyabr}{Em}
\newcommand{\txtabr}[1][]
	{\txt[\argdef{#1}{\txtstyabr}]}

\newcommandx{\mthfnt}[3][1=, 2=0]
	{{%
	\IfStrEqCase{#1}
		{%
		{}%
			{#3}%
		{Name}%
			{%
			\IfStrEqCase{#2}
				{%
				{0}{\mathcal{#3}}%
				{1}{\mathscr{#3}}%
				{2}{\mathfrak{#3}}%
				{3}{\mathbf{#3}}%
				}
				[\ensuremath{\clubsuit}]%
			}%
		{Set}%
			{%
			\IfStrEqCase{#2}
				{%
				{0}{\mathrm{#3}}%
				{1}{\mathsf{#3}}%
				{2}{\mathbb{#3}}%
				{3}{\mathtt{#3}}%
				}
				[\ensuremath{\clubsuit}]%
			}%
		{Fun}%
			{%
			\IfStrEqCase{#2}
				{%
				{0}{\mathsf{#3}}%
				{1}{\mathrm{#3}}%
				}
				[\ensuremath{\clubsuit}]%
			}%
		{Rel}%
			{%
			\IfStrEqCase{#2}
				{%
				{0}{\mathit{#3}}%
				{1}{\mathtt{#3}}%
				}
				[\ensuremath{\clubsuit}]%
			}%
		{Sym}%
			{%
			\IfStrEqCase{#2}
				{%
				{0}{\mathtt{#3}}%
				{1}{\mathbf{#3}}%
				}
				[\ensuremath{\clubsuit}]%
			}%
		{Elm}%
			{\mathnormal{#3}}
		}
		[\ensuremath{\clubsuit}]%
	}}

\newcommand{\mthsub}[1]
	{\argemp{#1}{\ensuremath{_{\mathnormal{#1}}}}}

\newcommand{\mthsup}[1]
	{\argemp{#1}{\ensuremath{^{\mathnormal{#1}}}}}

\newcommandx{\mth}[5][1=, 2=0, 4=, 5=]
	{{\ensuremath{\mthfnt[#1][#2]{#3}\mthsub{#4}\mthsup{#5}}}}

\newcommandx{\mtharg}[6][1=, 2=0, 4=, 5=]
	{{\mth[#1][#2]{#3}[#4][#5]\ensuremath{\argint{(}{#6}{)}}}}

\newcommand{\mthempty}
	{\mth[][]}

\newcommand{\mthstyname}{0}
\newcommand{\mthname}[1][]
	{\mth[Name][\argdef{#1}{\mthstyname}]}

\newcommand{\mthstyset}{0}
\newcommand{\mthset}[1][]
	{\mth[Set][\argdef{#1}{\mthstyset}]}
\newcommand{\mthargset}[1][]
	{\mtharg[Set][\argdef{#1}{\mthstyset}]}

\newcommand{\mthstyfun}{0}
\newcommand{\mthfun}[1][]
	{\mth[Fun][\argdef{#1}{\mthstyfun}]}
\newcommand{\mthargfun}[1][]
	{\mtharg[Fun][\argdef{#1}{\mthstyfun}]}

\newcommand{\mthstyrel}{0}
\newcommand{\mthrel}[1][]
	{\mth[Rel][\argdef{#1}{\mthstyrel}]}

\newcommand{\mthstysym}{0}
\newcommand{\mthsym}[1][]
	{\mth[Sym][\argdef{#1}{\mthstysym}]}

\newcommand{\mthstyelm}{0}
\newcommand{\mthelm}[1][]
	{\mth[Elm][\argdef{#1}{\mthstyelm}]}

\newcommandx{\AName}[4][1=, 2=, 3=, 4=]{\mthname[#4]{A#3}[#1][#2]}
\newcommandx{\BName}[4][1=, 2=, 3=, 4=]{\mthname[#4]{B#3}[#1][#2]}
\newcommandx{\CName}[4][1=, 2=, 3=, 4=]{\mthname[#4]{C#3}[#1][#2]}
\newcommandx{\DName}[4][1=, 2=, 3=, 4=]{\mthname[#4]{D#3}[#1][#2]}
\newcommandx{\EName}[4][1=, 2=, 3=, 4=]{\mthname[#4]{E#3}[#1][#2]}
\newcommandx{\FName}[4][1=, 2=, 3=, 4=]{\mthname[#4]{F#3}[#1][#2]}
\newcommandx{\GName}[4][1=, 2=, 3=, 4=]{\mthname[#4]{G#3}[#1][#2]}
\newcommandx{\HName}[4][1=, 2=, 3=, 4=]{\mthname[#4]{H#3}[#1][#2]}
\newcommandx{\IName}[4][1=, 2=, 3=, 4=]{\mthname[#4]{I#3}[#1][#2]}
\newcommandx{\JName}[4][1=, 2=, 3=, 4=]{\mthname[#4]{J#3}[#1][#2]}
\newcommandx{\KName}[4][1=, 2=, 3=, 4=]{\mthname[#4]{K#3}[#1][#2]}
\newcommandx{\LName}[4][1=, 2=, 3=, 4=]{\mthname[#4]{L#3}[#1][#2]}
\newcommandx{\MName}[4][1=, 2=, 3=, 4=]{\mthname[#4]{M#3}[#1][#2]}
\newcommandx{\NName}[4][1=, 2=, 3=, 4=]{\mthname[#4]{N#3}[#1][#2]}
\newcommandx{\OName}[4][1=, 2=, 3=, 4=]{\mthname[#4]{O#3}[#1][#2]}
\newcommandx{\PName}[4][1=, 2=, 3=, 4=]{\mthname[#4]{P#3}[#1][#2]}
\newcommandx{\QName}[4][1=, 2=, 3=, 4=]{\mthname[#4]{Q#3}[#1][#2]}
\newcommandx{\RName}[4][1=, 2=, 3=, 4=]{\mthname[#4]{R#3}[#1][#2]}
\newcommandx{\SName}[4][1=, 2=, 3=, 4=]{\mthname[#4]{S#3}[#1][#2]}
\newcommandx{\TName}[4][1=, 2=, 3=, 4=]{\mthname[#4]{T#3}[#1][#2]}
\newcommandx{\UName}[4][1=, 2=, 3=, 4=]{\mthname[#4]{U#3}[#1][#2]}
\newcommandx{\VName}[4][1=, 2=, 3=, 4=]{\mthname[#4]{V#3}[#1][#2]}
\newcommandx{\WName}[4][1=, 2=, 3=, 4=]{\mthname[#4]{W#3}[#1][#2]}
\newcommandx{\XName}[4][1=, 2=, 3=, 4=]{\mthname[#4]{X#3}[#1][#2]}
\newcommandx{\YName}[4][1=, 2=, 3=, 4=]{\mthname[#4]{Y#3}[#1][#2]}
\newcommandx{\ZName}[4][1=, 2=, 3=, 4=]{\mthname[#4]{Z#3}[#1][#2]}

\newcommandx{\ASet}[4][1=, 2=, 3=, 4=]{\mthset[#4]{A#3}[#1][#2]}
\newcommandx{\BSet}[4][1=, 2=, 3=, 4=]{\mthset[#4]{B#3}[#1][#2]}
\newcommandx{\CSet}[4][1=, 2=, 3=, 4=]{\mthset[#4]{C#3}[#1][#2]}
\newcommandx{\DSet}[4][1=, 2=, 3=, 4=]{\mthset[#4]{D#3}[#1][#2]}
\newcommandx{\ESet}[4][1=, 2=, 3=, 4=]{\mthset[#4]{E#3}[#1][#2]}
\newcommandx{\FSet}[4][1=, 2=, 3=, 4=]{\mthset[#4]{F#3}[#1][#2]}
\newcommandx{\GSet}[4][1=, 2=, 3=, 4=]{\mthset[#4]{G#3}[#1][#2]}
\newcommandx{\HSet}[4][1=, 2=, 3=, 4=]{\mthset[#4]{H#3}[#1][#2]}
\newcommandx{\ISet}[4][1=, 2=, 3=, 4=]{\mthset[#4]{I#3}[#1][#2]}
\newcommandx{\JSet}[4][1=, 2=, 3=, 4=]{\mthset[#4]{J#3}[#1][#2]}
\newcommandx{\KSet}[4][1=, 2=, 3=, 4=]{\mthset[#4]{K#3}[#1][#2]}
\newcommandx{\LSet}[4][1=, 2=, 3=, 4=]{\mthset[#4]{L#3}[#1][#2]}
\newcommandx{\MSet}[4][1=, 2=, 3=, 4=]{\mthset[#4]{M#3}[#1][#2]}
\newcommandx{\NSet}[4][1=, 2=, 3=, 4=]{\mthset[#4]{N#3}[#1][#2]}
\newcommandx{\OSet}[4][1=, 2=, 3=, 4=]{\mthset[#4]{O#3}[#1][#2]}
\newcommandx{\PSet}[4][1=, 2=, 3=, 4=]{\mthset[#4]{P#3}[#1][#2]}
\newcommandx{\QSet}[4][1=, 2=, 3=, 4=]{\mthset[#4]{Q#3}[#1][#2]}
\newcommandx{\RSet}[4][1=, 2=, 3=, 4=]{\mthset[#4]{R#3}[#1][#2]}
\newcommandx{\SSet}[4][1=, 2=, 3=, 4=]{\mthset[#4]{S#3}[#1][#2]}
\newcommandx{\TSet}[4][1=, 2=, 3=, 4=]{\mthset[#4]{T#3}[#1][#2]}
\newcommandx{\USet}[4][1=, 2=, 3=, 4=]{\mthset[#4]{U#3}[#1][#2]}
\newcommandx{\VSet}[4][1=, 2=, 3=, 4=]{\mthset[#4]{V#3}[#1][#2]}
\newcommandx{\WSet}[4][1=, 2=, 3=, 4=]{\mthset[#4]{W#3}[#1][#2]}
\newcommandx{\XSet}[4][1=, 2=, 3=, 4=]{\mthset[#4]{X#3}[#1][#2]}
\newcommandx{\YSet}[4][1=, 2=, 3=, 4=]{\mthset[#4]{Y#3}[#1][#2]}
\newcommandx{\ZSet}[4][1=, 2=, 3=, 4=]{\mthset[#4]{Z#3}[#1][#2]}

\newcommandx{\aSet}[4][1=, 2=, 3=, 4=]{\mthset[#4]{a#3}[#1][#2]}
\newcommandx{\bSet}[4][1=, 2=, 3=, 4=]{\mthset[#4]{b#3}[#1][#2]}
\newcommandx{\cSet}[4][1=, 2=, 3=, 4=]{\mthset[#4]{c#3}[#1][#2]}
\newcommandx{\dSet}[4][1=, 2=, 3=, 4=]{\mthset[#4]{d#3}[#1][#2]}
\newcommandx{\eSet}[4][1=, 2=, 3=, 4=]{\mthset[#4]{e#3}[#1][#2]}
\newcommandx{\fSet}[4][1=, 2=, 3=, 4=]{\mthset[#4]{f#3}[#1][#2]}
\newcommandx{\gSet}[4][1=, 2=, 3=, 4=]{\mthset[#4]{g#3}[#1][#2]}
\newcommandx{\hSet}[4][1=, 2=, 3=, 4=]{\mthset[#4]{h#3}[#1][#2]}
\newcommandx{\iSet}[4][1=, 2=, 3=, 4=]{\mthset[#4]{i#3}[#1][#2]}
\newcommandx{\jSet}[4][1=, 2=, 3=, 4=]{\mthset[#4]{j#3}[#1][#2]}
\newcommandx{\kSet}[4][1=, 2=, 3=, 4=]{\mthset[#4]{k#3}[#1][#2]}
\newcommandx{\lSet}[4][1=, 2=, 3=, 4=]{\mthset[#4]{l#3}[#1][#2]}
\newcommandx{\mSet}[4][1=, 2=, 3=, 4=]{\mthset[#4]{m#3}[#1][#2]}
\newcommandx{\nSet}[4][1=, 2=, 3=, 4=]{\mthset[#4]{n#3}[#1][#2]}
\newcommandx{\oSet}[4][1=, 2=, 3=, 4=]{\mthset[#4]{o#3}[#1][#2]}
\newcommandx{\pSet}[4][1=, 2=, 3=, 4=]{\mthset[#4]{p#3}[#1][#2]}
\newcommandx{\qSet}[4][1=, 2=, 3=, 4=]{\mthset[#4]{q#3}[#1][#2]}
\newcommandx{\rSet}[4][1=, 2=, 3=, 4=]{\mthset[#4]{r#3}[#1][#2]}
\newcommandx{\sSet}[4][1=, 2=, 3=, 4=]{\mthset[#4]{s#3}[#1][#2]}
\newcommandx{\tSet}[4][1=, 2=, 3=, 4=]{\mthset[#4]{t#3}[#1][#2]}
\newcommandx{\uSet}[4][1=, 2=, 3=, 4=]{\mthset[#4]{u#3}[#1][#2]}
\newcommandx{\vSet}[4][1=, 2=, 3=, 4=]{\mthset[#4]{v#3}[#1][#2]}
\newcommandx{\wSet}[4][1=, 2=, 3=, 4=]{\mthset[#4]{w#3}[#1][#2]}
\newcommandx{\xSet}[4][1=, 2=, 3=, 4=]{\mthset[#4]{x#3}[#1][#2]}
\newcommandx{\ySet}[4][1=, 2=, 3=, 4=]{\mthset[#4]{y#3}[#1][#2]}
\newcommandx{\zSet}[4][1=, 2=, 3=, 4=]{\mthset[#4]{z#3}[#1][#2]}

\newcommandx{\AFun}[4][1=, 2=, 3=, 4=]{\mthfun[#4]{A#3}[#1][#2]}
\newcommandx{\BFun}[4][1=, 2=, 3=, 4=]{\mthfun[#4]{B#3}[#1][#2]}
\newcommandx{\CFun}[4][1=, 2=, 3=, 4=]{\mthfun[#4]{C#3}[#1][#2]}
\newcommandx{\DFun}[4][1=, 2=, 3=, 4=]{\mthfun[#4]{D#3}[#1][#2]}
\newcommandx{\EFun}[4][1=, 2=, 3=, 4=]{\mthfun[#4]{E#3}[#1][#2]}
\newcommandx{\FFun}[4][1=, 2=, 3=, 4=]{\mthfun[#4]{F#3}[#1][#2]}
\newcommandx{\GFun}[4][1=, 2=, 3=, 4=]{\mthfun[#4]{G#3}[#1][#2]}
\newcommandx{\HFun}[4][1=, 2=, 3=, 4=]{\mthfun[#4]{H#3}[#1][#2]}
\newcommandx{\IFun}[4][1=, 2=, 3=, 4=]{\mthfun[#4]{I#3}[#1][#2]}
\newcommandx{\JFun}[4][1=, 2=, 3=, 4=]{\mthfun[#4]{J#3}[#1][#2]}
\newcommandx{\KFun}[4][1=, 2=, 3=, 4=]{\mthfun[#4]{K#3}[#1][#2]}
\newcommandx{\LFun}[4][1=, 2=, 3=, 4=]{\mthfun[#4]{L#3}[#1][#2]}
\newcommandx{\MFun}[4][1=, 2=, 3=, 4=]{\mthfun[#4]{M#3}[#1][#2]}
\newcommandx{\NFun}[4][1=, 2=, 3=, 4=]{\mthfun[#4]{N#3}[#1][#2]}
\newcommandx{\OFun}[4][1=, 2=, 3=, 4=]{\mthfun[#4]{O#3}[#1][#2]}
\newcommandx{\PFun}[4][1=, 2=, 3=, 4=]{\mthfun[#4]{P#3}[#1][#2]}
\newcommandx{\QFun}[4][1=, 2=, 3=, 4=]{\mthfun[#4]{Q#3}[#1][#2]}
\newcommandx{\RFun}[4][1=, 2=, 3=, 4=]{\mthfun[#4]{R#3}[#1][#2]}
\newcommandx{\SFun}[4][1=, 2=, 3=, 4=]{\mthfun[#4]{S#3}[#1][#2]}
\newcommandx{\TFun}[4][1=, 2=, 3=, 4=]{\mthfun[#4]{T#3}[#1][#2]}
\newcommandx{\UFun}[4][1=, 2=, 3=, 4=]{\mthfun[#4]{U#3}[#1][#2]}
\newcommandx{\VFun}[4][1=, 2=, 3=, 4=]{\mthfun[#4]{V#3}[#1][#2]}
\newcommandx{\WFun}[4][1=, 2=, 3=, 4=]{\mthfun[#4]{W#3}[#1][#2]}
\newcommandx{\XFun}[4][1=, 2=, 3=, 4=]{\mthfun[#4]{X#3}[#1][#2]}
\newcommandx{\YFun}[4][1=, 2=, 3=, 4=]{\mthfun[#4]{Y#3}[#1][#2]}
\newcommandx{\ZFun}[4][1=, 2=, 3=, 4=]{\mthfun[#4]{Z#3}[#1][#2]}

\newcommandx{\aFun}[4][1=, 2=, 3=, 4=]{\mthfun[#4]{a#3}[#1][#2]}
\newcommandx{\bFun}[4][1=, 2=, 3=, 4=]{\mthfun[#4]{b#3}[#1][#2]}
\newcommandx{\cFun}[4][1=, 2=, 3=, 4=]{\mthfun[#4]{c#3}[#1][#2]}
\newcommandx{\dFun}[4][1=, 2=, 3=, 4=]{\mthfun[#4]{d#3}[#1][#2]}
\newcommandx{\eFun}[4][1=, 2=, 3=, 4=]{\mthfun[#4]{e#3}[#1][#2]}
\newcommandx{\fFun}[4][1=, 2=, 3=, 4=]{\mthfun[#4]{f#3}[#1][#2]}
\newcommandx{\gFun}[4][1=, 2=, 3=, 4=]{\mthfun[#4]{g#3}[#1][#2]}
\newcommandx{\hFun}[4][1=, 2=, 3=, 4=]{\mthfun[#4]{h#3}[#1][#2]}
\newcommandx{\iFun}[4][1=, 2=, 3=, 4=]{\mthfun[#4]{i#3}[#1][#2]}
\newcommandx{\jFun}[4][1=, 2=, 3=, 4=]{\mthfun[#4]{j#3}[#1][#2]}
\newcommandx{\kFun}[4][1=, 2=, 3=, 4=]{\mthfun[#4]{k#3}[#1][#2]}
\newcommandx{\lFun}[4][1=, 2=, 3=, 4=]{\mthfun[#4]{l#3}[#1][#2]}
\newcommandx{\mFun}[4][1=, 2=, 3=, 4=]{\mthfun[#4]{m#3}[#1][#2]}
\newcommandx{\nFun}[4][1=, 2=, 3=, 4=]{\mthfun[#4]{n#3}[#1][#2]}
\newcommandx{\oFun}[4][1=, 2=, 3=, 4=]{\mthfun[#4]{o#3}[#1][#2]}
\newcommandx{\pFun}[4][1=, 2=, 3=, 4=]{\mthfun[#4]{p#3}[#1][#2]}
\newcommandx{\qFun}[4][1=, 2=, 3=, 4=]{\mthfun[#4]{q#3}[#1][#2]}
\newcommandx{\rFun}[4][1=, 2=, 3=, 4=]{\mthfun[#4]{r#3}[#1][#2]}
\newcommandx{\sFun}[4][1=, 2=, 3=, 4=]{\mthfun[#4]{s#3}[#1][#2]}
\newcommandx{\tFun}[4][1=, 2=, 3=, 4=]{\mthfun[#4]{t#3}[#1][#2]}
\newcommandx{\uFun}[4][1=, 2=, 3=, 4=]{\mthfun[#4]{u#3}[#1][#2]}
\newcommandx{\vFun}[4][1=, 2=, 3=, 4=]{\mthfun[#4]{v#3}[#1][#2]}
\newcommandx{\wFun}[4][1=, 2=, 3=, 4=]{\mthfun[#4]{w#3}[#1][#2]}
\newcommandx{\xFun}[4][1=, 2=, 3=, 4=]{\mthfun[#4]{x#3}[#1][#2]}
\newcommandx{\yFun}[4][1=, 2=, 3=, 4=]{\mthfun[#4]{y#3}[#1][#2]}
\newcommandx{\zFun}[4][1=, 2=, 3=, 4=]{\mthfun[#4]{z#3}[#1][#2]}

\newcommandx{\ARel}[4][1=, 2=, 3=, 4=]{\mthrel[#4]{A#3}[#1][#2]}
\newcommandx{\BRel}[4][1=, 2=, 3=, 4=]{\mthrel[#4]{B#3}[#1][#2]}
\newcommandx{\CRel}[4][1=, 2=, 3=, 4=]{\mthrel[#4]{C#3}[#1][#2]}
\newcommandx{\DRel}[4][1=, 2=, 3=, 4=]{\mthrel[#4]{D#3}[#1][#2]}
\newcommandx{\ERel}[4][1=, 2=, 3=, 4=]{\mthrel[#4]{E#3}[#1][#2]}
\newcommandx{\FRel}[4][1=, 2=, 3=, 4=]{\mthrel[#4]{F#3}[#1][#2]}
\newcommandx{\GRel}[4][1=, 2=, 3=, 4=]{\mthrel[#4]{G#3}[#1][#2]}
\newcommandx{\HRel}[4][1=, 2=, 3=, 4=]{\mthrel[#4]{H#3}[#1][#2]}
\newcommandx{\IRel}[4][1=, 2=, 3=, 4=]{\mthrel[#4]{I#3}[#1][#2]}
\newcommandx{\JRel}[4][1=, 2=, 3=, 4=]{\mthrel[#4]{J#3}[#1][#2]}
\newcommandx{\KRel}[4][1=, 2=, 3=, 4=]{\mthrel[#4]{K#3}[#1][#2]}
\newcommandx{\LRel}[4][1=, 2=, 3=, 4=]{\mthrel[#4]{L#3}[#1][#2]}
\newcommandx{\MRel}[4][1=, 2=, 3=, 4=]{\mthrel[#4]{M#3}[#1][#2]}
\newcommandx{\NRel}[4][1=, 2=, 3=, 4=]{\mthrel[#4]{N#3}[#1][#2]}
\newcommandx{\ORel}[4][1=, 2=, 3=, 4=]{\mthrel[#4]{O#3}[#1][#2]}
\newcommandx{\PRel}[4][1=, 2=, 3=, 4=]{\mthrel[#4]{P#3}[#1][#2]}
\newcommandx{\QRel}[4][1=, 2=, 3=, 4=]{\mthrel[#4]{Q#3}[#1][#2]}
\newcommandx{\RRel}[4][1=, 2=, 3=, 4=]{\mthrel[#4]{R#3}[#1][#2]}
\newcommandx{\SRel}[4][1=, 2=, 3=, 4=]{\mthrel[#4]{S#3}[#1][#2]}
\newcommandx{\TRel}[4][1=, 2=, 3=, 4=]{\mthrel[#4]{T#3}[#1][#2]}
\newcommandx{\URel}[4][1=, 2=, 3=, 4=]{\mthrel[#4]{U#3}[#1][#2]}
\newcommandx{\VRel}[4][1=, 2=, 3=, 4=]{\mthrel[#4]{V#3}[#1][#2]}
\newcommandx{\WRel}[4][1=, 2=, 3=, 4=]{\mthrel[#4]{W#3}[#1][#2]}
\newcommandx{\XRel}[4][1=, 2=, 3=, 4=]{\mthrel[#4]{X#3}[#1][#2]}
\newcommandx{\YRel}[4][1=, 2=, 3=, 4=]{\mthrel[#4]{Y#3}[#1][#2]}
\newcommandx{\ZRel}[4][1=, 2=, 3=, 4=]{\mthrel[#4]{Z#3}[#1][#2]}

\newcommandx{\aRel}[4][1=, 2=, 3=, 4=]{\mthrel[#4]{a#3}[#1][#2]}
\newcommandx{\bRel}[4][1=, 2=, 3=, 4=]{\mthrel[#4]{b#3}[#1][#2]}
\newcommandx{\cRel}[4][1=, 2=, 3=, 4=]{\mthrel[#4]{c#3}[#1][#2]}
\newcommandx{\dRel}[4][1=, 2=, 3=, 4=]{\mthrel[#4]{d#3}[#1][#2]}
\newcommandx{\eRel}[4][1=, 2=, 3=, 4=]{\mthrel[#4]{e#3}[#1][#2]}
\newcommandx{\fRel}[4][1=, 2=, 3=, 4=]{\mthrel[#4]{f#3}[#1][#2]}
\newcommandx{\gRel}[4][1=, 2=, 3=, 4=]{\mthrel[#4]{g#3}[#1][#2]}
\newcommandx{\hRel}[4][1=, 2=, 3=, 4=]{\mthrel[#4]{h#3}[#1][#2]}
\newcommandx{\iRel}[4][1=, 2=, 3=, 4=]{\mthrel[#4]{i#3}[#1][#2]}
\newcommandx{\jRel}[4][1=, 2=, 3=, 4=]{\mthrel[#4]{j#3}[#1][#2]}
\newcommandx{\kRel}[4][1=, 2=, 3=, 4=]{\mthrel[#4]{k#3}[#1][#2]}
\newcommandx{\lRel}[4][1=, 2=, 3=, 4=]{\mthrel[#4]{l#3}[#1][#2]}
\newcommandx{\mRel}[4][1=, 2=, 3=, 4=]{\mthrel[#4]{m#3}[#1][#2]}
\newcommandx{\nRel}[4][1=, 2=, 3=, 4=]{\mthrel[#4]{n#3}[#1][#2]}
\newcommandx{\oRel}[4][1=, 2=, 3=, 4=]{\mthrel[#4]{o#3}[#1][#2]}
\newcommandx{\pRel}[4][1=, 2=, 3=, 4=]{\mthrel[#4]{p#3}[#1][#2]}
\newcommandx{\qRel}[4][1=, 2=, 3=, 4=]{\mthrel[#4]{q#3}[#1][#2]}
\newcommandx{\rRel}[4][1=, 2=, 3=, 4=]{\mthrel[#4]{r#3}[#1][#2]}
\newcommandx{\sRel}[4][1=, 2=, 3=, 4=]{\mthrel[#4]{s#3}[#1][#2]}
\newcommandx{\tRel}[4][1=, 2=, 3=, 4=]{\mthrel[#4]{t#3}[#1][#2]}
\newcommandx{\uRel}[4][1=, 2=, 3=, 4=]{\mthrel[#4]{u#3}[#1][#2]}
\newcommandx{\vRel}[4][1=, 2=, 3=, 4=]{\mthrel[#4]{v#3}[#1][#2]}
\newcommandx{\wRel}[4][1=, 2=, 3=, 4=]{\mthrel[#4]{w#3}[#1][#2]}
\newcommandx{\xRel}[4][1=, 2=, 3=, 4=]{\mthrel[#4]{x#3}[#1][#2]}
\newcommandx{\yRel}[4][1=, 2=, 3=, 4=]{\mthrel[#4]{y#3}[#1][#2]}
\newcommandx{\zRel}[4][1=, 2=, 3=, 4=]{\mthrel[#4]{z#3}[#1][#2]}

\newcommandx{\ASym}[4][1=, 2=, 3=, 4=]{\mthsym[#4]{A#3}[#1][#2]}
\newcommandx{\BSym}[4][1=, 2=, 3=, 4=]{\mthsym[#4]{B#3}[#1][#2]}
\newcommandx{\CSym}[4][1=, 2=, 3=, 4=]{\mthsym[#4]{C#3}[#1][#2]}
\newcommandx{\DSym}[4][1=, 2=, 3=, 4=]{\mthsym[#4]{D#3}[#1][#2]}
\newcommandx{\ESym}[4][1=, 2=, 3=, 4=]{\mthsym[#4]{E#3}[#1][#2]}
\newcommandx{\FSym}[4][1=, 2=, 3=, 4=]{\mthsym[#4]{F#3}[#1][#2]}
\newcommandx{\GSym}[4][1=, 2=, 3=, 4=]{\mthsym[#4]{G#3}[#1][#2]}
\newcommandx{\HSym}[4][1=, 2=, 3=, 4=]{\mthsym[#4]{H#3}[#1][#2]}
\newcommandx{\ISym}[4][1=, 2=, 3=, 4=]{\mthsym[#4]{I#3}[#1][#2]}
\newcommandx{\JSym}[4][1=, 2=, 3=, 4=]{\mthsym[#4]{J#3}[#1][#2]}
\newcommandx{\KSym}[4][1=, 2=, 3=, 4=]{\mthsym[#4]{K#3}[#1][#2]}
\newcommandx{\LSym}[4][1=, 2=, 3=, 4=]{\mthsym[#4]{L#3}[#1][#2]}
\newcommandx{\MSym}[4][1=, 2=, 3=, 4=]{\mthsym[#4]{M#3}[#1][#2]}
\newcommandx{\NSym}[4][1=, 2=, 3=, 4=]{\mthsym[#4]{N#3}[#1][#2]}
\newcommandx{\OSym}[4][1=, 2=, 3=, 4=]{\mthsym[#4]{O#3}[#1][#2]}
\newcommandx{\PSym}[4][1=, 2=, 3=, 4=]{\mthsym[#4]{P#3}[#1][#2]}
\newcommandx{\QSym}[4][1=, 2=, 3=, 4=]{\mthsym[#4]{Q#3}[#1][#2]}
\newcommandx{\RSym}[4][1=, 2=, 3=, 4=]{\mthsym[#4]{R#3}[#1][#2]}
\newcommandx{\SSym}[4][1=, 2=, 3=, 4=]{\mthsym[#4]{S#3}[#1][#2]}
\newcommandx{\TSym}[4][1=, 2=, 3=, 4=]{\mthsym[#4]{T#3}[#1][#2]}
\newcommandx{\USym}[4][1=, 2=, 3=, 4=]{\mthsym[#4]{U#3}[#1][#2]}
\newcommandx{\VSym}[4][1=, 2=, 3=, 4=]{\mthsym[#4]{V#3}[#1][#2]}
\newcommandx{\WSym}[4][1=, 2=, 3=, 4=]{\mthsym[#4]{W#3}[#1][#2]}
\newcommandx{\XSym}[4][1=, 2=, 3=, 4=]{\mthsym[#4]{X#3}[#1][#2]}
\newcommandx{\YSym}[4][1=, 2=, 3=, 4=]{\mthsym[#4]{Y#3}[#1][#2]}
\newcommandx{\ZSym}[4][1=, 2=, 3=, 4=]{\mthsym[#4]{Z#3}[#1][#2]}

\newcommandx{\aSym}[4][1=, 2=, 3=, 4=]{\mthsym[#4]{a#3}[#1][#2]}
\newcommandx{\bSym}[4][1=, 2=, 3=, 4=]{\mthsym[#4]{b#3}[#1][#2]}
\newcommandx{\cSym}[4][1=, 2=, 3=, 4=]{\mthsym[#4]{c#3}[#1][#2]}
\newcommandx{\dSym}[4][1=, 2=, 3=, 4=]{\mthsym[#4]{d#3}[#1][#2]}
\newcommandx{\eSym}[4][1=, 2=, 3=, 4=]{\mthsym[#4]{e#3}[#1][#2]}
\newcommandx{\fSym}[4][1=, 2=, 3=, 4=]{\mthsym[#4]{f#3}[#1][#2]}
\newcommandx{\gSym}[4][1=, 2=, 3=, 4=]{\mthsym[#4]{g#3}[#1][#2]}
\newcommandx{\hSym}[4][1=, 2=, 3=, 4=]{\mthsym[#4]{h#3}[#1][#2]}
\newcommandx{\iSym}[4][1=, 2=, 3=, 4=]{\mthsym[#4]{i#3}[#1][#2]}
\newcommandx{\jSym}[4][1=, 2=, 3=, 4=]{\mthsym[#4]{j#3}[#1][#2]}
\newcommandx{\kSym}[4][1=, 2=, 3=, 4=]{\mthsym[#4]{k#3}[#1][#2]}
\newcommandx{\lSym}[4][1=, 2=, 3=, 4=]{\mthsym[#4]{l#3}[#1][#2]}
\newcommandx{\mSym}[4][1=, 2=, 3=, 4=]{\mthsym[#4]{m#3}[#1][#2]}
\newcommandx{\nSym}[4][1=, 2=, 3=, 4=]{\mthsym[#4]{n#3}[#1][#2]}
\newcommandx{\oSym}[4][1=, 2=, 3=, 4=]{\mthsym[#4]{o#3}[#1][#2]}
\newcommandx{\pSym}[4][1=, 2=, 3=, 4=]{\mthsym[#4]{p#3}[#1][#2]}
\newcommandx{\qSym}[4][1=, 2=, 3=, 4=]{\mthsym[#4]{q#3}[#1][#2]}
\newcommandx{\rSym}[4][1=, 2=, 3=, 4=]{\mthsym[#4]{r#3}[#1][#2]}
\newcommandx{\sSym}[4][1=, 2=, 3=, 4=]{\mthsym[#4]{s#3}[#1][#2]}
\newcommandx{\tSym}[4][1=, 2=, 3=, 4=]{\mthsym[#4]{t#3}[#1][#2]}
\newcommandx{\uSym}[4][1=, 2=, 3=, 4=]{\mthsym[#4]{u#3}[#1][#2]}
\newcommandx{\vSym}[4][1=, 2=, 3=, 4=]{\mthsym[#4]{v#3}[#1][#2]}
\newcommandx{\wSym}[4][1=, 2=, 3=, 4=]{\mthsym[#4]{w#3}[#1][#2]}
\newcommandx{\xSym}[4][1=, 2=, 3=, 4=]{\mthsym[#4]{x#3}[#1][#2]}
\newcommandx{\ySym}[4][1=, 2=, 3=, 4=]{\mthsym[#4]{y#3}[#1][#2]}
\newcommandx{\zSym}[4][1=, 2=, 3=, 4=]{\mthsym[#4]{z#3}[#1][#2]}

\newcommandx{\AElm}[4][1=, 2=, 3=, 4=]{\mthelm[#4]{A#3}[#1][#2]}
\newcommandx{\BElm}[4][1=, 2=, 3=, 4=]{\mthelm[#4]{B#3}[#1][#2]}
\newcommandx{\CElm}[4][1=, 2=, 3=, 4=]{\mthelm[#4]{C#3}[#1][#2]}
\newcommandx{\DElm}[4][1=, 2=, 3=, 4=]{\mthelm[#4]{D#3}[#1][#2]}
\newcommandx{\EElm}[4][1=, 2=, 3=, 4=]{\mthelm[#4]{E#3}[#1][#2]}
\newcommandx{\FElm}[4][1=, 2=, 3=, 4=]{\mthelm[#4]{F#3}[#1][#2]}
\newcommandx{\GElm}[4][1=, 2=, 3=, 4=]{\mthelm[#4]{G#3}[#1][#2]}
\newcommandx{\HElm}[4][1=, 2=, 3=, 4=]{\mthelm[#4]{H#3}[#1][#2]}
\newcommandx{\IElm}[4][1=, 2=, 3=, 4=]{\mthelm[#4]{I#3}[#1][#2]}
\newcommandx{\JElm}[4][1=, 2=, 3=, 4=]{\mthelm[#4]{J#3}[#1][#2]}
\newcommandx{\KElm}[4][1=, 2=, 3=, 4=]{\mthelm[#4]{K#3}[#1][#2]}
\newcommandx{\LElm}[4][1=, 2=, 3=, 4=]{\mthelm[#4]{L#3}[#1][#2]}
\newcommandx{\MElm}[4][1=, 2=, 3=, 4=]{\mthelm[#4]{M#3}[#1][#2]}
\newcommandx{\NElm}[4][1=, 2=, 3=, 4=]{\mthelm[#4]{N#3}[#1][#2]}
\newcommandx{\OElm}[4][1=, 2=, 3=, 4=]{\mthelm[#4]{O#3}[#1][#2]}
\newcommandx{\PElm}[4][1=, 2=, 3=, 4=]{\mthelm[#4]{P#3}[#1][#2]}
\newcommandx{\QElm}[4][1=, 2=, 3=, 4=]{\mthelm[#4]{Q#3}[#1][#2]}
\newcommandx{\RElm}[4][1=, 2=, 3=, 4=]{\mthelm[#4]{R#3}[#1][#2]}
\newcommandx{\SElm}[4][1=, 2=, 3=, 4=]{\mthelm[#4]{S#3}[#1][#2]}
\newcommandx{\TElm}[4][1=, 2=, 3=, 4=]{\mthelm[#4]{T#3}[#1][#2]}
\newcommandx{\UElm}[4][1=, 2=, 3=, 4=]{\mthelm[#4]{U#3}[#1][#2]}
\newcommandx{\VElm}[4][1=, 2=, 3=, 4=]{\mthelm[#4]{V#3}[#1][#2]}
\newcommandx{\WElm}[4][1=, 2=, 3=, 4=]{\mthelm[#4]{W#3}[#1][#2]}
\newcommandx{\XElm}[4][1=, 2=, 3=, 4=]{\mthelm[#4]{X#3}[#1][#2]}
\newcommandx{\YElm}[4][1=, 2=, 3=, 4=]{\mthelm[#4]{Y#3}[#1][#2]}
\newcommandx{\ZElm}[4][1=, 2=, 3=, 4=]{\mthelm[#4]{Z#3}[#1][#2]}

\newcommandx{\aElm}[4][1=, 2=, 3=, 4=]{\mthelm[#4]{a#3}[#1][#2]}
\newcommandx{\bElm}[4][1=, 2=, 3=, 4=]{\mthelm[#4]{b#3}[#1][#2]}
\newcommandx{\cElm}[4][1=, 2=, 3=, 4=]{\mthelm[#4]{c#3}[#1][#2]}
\newcommandx{\dElm}[4][1=, 2=, 3=, 4=]{\mthelm[#4]{d#3}[#1][#2]}
\newcommandx{\eElm}[4][1=, 2=, 3=, 4=]{\mthelm[#4]{e#3}[#1][#2]}
\newcommandx{\fElm}[4][1=, 2=, 3=, 4=]{\mthelm[#4]{f#3}[#1][#2]}
\newcommandx{\gElm}[4][1=, 2=, 3=, 4=]{\mthelm[#4]{g#3}[#1][#2]}
\newcommandx{\hElm}[4][1=, 2=, 3=, 4=]{\mthelm[#4]{h#3}[#1][#2]}
\newcommandx{\iElm}[4][1=, 2=, 3=, 4=]{\mthelm[#4]{i#3}[#1][#2]}
\newcommandx{\jElm}[4][1=, 2=, 3=, 4=]{\mthelm[#4]{j#3}[#1][#2]}
\newcommandx{\kElm}[4][1=, 2=, 3=, 4=]{\mthelm[#4]{k#3}[#1][#2]}
\newcommandx{\lElm}[4][1=, 2=, 3=, 4=]{\mthelm[#4]{l#3}[#1][#2]}
\newcommandx{\mElm}[4][1=, 2=, 3=, 4=]{\mthelm[#4]{m#3}[#1][#2]}
\newcommandx{\nElm}[4][1=, 2=, 3=, 4=]{\mthelm[#4]{n#3}[#1][#2]}
\newcommandx{\oElm}[4][1=, 2=, 3=, 4=]{\mthelm[#4]{o#3}[#1][#2]}
\newcommandx{\pElm}[4][1=, 2=, 3=, 4=]{\mthelm[#4]{p#3}[#1][#2]}
\newcommandx{\qElm}[4][1=, 2=, 3=, 4=]{\mthelm[#4]{q#3}[#1][#2]}
\newcommandx{\rElm}[4][1=, 2=, 3=, 4=]{\mthelm[#4]{r#3}[#1][#2]}
\newcommandx{\sElm}[4][1=, 2=, 3=, 4=]{\mthelm[#4]{s#3}[#1][#2]}
\newcommandx{\tElm}[4][1=, 2=, 3=, 4=]{\mthelm[#4]{t#3}[#1][#2]}
\newcommandx{\uElm}[4][1=, 2=, 3=, 4=]{\mthelm[#4]{u#3}[#1][#2]}
\newcommandx{\vElm}[4][1=, 2=, 3=, 4=]{\mthelm[#4]{v#3}[#1][#2]}
\newcommandx{\wElm}[4][1=, 2=, 3=, 4=]{\mthelm[#4]{w#3}[#1][#2]}
\newcommandx{\xElm}[4][1=, 2=, 3=, 4=]{\mthelm[#4]{x#3}[#1][#2]}
\newcommandx{\yElm}[4][1=, 2=, 3=, 4=]{\mthelm[#4]{y#3}[#1][#2]}
\newcommandx{\zElm}[4][1=, 2=, 3=, 4=]{\mthelm[#4]{z#3}[#1][#2]}

\newcommand{\adhoc}
	{\txtabr{ad hoc}\xspace}

\newcommand{\ie}
	{\txtabr{i.e.}\xspace}

\newcommand{\wrt}
	{\txtabr{w.r.t.}\xspace}

\newcommand{\Wlogx}
	{\txtabr{W.l.o.g.}\xspace}

\newcommand{\defeq}
	{\ensuremath{\triangleq}}

\newcommand{\fst}
	{\mthargfun{fst}}
\newcommand{\lst}
	{\mthargfun{lst}}

\newcommand{\dual}[1]
	{\mthempty{\overline{#1}}}

\newcommand{\tuple}[1]
	{\ensuremath{\!\argint{\langle}{#1}{\rangle}}}

\newcommand{\tupleb}[2]
	{\tuple{\argb{#1}{#2}}}
\newcommand{\tuplec}[3]
	{\tuple{\argc{#1}{#2}{#3}}}
\newcommand{\tupled}[4]
	{\tuple{\argd{#1}{#2}{#3}{#4}}}
\newcommand{\tuplee}[5]
	{\tuple{\arge{#1}{#2}{#3}{#4}{#5}}}
\newcommand{\tuplef}[6]
	{\tuple{\argf{#1}{#2}{#3}{#4}{#5}{#6}}}
\newcommand{\tupleg}[7]
	{\tuple{\argg{#1}{#2}{#3}{#4}{#5}{#6}{#7}}}
\newcommand{\tupleh}[8]
	{\tuple{\argh{#1}{#2}{#3}{#4}{#5}{#6}{#7}{#8}}}
\newcommand{\tuplei}[9]
	{\tuple{\argi{#1}{#2}{#3}{#4}{#5}{#6}{#7}{#8}{#9}}}

\newcommand{\tuplej}[9]
	{%
	\def\defarga{#1}%
	\def\defargb{#2}%
	\def\defargc{#3}%
	\def\defargd{#4}%
	\def\defarge{#5}%
	\def\defargf{#6}%
	\def\defargg{#7}%
	\def\defargh{#8}%
	\def\defargi{#9}%
	\tupleauxj%
	}
\newcommand{\tuplek}[9]
	{%
	\def\defarga{#1}%
	\def\defargb{#2}%
	\def\defargc{#3}%
	\def\defargd{#4}%
	\def\defarge{#5}%
	\def\defargf{#6}%
	\def\defargg{#7}%
	\def\defargh{#8}%
	\def\defargi{#9}%
	\tupleauxk%
	}
\newcommand{\tuplel}[9]
	{%
	\def\defarga{#1}%
	\def\defargb{#2}%
	\def\defargc{#3}%
	\def\defargd{#4}%
	\def\defarge{#5}%
	\def\defargf{#6}%
	\def\defargg{#7}%
	\def\defargh{#8}%
	\def\defargi{#9}%
	\tupleauxl%
	}
\newcommand{\tuplem}[9]
	{%
	\def\defarga{#1}%
	\def\defargb{#2}%
	\def\defargc{#3}%
	\def\defargd{#4}%
	\def\defarge{#5}%
	\def\defargf{#6}%
	\def\defargg{#7}%
	\def\defargh{#8}%
	\def\defargi{#9}%
	\tupleauxm%
	}
\newcommand{\tuplen}[9]
	{%
	\def\defarga{#1}%
	\def\defargb{#2}%
	\def\defargc{#3}%
	\def\defargd{#4}%
	\def\defarge{#5}%
	\def\defargf{#6}%
	\def\defargg{#7}%
	\def\defargh{#8}%
	\def\defargi{#9}%
	\tupleauxn%
	}
\newcommand{\tupleo}[9]
	{%
	\def\defarga{#1}%
	\def\defargb{#2}%
	\def\defargc{#3}%
	\def\defargd{#4}%
	\def\defarge{#5}%
	\def\defargf{#6}%
	\def\defargg{#7}%
	\def\defargh{#8}%
	\def\defargi{#9}%
	\tupleauxo%
	}
\newcommand{\tuplep}[9]
	{%
	\def\defarga{#1}%
	\def\defargb{#2}%
	\def\defargc{#3}%
	\def\defargd{#4}%
	\def\defarge{#5}%
	\def\defargf{#6}%
	\def\defargg{#7}%
	\def\defargh{#8}%
	\def\defargi{#9}%
	\tupleauxp%
	}
\newcommand{\tupleq}[9]
	{%
	\def\defarga{#1}%
	\def\defargb{#2}%
	\def\defargc{#3}%
	\def\defargd{#4}%
	\def\defarge{#5}%
	\def\defargf{#6}%
	\def\defargg{#7}%
	\def\defargh{#8}%
	\def\defargi{#9}%
	\tupleauxq%
	}
\newcommand{\tupler}[9]
	{%
	\def\defarga{#1}%
	\def\defargb{#2}%
	\def\defargc{#3}%
	\def\defargd{#4}%
	\def\defarge{#5}%
	\def\defargf{#6}%
	\def\defargg{#7}%
	\def\defargh{#8}%
	\def\defargi{#9}%
	\tupleauxr%
	}

\newcommand{\tupleauxj}[1]
	{%
	\tuple{\argj{\defarga}{\defargb}{\defargc}{\defargd}{\defarge}{\defargf}%
		{\defargg}{\defargh}{\defargi}{#1}}%
	}
\newcommand{\tupleauxk}[2]
	{%
	\tuple{\argk{\defarga}{\defargb}{\defargc}{\defargd}{\defarge}{\defargf}%
		{\defargg}{\defargh}{\defargi}{#1}{#2}}%
	}
\newcommand{\tupleauxl}[3]
	{%
	\tuple{\argl{\defarga}{\defargb}{\defargc}{\defargd}{\defarge}{\defargf}%
		{\defargg}{\defargh}{\defargi}{#1}{#2}{#3}}%
	}
\newcommand{\tupleauxm}[4]
	{%
	\tuple{\argm{\defarga}{\defargb}{\defargc}{\defargd}{\defarge}{\defargf}%
		{\defargg}{\defargh}{\defargi}{#1}{#2}{#3}{#4}}%
	}
\newcommand{\tupleauxn}[5]
	{%
	\tuple{\argn{\defarga}{\defargb}{\defargc}{\defargd}{\defarge}{\defargf}%
		{\defargg}{\defargh}{\defargi}{#1}{#2}{#3}{#4}{#5}}%
	}
\newcommand{\tupleauxo}[6]
	{%
	\tuple{\argo{\defarga}{\defargb}{\defargc}{\defargd}{\defarge}{\defargf}%
		{\defargg}{\defargh}{\defargi}{#1}{#2}{#3}{#4}{#5}{#6}}%
	}
\newcommand{\tupleauxp}[7]
	{%
	\tuple{\argp{\defarga}{\defargb}{\defargc}{\defargd}{\defarge}{\defargf}%
		{\defargg}{\defargh}{\defargi}{#1}{#2}{#3}{#4}{#5}{#6}{#7}}%
	}
\newcommand{\tupleauxq}[8]
	{%
	\tuple{\argq{\defarga}{\defargb}{\defargc}{\defargd}{\defarge}{\defargf}%
		{\defargg}{\defargh}{\defargi}{#1}{#2}{#3}{#4}{#5}{#6}{#7}{#8}}%
	}
\newcommand{\tupleauxr}[9]
	{%
	\tuple{\argr{\defarga}{\defargb}{\defargc}{\defargd}{\defarge}{\defargf}%
		{\defargg}{\defargh}{\defargi}{#1}{#2}{#3}{#4}{#5}{#6}{#7}{#8}{#9}}%
	}

\newcommand{\tuplecx}[3]
	{%
	\def\defarga{#1}%
	\def\defargb{#2}%
	\def\defargc{#3}%
	\argsubsup{\tupleauxcx}%
	}
\newcommand{\tupledx}[4]
	{%
	\def\defarga{#1}%
	\def\defargb{#2}%
	\def\defargc{#3}%
	\def\defargd{#4}%
	\argsubsup{\tupleauxdx}%
	}
\newcommand{\tupleex}[5]
	{%
	\def\defarga{#1}%
	\def\defargb{#2}%
	\def\defargc{#3}%
	\def\defargd{#4}%
	\def\defarge{#5}%
	\argsubsup{\tupleauxex}%
	}
\newcommand{\tuplefx}[6]
	{%
	\def\defarga{#1}%
	\def\defargb{#2}%
	\def\defargc{#3}%
	\def\defargd{#4}%
	\def\defarge{#5}%
	\def\defargf{#6}%
	\argsubsup{\tupleauxfx}%
	}
\newcommand{\tuplegx}[7]
	{%
	\def\defarga{#1}%
	\def\defargb{#2}%
	\def\defargc{#3}%
	\def\defargd{#4}%
	\def\defarge{#5}%
	\def\defargf{#6}%
	\def\defargg{#7}%
	\argsubsup{\tupleauxgx}%
	}
\newcommand{\tuplehx}[8]
	{%
	\def\defarga{#1}%
	\def\defargb{#2}%
	\def\defargc{#3}%
	\def\defargd{#4}%
	\def\defarge{#5}%
	\def\defargf{#6}%
	\def\defargg{#7}%
	\def\defargh{#8}%
	\argsubsup{\tupleauxhx}%
	}
\newcommand{\tupleix}[9]
	{%
	\def\defarga{#1}%
	\def\defargb{#2}%
	\def\defargc{#3}%
	\def\defargd{#4}%
	\def\defarge{#5}%
	\def\defargf{#6}%
	\def\defargg{#7}%
	\def\defargh{#8}%
	\def\defargi{#9}%
	\argsubsup{\tupleauxix}%
	}

\newcommandx{\tupleauxbx}[2][1=, 2=]
	{%
	\tupleb
		{\argdef{#1}{\defarga[\argsubscript][\argsuperscript]}}
		{\argdef{#2}{\defargb[\argsubscript][\argsuperscript]}}%
	}
\newcommandx{\tupleauxcx}[3][1=, 2=, 3=]
	{%
	\tuplec
		{\argdef{#1}{\defarga[\argsubscript][\argsuperscript]}}
		{\argdef{#2}{\defargb[\argsubscript][\argsuperscript]}}
		{\argdef{#3}{\defargc[\argsubscript][\argsuperscript]}}%
	}
\newcommandx{\tupleauxdx}[4][1=, 2=, 3=, 4=]
	{%
	\tupled
		{\argdef{#1}{\defarga[\argsubscript][\argsuperscript]}}
		{\argdef{#2}{\defargb[\argsubscript][\argsuperscript]}}
		{\argdef{#3}{\defargc[\argsubscript][\argsuperscript]}}
		{\argdef{#4}{\defargd[\argsubscript][\argsuperscript]}}%
	}
\newcommandx{\tupleauxex}[5][1=, 2=, 3=, 4=, 5=]
	{%
	\tuplee
		{\argdef{#1}{\defarga[\argsubscript][\argsuperscript]}}
		{\argdef{#2}{\defargb[\argsubscript][\argsuperscript]}}
		{\argdef{#3}{\defargc[\argsubscript][\argsuperscript]}}
		{\argdef{#4}{\defargd[\argsubscript][\argsuperscript]}}
		{\argdef{#5}{\defarge[\argsubscript][\argsuperscript]}}%
	}
\newcommandx{\tupleauxfx}[6][1=, 2=, 3=, 4=, 5=, 6=]
	{%
	\tuplef
		{\argdef{#1}{\defarga[\argsubscript][\argsuperscript]}}
		{\argdef{#2}{\defargb[\argsubscript][\argsuperscript]}}
		{\argdef{#3}{\defargc[\argsubscript][\argsuperscript]}}
		{\argdef{#4}{\defargd[\argsubscript][\argsuperscript]}}
		{\argdef{#5}{\defarge[\argsubscript][\argsuperscript]}}
		{\argdef{#6}{\defargf[\argsubscript][\argsuperscript]}}%
	}
\newcommandx{\tupleauxgx}[7][1=, 2=, 3=, 4=, 5=, 6=, 7=]
	{%
	\tupleg
		{\argdef{#1}{\defarga[\argsubscript][\argsuperscript]}}
		{\argdef{#2}{\defargb[\argsubscript][\argsuperscript]}}
		{\argdef{#3}{\defargc[\argsubscript][\argsuperscript]}}
		{\argdef{#4}{\defargd[\argsubscript][\argsuperscript]}}
		{\argdef{#5}{\defarge[\argsubscript][\argsuperscript]}}
		{\argdef{#6}{\defargf[\argsubscript][\argsuperscript]}}
		{\argdef{#7}{\defargg[\argsubscript][\argsuperscript]}}%
	}
\newcommandx{\tupleauxhx}[8][1=, 2=, 3=, 4=, 5=, 6=, 7=, 8=]
	{%
	\tupleh
		{\argdef{#1}{\defarga[\argsubscript][\argsuperscript]}}
		{\argdef{#2}{\defargb[\argsubscript][\argsuperscript]}}
		{\argdef{#3}{\defargc[\argsubscript][\argsuperscript]}}
		{\argdef{#4}{\defargd[\argsubscript][\argsuperscript]}}
		{\argdef{#5}{\defarge[\argsubscript][\argsuperscript]}}
		{\argdef{#6}{\defargf[\argsubscript][\argsuperscript]}}
		{\argdef{#7}{\defargg[\argsubscript][\argsuperscript]}}
		{\argdef{#8}{\defargh[\argsubscript][\argsuperscript]}}%
	}
\newcommandx{\tupleauxix}[9][1=, 2=, 3=, 4=, 5=, 6=, 7=, 8=, 9=]
	{%
	\tuplei
		{\argdef{#1}{\defarga[\argsubscript][\argsuperscript]}}
		{\argdef{#2}{\defargb[\argsubscript][\argsuperscript]}}
		{\argdef{#3}{\defargc[\argsubscript][\argsuperscript]}}
		{\argdef{#4}{\defargd[\argsubscript][\argsuperscript]}}
		{\argdef{#5}{\defarge[\argsubscript][\argsuperscript]}}
		{\argdef{#6}{\defargf[\argsubscript][\argsuperscript]}}
		{\argdef{#7}{\defargg[\argsubscript][\argsuperscript]}}
		{\argdef{#8}{\defargh[\argsubscript][\argsuperscript]}}
		{\argdef{#9}{\defargi[\argsubscript][\argsuperscript]}}%
	}

\newcommand{\tuplejx}[9]
	{%
	\def\tuplearga{#1}%
	\def\tupleargb{#2}%
	\def\tupleargc{#3}%
	\def\tupleargd{#4}%
	\def\tuplearge{#5}%
	\def\tupleargf{#6}%
	\def\tupleargg{#7}%
	\def\tupleargh{#8}%
	\def\tupleargi{#9}%
	\argsubsup{\tupleauxjx}%
	}

\newcommand{\tupleauxjx}[1]
	{%
	\def\tupleargj{#1}%
	\argsubsup{\tupleauxxjx}%
	}

\newcommandx{\tupleauxxjx}[9][1=, 2=, 3=, 4=, 5=, 6=, 7=, 8=, 9=]
	{%
	\def\optarga{#1}%
	\def\optargb{#2}%
	\def\optargc{#3}%
	\def\optargd{#4}%
	\def\optarge{#5}%
	\def\optargf{#6}%
	\def\optargg{#7}%
	\def\optargh{#8}%
	\def\optargi{#9}%
	\tupleauxxxjx%
	}
\newcommandx{\tupleauxxkx}[9][1=, 2=, 3=, 4=, 5=, 6=, 7=, 8=, 9=]
	{%
	\def\optarga{#1}%
	\def\optargb{#2}%
	\def\optargc{#3}%
	\def\optargd{#4}%
	\def\optarge{#5}%
	\def\optargf{#6}%
	\def\optargg{#7}%
	\def\optargh{#8}%
	\def\optargi{#9}%
	\tupleauxxxkx%
	}
\newcommandx{\tupleauxxlx}[9][1=, 2=, 3=, 4=, 5=, 6=, 7=, 8=, 9=]
	{%
	\def\optarga{#1}%
	\def\optargb{#2}%
	\def\optargc{#3}%
	\def\optargd{#4}%
	\def\optarge{#5}%
	\def\optargf{#6}%
	\def\optargg{#7}%
	\def\optargh{#8}%
	\def\optargi{#9}%
	\tupleauxxxlx%
	}
\newcommandx{\tupleauxxmx}[9][1=, 2=, 3=, 4=, 5=, 6=, 7=, 8=, 9=]
	{%
	\def\optarga{#1}%
	\def\optargb{#2}%
	\def\optargc{#3}%
	\def\optargd{#4}%
	\def\optarge{#5}%
	\def\optargf{#6}%
	\def\optargg{#7}%
	\def\optargh{#8}%
	\def\optargi{#9}%
	\tupleauxxxmx%
	}
\newcommandx{\tupleauxxnx}[9][1=, 2=, 3=, 4=, 5=, 6=, 7=, 8=, 9=]
	{%
	\def\optarga{#1}%
	\def\optargb{#2}%
	\def\optargc{#3}%
	\def\optargd{#4}%
	\def\optarge{#5}%
	\def\optargf{#6}%
	\def\optargg{#7}%
	\def\optargh{#8}%
	\def\optargi{#9}%
	\tupleauxxxnx%
	}
\newcommandx{\tupleauxxox}[9][1=, 2=, 3=, 4=, 5=, 6=, 7=, 8=, 9=]
	{%
	\def\optarga{#1}%
	\def\optargb{#2}%
	\def\optargc{#3}%
	\def\optargd{#4}%
	\def\optarge{#5}%
	\def\optargf{#6}%
	\def\optargg{#7}%
	\def\optargh{#8}%
	\def\optargi{#9}%
	\tupleauxxxox%
	}
\newcommandx{\tupleauxxpx}[9][1=, 2=, 3=, 4=, 5=, 6=, 7=, 8=, 9=]
	{%
	\def\optarga{#1}%
	\def\optargb{#2}%
	\def\optargc{#3}%
	\def\optargd{#4}%
	\def\optarge{#5}%
	\def\optargf{#6}%
	\def\optargg{#7}%
	\def\optargh{#8}%
	\def\optargi{#9}%
	\tupleauxxxpx%
	}
\newcommandx{\tupleauxxqx}[9][1=, 2=, 3=, 4=, 5=, 6=, 7=, 8=, 9=]
	{%
	\def\optarga{#1}%
	\def\optargb{#2}%
	\def\optargc{#3}%
	\def\optargd{#4}%
	\def\optarge{#5}%
	\def\optargf{#6}%
	\def\optargg{#7}%
	\def\optargh{#8}%
	\def\optargi{#9}%
	\tupleauxxxqx%
	}
\newcommandx{\tupleauxxrx}[9][1=, 2=, 3=, 4=, 5=, 6=, 7=, 8=, 9=]
	{%
	\def\optarga{#1}%
	\def\optargb{#2}%
	\def\optargc{#3}%
	\def\optargd{#4}%
	\def\optarge{#5}%
	\def\optargf{#6}%
	\def\optargg{#7}%
	\def\optargh{#8}%
	\def\optargi{#9}%
	\tupleauxxxrx%
	}

\newcommandx{\tupleauxxxjx}[1][1=]
	{%
	\tuplej
		{\argdef{\optarga}{\tuplearga[\argsubscript][\argsuperscript]}}
		{\argdef{\optargb}{\tupleargb[\argsubscript][\argsuperscript]}}
		{\argdef{\optargc}{\tupleargc[\argsubscript][\argsuperscript]}}
		{\argdef{\optargd}{\tupleargd[\argsubscript][\argsuperscript]}}
		{\argdef{\optarge}{\tuplearge[\argsubscript][\argsuperscript]}}
		{\argdef{\optargf}{\tupleargf[\argsubscript][\argsuperscript]}}
		{\argdef{\optargg}{\tupleargg[\argsubscript][\argsuperscript]}}
		{\argdef{\optargh}{\tupleargh[\argsubscript][\argsuperscript]}}
		{\argdef{\optargi}{\tupleargi[\argsubscript][\argsuperscript]}}
		{\argdef{#1}{\tupleargj[\argsubscript][\argsuperscript]}}%
	}
\newcommandx{\tupleauxxxkx}[2][1=, 2=]
	{%
	\tuplek
		{\argdef{\optarga}{\tuplearga[\argsubscript][\argsuperscript]}}
		{\argdef{\optargb}{\tupleargb[\argsubscript][\argsuperscript]}}
		{\argdef{\optargc}{\tupleargc[\argsubscript][\argsuperscript]}}
		{\argdef{\optargd}{\tupleargd[\argsubscript][\argsuperscript]}}
		{\argdef{\optarge}{\tuplearge[\argsubscript][\argsuperscript]}}
		{\argdef{\optargf}{\tupleargf[\argsubscript][\argsuperscript]}}
		{\argdef{\optargg}{\tupleargg[\argsubscript][\argsuperscript]}}
		{\argdef{\optargh}{\tupleargh[\argsubscript][\argsuperscript]}}
		{\argdef{\optargi}{\tupleargi[\argsubscript][\argsuperscript]}}
		{\argdef{#1}{\tupleargj[\argsubscript][\argsuperscript]}}
		{\argdef{#2}{\tupleargk[\argsubscript][\argsuperscript]}}
	}
\newcommandx{\tupleauxxxlx}[3][1=, 2=, 3=]
	{%
	\tuplel
		{\argdef{\optarga}{\tuplearga[\argsubscript][\argsuperscript]}}
		{\argdef{\optargb}{\tupleargb[\argsubscript][\argsuperscript]}}
		{\argdef{\optargc}{\tupleargc[\argsubscript][\argsuperscript]}}
		{\argdef{\optargd}{\tupleargd[\argsubscript][\argsuperscript]}}
		{\argdef{\optarge}{\tuplearge[\argsubscript][\argsuperscript]}}
		{\argdef{\optargf}{\tupleargf[\argsubscript][\argsuperscript]}}
		{\argdef{\optargg}{\tupleargg[\argsubscript][\argsuperscript]}}
		{\argdef{\optargh}{\tupleargh[\argsubscript][\argsuperscript]}}
		{\argdef{\optargi}{\tupleargi[\argsubscript][\argsuperscript]}}
		{\argdef{#1}{\tupleargj[\argsubscript][\argsuperscript]}}
		{\argdef{#2}{\tupleargk[\argsubscript][\argsuperscript]}}
		{\argdef{#3}{\tupleargl[\argsubscript][\argsuperscript]}}
	}
\newcommandx{\tupleauxxxmx}[4][1=, 2=, 3=, 4=]
	{%
	\tuplem
		{\argdef{\optarga}{\tuplearga[\argsubscript][\argsuperscript]}}
		{\argdef{\optargb}{\tupleargb[\argsubscript][\argsuperscript]}}
		{\argdef{\optargc}{\tupleargc[\argsubscript][\argsuperscript]}}
		{\argdef{\optargd}{\tupleargd[\argsubscript][\argsuperscript]}}
		{\argdef{\optarge}{\tuplearge[\argsubscript][\argsuperscript]}}
		{\argdef{\optargf}{\tupleargf[\argsubscript][\argsuperscript]}}
		{\argdef{\optargg}{\tupleargg[\argsubscript][\argsuperscript]}}
		{\argdef{\optargh}{\tupleargh[\argsubscript][\argsuperscript]}}
		{\argdef{\optargi}{\tupleargi[\argsubscript][\argsuperscript]}}
		{\argdef{#1}{\tupleargj[\argsubscript][\argsuperscript]}}
		{\argdef{#2}{\tupleargk[\argsubscript][\argsuperscript]}}
		{\argdef{#3}{\tupleargl[\argsubscript][\argsuperscript]}}
		{\argdef{#4}{\tupleargm[\argsubscript][\argsuperscript]}}
	}
\newcommandx{\tupleauxxxnx}[5][1=, 2=, 3=, 4=, 5=]
	{%
	\tuplen
		{\argdef{\optarga}{\tuplearga[\argsubscript][\argsuperscript]}}
		{\argdef{\optargb}{\tupleargb[\argsubscript][\argsuperscript]}}
		{\argdef{\optargc}{\tupleargc[\argsubscript][\argsuperscript]}}
		{\argdef{\optargd}{\tupleargd[\argsubscript][\argsuperscript]}}
		{\argdef{\optarge}{\tuplearge[\argsubscript][\argsuperscript]}}
		{\argdef{\optargf}{\tupleargf[\argsubscript][\argsuperscript]}}
		{\argdef{\optargg}{\tupleargg[\argsubscript][\argsuperscript]}}
		{\argdef{\optargh}{\tupleargh[\argsubscript][\argsuperscript]}}
		{\argdef{\optargi}{\tupleargi[\argsubscript][\argsuperscript]}}
		{\argdef{#1}{\tupleargj[\argsubscript][\argsuperscript]}}
		{\argdef{#2}{\tupleargk[\argsubscript][\argsuperscript]}}
		{\argdef{#3}{\tupleargl[\argsubscript][\argsuperscript]}}
		{\argdef{#4}{\tupleargm[\argsubscript][\argsuperscript]}}
		{\argdef{#5}{\tupleargn[\argsubscript][\argsuperscript]}}
	}
\newcommandx{\tupleauxxxox}[6][1=, 2=, 3=, 4=, 5=, 6=]
	{%
	\tupleo
		{\argdef{\optarga}{\tuplearga[\argsubscript][\argsuperscript]}}
		{\argdef{\optargb}{\tupleargb[\argsubscript][\argsuperscript]}}
		{\argdef{\optargc}{\tupleargc[\argsubscript][\argsuperscript]}}
		{\argdef{\optargd}{\tupleargd[\argsubscript][\argsuperscript]}}
		{\argdef{\optarge}{\tuplearge[\argsubscript][\argsuperscript]}}
		{\argdef{\optargf}{\tupleargf[\argsubscript][\argsuperscript]}}
		{\argdef{\optargg}{\tupleargg[\argsubscript][\argsuperscript]}}
		{\argdef{\optargh}{\tupleargh[\argsubscript][\argsuperscript]}}
		{\argdef{\optargi}{\tupleargi[\argsubscript][\argsuperscript]}}
		{\argdef{#1}{\tupleargj[\argsubscript][\argsuperscript]}}
		{\argdef{#2}{\tupleargk[\argsubscript][\argsuperscript]}}
		{\argdef{#3}{\tupleargl[\argsubscript][\argsuperscript]}}
		{\argdef{#4}{\tupleargm[\argsubscript][\argsuperscript]}}
		{\argdef{#5}{\tupleargn[\argsubscript][\argsuperscript]}}
		{\argdef{#6}{\tupleargo[\argsubscript][\argsuperscript]}}
	}
\newcommandx{\tupleauxxxpx}[7][1=, 2=, 3=, 4=, 5=, 6=, 7=]
	{%
	\tuplep
		{\argdef{\optarga}{\tuplearga[\argsubscript][\argsuperscript]}}
		{\argdef{\optargb}{\tupleargb[\argsubscript][\argsuperscript]}}
		{\argdef{\optargc}{\tupleargc[\argsubscript][\argsuperscript]}}
		{\argdef{\optargd}{\tupleargd[\argsubscript][\argsuperscript]}}
		{\argdef{\optarge}{\tuplearge[\argsubscript][\argsuperscript]}}
		{\argdef{\optargf}{\tupleargf[\argsubscript][\argsuperscript]}}
		{\argdef{\optargg}{\tupleargg[\argsubscript][\argsuperscript]}}
		{\argdef{\optargh}{\tupleargh[\argsubscript][\argsuperscript]}}
		{\argdef{\optargi}{\tupleargi[\argsubscript][\argsuperscript]}}
		{\argdef{#1}{\tupleargj[\argsubscript][\argsuperscript]}}
		{\argdef{#2}{\tupleargk[\argsubscript][\argsuperscript]}}
		{\argdef{#3}{\tupleargl[\argsubscript][\argsuperscript]}}
		{\argdef{#4}{\tupleargm[\argsubscript][\argsuperscript]}}
		{\argdef{#5}{\tupleargn[\argsubscript][\argsuperscript]}}
		{\argdef{#6}{\tupleargo[\argsubscript][\argsuperscript]}}
		{\argdef{#7}{\tupleargp[\argsubscript][\argsuperscript]}}
	}
\newcommandx{\tupleauxxxqx}[8][1=, 2=, 3=, 4=, 5=, 6=, 7=, 8=]
	{%
	\tupleq
		{\argdef{\optarga}{\tuplearga[\argsubscript][\argsuperscript]}}
		{\argdef{\optargb}{\tupleargb[\argsubscript][\argsuperscript]}}
		{\argdef{\optargc}{\tupleargc[\argsubscript][\argsuperscript]}}
		{\argdef{\optargd}{\tupleargd[\argsubscript][\argsuperscript]}}
		{\argdef{\optarge}{\tuplearge[\argsubscript][\argsuperscript]}}
		{\argdef{\optargf}{\tupleargf[\argsubscript][\argsuperscript]}}
		{\argdef{\optargg}{\tupleargg[\argsubscript][\argsuperscript]}}
		{\argdef{\optargh}{\tupleargh[\argsubscript][\argsuperscript]}}
		{\argdef{\optargi}{\tupleargi[\argsubscript][\argsuperscript]}}
		{\argdef{#1}{\tupleargj[\argsubscript][\argsuperscript]}}
		{\argdef{#2}{\tupleargk[\argsubscript][\argsuperscript]}}
		{\argdef{#3}{\tupleargl[\argsubscript][\argsuperscript]}}
		{\argdef{#4}{\tupleargm[\argsubscript][\argsuperscript]}}
		{\argdef{#5}{\tupleargn[\argsubscript][\argsuperscript]}}
		{\argdef{#6}{\tupleargo[\argsubscript][\argsuperscript]}}
		{\argdef{#7}{\tupleargp[\argsubscript][\argsuperscript]}}
		{\argdef{#8}{\tupleargq[\argsubscript][\argsuperscript]}}
	}
\newcommandx{\tupleauxxxrx}[9][1=, 2=, 3=, 4=, 5=, 6=, 7=, 8=, 9=]
	{%
	\tupler
		{\argdef{\optarga}{\tuplearga[\argsubscript][\argsuperscript]}}
		{\argdef{\optargb}{\tupleargb[\argsubscript][\argsuperscript]}}
		{\argdef{\optargc}{\tupleargc[\argsubscript][\argsuperscript]}}
		{\argdef{\optargd}{\tupleargd[\argsubscript][\argsuperscript]}}
		{\argdef{\optarge}{\tuplearge[\argsubscript][\argsuperscript]}}
		{\argdef{\optargf}{\tupleargf[\argsubscript][\argsuperscript]}}
		{\argdef{\optargg}{\tupleargg[\argsubscript][\argsuperscript]}}
		{\argdef{\optargh}{\tupleargh[\argsubscript][\argsuperscript]}}
		{\argdef{\optargi}{\tupleargi[\argsubscript][\argsuperscript]}}
		{\argdef{#1}{\tupleargj[\argsubscript][\argsuperscript]}}
		{\argdef{#2}{\tupleargk[\argsubscript][\argsuperscript]}}
		{\argdef{#3}{\tupleargl[\argsubscript][\argsuperscript]}}
		{\argdef{#4}{\tupleargm[\argsubscript][\argsuperscript]}}
		{\argdef{#5}{\tupleargn[\argsubscript][\argsuperscript]}}
		{\argdef{#6}{\tupleargo[\argsubscript][\argsuperscript]}}
		{\argdef{#7}{\tupleargp[\argsubscript][\argsuperscript]}}
		{\argdef{#8}{\tupleargq[\argsubscript][\argsuperscript]}}
		{\argdef{#9}{\tupleargr[\argsubscript][\argsuperscript]}}%
	}

\newcommand{\set}[2]
	{\ensuremath{\argint{\{}{\argext{#1}{\allowbreak:\allowbreak}{#2}}{\}}}}

\newcommand{\pow}[1]
	{\ensuremath{2^{#1}}}

\newcommand{\card}[1]
	{\mthempty{\argint{\vert}{#1}{\vert}}}

\newcommand{\rng}
	{\mthargfun{rng}}

\newcommandx{\pto}[2][1=, 2=]
	{\ensuremath{\rightharpoonup}}

\newcommandx{\cto}[2][1=, 2=]
	{\:\mthempty{\to}[#1][#2]\:}
\newcommandx{\cpto}[2][1=, 2=]
	{\:\mthempty{\pto}[#1][#2]\:}

\newcommand{\AOmicron}
	{\mthargset{O}}

\newcommand{\SetN}
	{\mthset[2]{N}}

\newcommand{\numcc}[2]
	{\mthempty{[\argb{#1}{#2}]}}
\newcommand{\numco}[2]
	{\mthempty{[\argb{#1}{#2}[\:\!}}

\newcommand{\argset}{Ar}
\newcommandx{\ArgSet}[3][1=, 2=, 3=]
	{\mthset{\argset#3}[#1][#2]}

\newcommand{\argsym}{a}
\newcommandx{\argSym}[3][1=, 2=, 3=]
	{\mthsym{\argsym#3}[#1][#2]}

\newcommand{\argelm}{a}
\newcommandx{\argElm}[3][1=, 2=, 3=]
	{\mthelm{\argelm#3}[#1][#2]}

\newcommand{\relset}{Rl}
\newcommandx{\RelSet}[3][1=, 2=, 3=]
	{\mthset{\relset#3}[#1][#2]}

\newcommand{\relsym}{r}
\newcommandx{\relSym}[3][1=, 2=, 3=]
	{\mthsym{\relsym#3}[#1][#2]}

\newcommand{\relelm}{r}
\newcommandx{\relElm}[3][1=, 2=, 3=]
	{\mthelm{\relelm#3}[#1][#2]}

\newcommand{\argfun}{ar}
\newcommandx{\argFun}[4][1=, 2=, 3=, 4=]
	{\mthargfun{\argfun#4}[#1][#2]{#3}}

\newcommand{\lansig}{LS}
\newcommandx{\LanSig}[5][1=, 2=, 3=, 4=, 5=]
	{\txtargname{\lansig#5{\small\argint{$[$}{#1}{$]$}}}[#2][#3]{#4}\xspace}

\newcommand{\lansigcls}{LS}
\newcommandx{\LanSigCls}[5][1=, 2=, 3=, 4=, 5=]
	{\mthset[#5]{\lansigcls#4\text{\txtname{\small\argint{$[$}{#1}{$]$}}}}[#2]%
	[#3]}

\newcommand{\domset}{Dm}
\newcommandx{\DomSet}[3][1=, 2=, 3=]
	{\mthset{\domset#3}[#1][#2]}

\newcommand{\domsym}{d}
\newcommandx{\domSym}[3][1=, 2=, 3=]
	{\mthsym{\domsym#3}[#1][#2]}

\newcommand{\domelm}{d}
\newcommandx{\domElm}[3][1=, 2=, 3=]
	{\mthelm{\domelm#3}[#1][#2]}

\newcommand{\relfun}{rl}
\newcommandx{\relFun}[4][1=, 2=, 3=, 4=]
	{\mthargfun{\relfun#4}[#1][#2]{#3}}

\newcommand{\relstr}{RS}
\newcommandx{\RelStr}[5][1=, 2=, 3=, 4=, 5=]
	{\txtargname{\relstr#5{\small\argint{$[$}{#1}{$]$}}}[#2][#3]{#4}\xspace}

\newcommand{\relstrcls}{RS}
\newcommandx{\RelStrCls}[5][1=, 2=, 3=, 4=, 5=]
	{\mthset[#5]{\relstrcls#4\text{\txtname{\small\argint{$[$}{#1}{$]$}}}}[#2]%
	[#3]}

\newcommandx{\ordFun}[3][1=, 2=, 3=]
	{\mthempty{\argint{\left\vert}{#3}{\right\vert}}[#1][#2]}

\newcommandx{\sizFun}[3][1=, 2=, 3=]
	{\mthempty{\argint{\left\Vert}{#3}{\right\Vert}}[#1][#2]}

\newcommand{\verset}{Vr}
\newcommandx{\VerSet}[3][1=, 2=, 3=]
	{\mthset{\verset#3}[#1][#2]}

\newcommand{\versym}{v}
\newcommandx{\verSym}[3][1=, 2=, 3=]
	{\mthsym{\versym#3}[#1][#2]}

\newcommand{\verelm}{v}
\newcommandx{\verElm}[3][1=, 2=, 3=]
	{\mthelm{\verelm#3}[#1][#2]}

\newcommand{\edgrel}{Ed}
\newcommandx{\EdgRel}[3][1=, 2=, 3=]
	{\mthrel{\edgrel#3}[#1][#2]}

\newcommand{\edgsym}{e}
\newcommandx{\edgSym}[3][1=, 2=, 3=]
	{\mthsym{\edgsym#3}[#1][#2]}

\newcommand{\edgelm}{e}
\newcommandx{\edgElm}[3][1=, 2=, 3=]
	{\mthelm{\edgelm#3}[#1][#2]}

\newcommand{\orgfun}{or}
\newcommandx{\orgFun}[4][1=, 2=, 3=, 4=]
	{\mthargfun{\orgfun#4}[#1][#2]{#3}}

\newcommand{\desfun}{ds}
\newcommandx{\desFun}[4][1=, 2=, 3=, 4=]
	{\mthargfun{\desfun#4}[#1][#2]{#3}}

\newcommand{\grp}{Gr}
\newcommandx{\Grp}[5][1=, 2=, 3=, 4=, 5=]
	{\txtargname{\grp#5{\small\argint{$[$}{#1}{$]$}}}[#2][#3]{#4}\xspace}

\newcommand{\grpcls}{Gr}
\newcommandx{\GrpCls}[5][1=, 2=, 3=, 4=, 5=]
	{\mthset[#5]{\grpcls#4\text{\small\txtname{\argint{$[$}{#1}{$]$}}}}[#2][#3]}

\newcommand{\pthset}{Pth}
\newcommandx{\PthSet}[3][1=, 2=, 3=]
	{\mthset{\pthset#3}[#1][#2]}

\newcommand{\pthsym}{\pi}
\newcommandx{\pthSym}[3][1=, 2=, 3=]
	{\mthsym{\pthsym#3}[#1][#2]}

\newcommand{\pthelm}{\pi}
\newcommandx{\pthElm}[3][1=, 2=, 3=]
	{\mthelm{\pthelm#3}[#1][#2]}

\newcommand{\apset}{AP}
\newcommandx{\APSet}[3][1=, 2=, 3=]
	{\mthset{\apset#3}[#1][#2]}

\newcommand{\apsym}{p}
\newcommandx{\apSym}[3][1=, 2=, 3=]
	{\mthsym{\apsym#3}[#1][#2]}

\newcommand{\apelm}{p}
\newcommandx{\apElm}[3][1=, 2=, 3=]
	{\mthelm{\apelm#3}[#1][#2]}

\newcommand{\apfun}{ap}
\newcommandx{\apFun}[4][1=, 2=, 3=, 4=]
	{\mthargfun{\apfun#4}[#1][#2]{#3}}

\newcommand{\labgrp}{L\grp}
\newcommandx{\LabGrp}[5][1=, 2=, 3=, 4=, 5=]
	{\txtargname{\labgrp#5{\small\argint{$[$}{#1}{$]$}}}[#2][#3]{#4}\xspace}

\newcommand{\labgrpcls}{L\grpcls}
\newcommandx{\LabGrpCls}[5][1=, 2=, 3=, 4=, 5=]
	{\mthset[#5]{\labgrpcls#4\text{\small\txtname{\argint{$[$}{#1}{$]$}}}}[#2]%
	[#3]}

\newcommand{\trcset}{Trc}
\newcommandx{\TrcSet}[3][1=, 2=, 3=]
	{\mthset{\trcset#3}[#1][#2]}

\newcommand{\trcsym}{\varrho}
\newcommandx{\trcSym}[3][1=, 2=, 3=]
	{\mthsym{\trcsym#3}[#1][#2]}

\newcommand{\trcelm}{\varrho}
\newcommandx{\trcElm}[3][1=, 2=, 3=]
	{\mthelm{\trcelm#3}[#1][#2]}

\newcommand{\colset}{Cl}
\newcommandx{\ColSet}[3][1=, 2=, 3=]
	{\mthset{\colset#3}[#1][#2]}

\newcommand{\colsym}{c}
\newcommandx{\colSym}[3][1=, 2=, 3=]
	{\mthsym{\colsym#3}[#1][#2]}

\newcommand{\colelm}{c}
\newcommandx{\colElm}[3][1=, 2=, 3=]
	{\mthelm{\colelm#3}[#1][#2]}

\newcommand{\colfun}{cl}
\newcommandx{\colFun}[4][1=, 2=, 3=, 4=]
	{\mthargfun{\colfun#4}[#1][#2]{#3}}

\newcommand{\colgrp}{C\grp}
\newcommandx{\ColGrp}[5][1=, 2=, 3=, 4=, 5=]
	{\txtargname{\colgrp#5{\small\argint{$[$}{#1}{$]$}}}[#2][#3]{#4}\xspace}

\newcommand{\colgrpcls}{C\grpcls}
\newcommandx{\ColGrpCls}[5][1=, 2=, 3=, 4=, 5=]
	{\mthset[#5]{\colgrpcls#4\text{\small\txtname{\argint{$[$}{#1}{$]$}}}}[#2]%
	[#3]}

\newcommand{\wghset}{Wg}
\newcommandx{\WghSet}[3][1=, 2=, 3=]
	{\mthset{\wghset#3}[#1][#2]}

\newcommand{\wghsym}{w}
\newcommandx{\wghSym}[3][1=, 2=, 3=]
	{\mthsym{\wghsym#3}[#1][#2]}

\newcommand{\wghelm}{w}
\newcommandx{\wghElm}[3][1=, 2=, 3=]
	{\mthelm{\wghelm#3}[#1][#2]}

\newcommand{\wghfun}{wg}
\newcommandx{\wghFun}[4][1=, 2=, 3=, 4=]
	{\mthargfun{\wghfun#4}[#1][#2]{#3}}

\newcommand{\wghgrp}{W\grp}
\newcommandx{\WghGrp}[5][1=, 2=, 3=, 4=, 5=]
	{\txtargname{\wghgrp#5{\small\argint{$[$}{#1}{$]$}}}[#2][#3]{#4}\xspace}

\newcommand{\wghgrpcls}{W\grpcls}
\newcommandx{\WghGrpCls}[5][1=, 2=, 3=, 4=, 5=]
	{\mthset[#5]{\wghgrpcls#4\text{\small\txtname{\argint{$[$}{#1}{$]$}}}}[#2]%
	[#3]}

\newcommand{\gamkin}{2PT}

\newcommand{\plrset}{Pl}
\newcommandx{\PlrSet}[3][1=, 2=, 3=]
	{\mthset{\plrset#3}[#1][#2]}

\newcommand{\plrsym}{p}
\newcommandx{\plrSym}[3][1=, 2=, 3=]
	{\mthsym{\plrsym#3}[#1][#2]}

\newcommand{\plrelm}{p}
\newcommandx{\plrElm}[3][1=, 2=, 3=]
	{\mthelm{\plrelm#3}[#1][#2]}

\newcommand{\agnset}{Ag}
\newcommandx{\AgnSet}[3][1=, 2=, 3=]
	{\mthset{\agnset#3}[#1][#2]}

\newcommand{\agnsym}{a}
\newcommandx{\agnSym}[3][1=, 2=, 3=]
	{\mthsym{\agnsym#3}[#1][#2]}

\newcommand{\agnelm}{a}
\newcommandx{\agnElm}[3][1=, 2=, 3=]
	{\mthelm{\agnelm#3}[#1][#2]}

\newcommand{\movset}{Mv}
\newcommandx{\MovSet}[3][1=, 2=, 3=]
	{\mthset{\movset#3}[#1][#2]}

\newcommand{\movrel}{Mv}
\newcommandx{\MovRel}[3][1=, 2=, 3=]
	{\mthrel{\movrel#3}[#1][#2]}

\newcommand{\movsym}{m}
\newcommandx{\movSym}[3][1=, 2=, 3=]
	{\mthsym{\movsym#3}[#1][#2]}

\newcommand{\movelm}{m}
\newcommandx{\movElm}[3][1=, 2=, 3=]
	{\mthelm{\movelm#3}[#1][#2]}

\newcommand{\actset}{Ac}
\newcommandx{\ActSet}[3][1=, 2=, 3=]
	{\mthset{\actset#3}[#1][#2]}

\newcommand{\actrel}{Ac}
\newcommandx{\ActRel}[3][1=, 2=, 3=]
	{\mthrel{\actrel#3}[#1][#2]}

\newcommand{\actsym}{c}
\newcommandx{\actSym}[3][1=, 2=, 3=]
	{\mthsym{\actsym#3}[#1][#2]}

\newcommand{\actelm}{c}
\newcommandx{\actElm}[3][1=, 2=, 3=]
	{\mthelm{\actelm#3}[#1][#2]}

\newcommand{\decset}{Dc}
\newcommandx{\DecSet}[3][1=, 2=, 3=]
	{\mthset{\decset#3}[#1][#2]}

\newcommand{\decsym}{\delta}
\newcommandx{\decSym}[4][1=, 2=, 3=, 4=]
	{\mthargfun{\decsym#4}[#1][#2]{#3}}

\newcommand{\decelm}{\delta}
\newcommandx{\decElm}[4][1=, 2=, 3=, 4=]
	{\mthargfun{\decelm#4}[#1][#2]{#3}}

\newcommand{\posset}{Ps}
\newcommandx{\PosSet}[3][1=, 2=, 3=]
	{\mthset{\posset#3}[#1][#2]}
\newcommand{\fpossub}{0}
\newcommandx{\FPosSet}[3][1=, 2=, 3=]
	{\mthset{\posset#3}[\fpossub#1][#2]}
\newcommand{\spossub}{1}
\newcommandx{\SPosSet}[3][1=, 2=, 3=]
	{\mthset{\posset#3}[\spossub#1][#2]}

\newcommand{\possym}{v}
\newcommandx{\posSym}[3][1=, 2=, 3=]
	{\mthsym{\possym#3}[#1][#2]}
\newcommandx{\fposSym}[1][1=]
	{\posSym[\fpossub#1]}
\newcommandx{\sposSym}[1][1=]
	{\posSym[\spossub#1]}
\newcommand{\ipossub}{I}
\newcommandx{\iposSym}[1][1=]
	{\posSym[\ipossub#1]}

\newcommand{\poselm}{v}
\newcommandx{\posElm}[3][1=, 2=, 3=]
	{\mthelm{\poselm#3}[#1][#2]}
\newcommandx{\fposElm}[1][1=]
	{\posElm[\fpossub#1]}
\newcommandx{\sposElm}[1][1=]
	{\posElm[\spossub#1]}
\newcommandx{\iposElm}[1][1=]
	{\posElm[\ipossub#1]}

\newcommand{\sttset}{St}
\newcommandx{\SttSet}[3][1=, 2=, 3=]
	{\mthset{\sttset#3}[#1][#2]}
\newcommand{\fsttsub}{0}
\newcommandx{\FSttSet}[3][1=, 2=, 3=]
	{\mthset{\sttset#3}[\fsttsub#1][#2]}
\newcommand{\ssttsub}{1}
\newcommandx{\SSttSet}[3][1=, 2=, 3=]
	{\mthset{\sttset#3}[\ssttsub#1][#2]}

\newcommand{\sttsym}{s}
\newcommandx{\sttSym}[3][1=, 2=, 3=]
	{\mthsym{\sttsym#3}[#1][#2]}
\newcommandx{\fsttSym}[1][1=]
	{\sttSym[\fsttsub#1]}
\newcommandx{\ssttSym}[1][1=]
	{\sttSym[\ssttsub#1]}
\newcommand{\isttsub}{I}
\newcommandx{\isttSym}[1][1=]
	{\sttSym[\isttsub#1]}

\newcommand{\sttelm}{s}
\newcommandx{\sttElm}[3][1=, 2=, 3=]
	{\mthelm{\sttelm#3}[#1][#2]}
\newcommandx{\fsttElm}[1][1=]
	{\sttElm[\fsttsub#1]}
\newcommandx{\ssttElm}[1][1=]
	{\sttElm[\ssttsub#1]}
\newcommandx{\isttElm}[1][1=]
	{\sttElm[\isttsub#1]}

\newcommand{\plrfun}{pl}
\newcommandx{\plrFun}[4][1=, 2=, 3=, 4=]
	{\mthargfun{\plrfun#4}[#1][#2]{#3}}

\newcommand{\agnfun}{ag}
\newcommandx{\agnFun}[4][1=, 2=, 3=, 4=]
	{\mthargfun{\agnfun#4}[#1][#2]{#3}}

\newcommand{\movfun}{mv}
\newcommandx{\movFun}[4][1=, 2=, 3=, 4=]
	{\mthargfun{\movfun#4}[#1][#2]{#3}}

\newcommand{\actfun}{ac}
\newcommandx{\actFun}[4][1=, 2=, 3=, 4=]
	{\mthargfun{\actfun#4}[#1][#2]{#3}}

\newcommand{\decfun}{dc}
\newcommandx{\decFun}[4][1=, 2=, 3=, 4=]
	{\mthargfun{\decfun#4}[#1][#2]{#3}}

\newcommand{\trnfun}{tr}
\newcommandx{\trnFun}[4][1=, 2=, 3=, 4=]
	{\mthargfun{\trnfun#4}[#1][#2]{#3}}

\newcommand{\arn}{Ar}
\newcommandx{\Arn}[5][1=, 2=, 3=, 4=, 5=]
	{\txtargname{\arn#5{\small\argint{$[$}{#1}{$]$}}}[#2][#3]{#4}\xspace}

\newcommand{\arnname}{A}
\newcommand{\ArnName}
	{\mthname{\arnname}}

\newcommand{\arncls}{Ar}
\newcommandx{\ArnCls}[5][1=, 2=, 3=, 4=, 5=]
	{\mthset[#5]{\arncls#4\text{\small\txtname{\argint{$[$}{#1}{$]$}}}}[#2][#3]}

\newcommand{\ArnStr}[1][]
	{%
	\IfStrEqCase{\argdef{#1}{\gamkin}}
		{%
		{2PT}
			{\tuplecx{\FPosSet}{\SPosSet}{\MovRel}}%
		{MPC0}
			{\tupledx{\PlrSet}{\MovSet}{\PosSet}{\trnFun}}%
		{MPC1}
			{\tupleex{\PlrSet}{\MovSet}{\PosSet}{\decFun}{\trnFun}}%
		{MPC2}
			{\tuplefx{\PlrSet}{\MovSet}{\PosSet}{\plrFun}{\movFun}{\trnFun}}%
		{MPC3}
			{\tuplegx{\PlrSet}{\MovSet}{\PosSet}{\plrFun}{\movFun}{\decFun}{\trnFun}}%
		{2AT}
			{\tuplecx{\FSttSet}{\SSttSet}{\ActRel}}%
		{MAC0}
			{\tupledx{\AgnSet}{\ActSet}{\SttSet}{\trnFun}}%
		{MAC1}
			{\tupleex{\AgnSet}{\ActSet}{\SttSet}{\decFun}{\trnFun}}%
		{MAC2}
			{\tuplefx{\AgnSet}{\ActSet}{\SttSet}{\agnFun}{\actFun}{\trnFun}}%
		{MAC3}
			{\tuplegx{\AgnSet}{\ActSet}{\SttSet}{\agnFun}{\actFun}{\decFun}{\trnFun}}%
		}
		[\ensuremath{\clubsuit}]%
	}

\newcommand{\hstset}{Hst}
\newcommandx{\HstSet}[3][1=, 2=, 3=]
	{\mthset{\hstset#3}[#1][#2]}

\newcommand{\hstsym}{\rho}
\newcommandx{\hstSym}[3][1=, 2=, 3=]
	{\mthsym{\hstsym#3}[#1][#2]}

\newcommand{\hstelm}{\rho}
\newcommandx{\hstElm}[3][1=, 2=, 3=]
	{\mthelm{\hstelm#3}[#1][#2]}

\newcommand{\strset}{Str}
\newcommandx{\StrSet}[3][1=, 2=, 3=]
	{\mthset{\strset#3}[#1][#2]}

\newcommand{\strsym}{\sigma}
\newcommandx{\strSym}[4][1=, 2=, 3=, 4=]
	{\mthargfun{\strsym#4}[#1][#2]{#3}}

\newcommand{\strelm}{\sigma}
\newcommandx{\strElm}[4][1=, 2=, 3=, 4=]
	{\mthargfun{\strelm#4}[#1][#2]{#3}}

\newcommand{\prfset}{Prf}
\newcommandx{\PrfSet}[3][1=, 2=, 3=]
	{\mthset{\prfset#3}[#1][#2]}

\newcommand{\prfsym}{\xi}
\newcommandx{\prfSym}[4][1=, 2=, 3=, 4=]
	{\mthargfun{\prfsym#4}[#1][#2]{#3}}

\newcommandx{\prfElm}[4][1=, 2=, 3=, 4=]
	{\mthargfun{\prfsym#4}[#1][#2]{#3}}

\newcommand{\playfun}{play}
\newcommandx{\playFun}[4][1=, 2=, 3=, 4=]
	{\mthargfun{\playfun#4}[#1][#2]{#3}}

\newcommand{\labarn}{L\arn}
\newcommandx{\LabArn}[5][1=, 2=, 3=, 4=, 5=]
	{\txtargname{\labarn#5{\small\argint{$[$}{#1}{$]$}}}[#2][#3]{#4}\xspace}

\newcommand{\labarncls}{L\arncls}
\newcommandx{\LabArnCls}[5][1=, 2=, 3=, 4=, 5=]
	{\mthset[#5]{\labarncls#4\text{\small\txtname{\argint{$[$}{#1}{$]$}}}}[#2]%
	[#3]}

\newcommand{\colarn}{C\arn}
\newcommandx{\ColArn}[5][1=, 2=, 3=, 4=, 5=]
	{\txtargname{\colarn#5{\small\argint{$[$}{#1}{$]$}}}[#2][#3]{#4}\xspace}

\newcommand{\colarncls}{C\arncls}
\newcommandx{\ColArnCls}[5][1=, 2=, 3=, 4=, 5=]
	{\mthset[#5]{\colarncls#4\text{\small\txtname{\argint{$[$}{#1}{$]$}}}}[#2]%
	[#3]}

\newcommand{\wgharn}{W\arn}
\newcommandx{\WghArn}[5][1=, 2=, 3=, 4=, 5=]
	{\txtargname{\wgharn#5{\small\argint{$[$}{#1}{$]$}}}[#2][#3]{#4}\xspace}

\newcommand{\wgharncls}{W\arncls}
\newcommandx{\WghArnCls}[5][1=, 2=, 3=, 4=, 5=]
	{\mthset[#5]{\wgharncls#4\text{\small\txtname{\argint{$[$}{#1}{$]$}}}}[#2]%
	[#3]}

\newcommand{\winset}{Wn}
\newcommandx{\WinSet}[3][1=, 2=, 3=]
	{\mthset{\winset#3}[#1][#2]}

\newcommand{\prdset}{Pr}
\newcommandx{\PrdSet}[3][1=, 2=, 3=]
	{\mthset{\prdset#3}[#1][#2]}

\newcommand{\prdsym}{p}
\newcommandx{\prdSym}[3][1=, 2=, 3=]
	{\mthsym{\prdsym#3}[#1][#2]}

\newcommand{\prdelm}{p}
\newcommandx{\prdElm}[3][1=, 2=, 3=]
	{\mthelm{\prdelm#3}[#1][#2]}

\newcommand{\prdfun}{pr}
\newcommandx{\prdFun}[4][1=, 2=, 3=, 4=]
	{\mthargfun{\prdfun#4}[#1][#2]{#3}}

\newcommand{\extname}{E}
\newcommand{\ExtName}
	{\mthname{\extname}}

\newcommand{\extcls}{Ex}
\newcommandx{\ExtCls}[5][1=, 2=, 3=, 4=, 5=]
	{\mthset[#5]{\extcls#4\text{\small\txtname{\argint{$[$}{#1}{$]$}}}}[#2][#3]}

\newcommand{\conset}{Cn}
\newcommandx{\ConSet}[3][1=, 2=, 3=]
	{\mthset{\conset#3}[#1][#2]}

\newcommand{\consym}{\varphi}
\newcommandx{\conSym}[3][1=, 2=, 3=]
	{\mthsym{\consym#3}[#1][#2]}

\newcommand{\conelm}{\varphi}
\newcommandx{\conElm}[3][1=, 2=, 3=]
	{\mthelm{\conelm#3}[#1][#2]}

\newcommand{\schrel}{\models}
\newcommandx{\schRel}[4][1=, 2=, 3=, 4=]
	{\mthrel{\schrel#3}[#1][#2]}

\newcommand{\schcls}{Sc}
\newcommandx{\SchCls}[5][1=, 2=, 3=, 4=, 5=]
	{\mthset[#5]{\schcls#4\text{\small\txtname{\argint{$[$}{#1}{$]$}}}}[#2][#3]}

\newcommand{\gamcls}{Gm}
\newcommandx{\GamCls}[5][1=, 2=, 3=, 4=, 5=]
	{\mthset[#5]{\gamcls#4\text{\small\txtname{\argint{$[$}{#1}{$]$}}}}[#2][#3]}

\newcommandx{\GamStr}[1][1=]
	{%
	\StrLeft{\argdef{#1}{\gamkin}}{2}[\optgamkin]%
	\IfStrEqCase{\optgamkin}
		{%
		{2P}
			{\gamstrauxtp}%
		{MP}
			{\gamstrauxmp}%
		{2A}
			{\gamstrauxta}%
		{MA}
			{\gamstrauxma}%
		}
		[\ensuremath{\clubsuit}]%
	}
\newcommandx{\gamstrauxtp}[5][1=, 2=, 3=, 4=, 5=]
	{\tuplecx{\ArnName}{\iposElm}{\WinSet}[#3][#4][#5][#1][#2]}
\newcommandx{\gamstrauxmp}[5][1=, 2=, 3=, 4=, 5=]
	{\tuplecx{\ExtName}{\iposElm}{\conElm}[#3][#4][#5][#1][#2]}
\newcommandx{\gamstrauxta}[5][1=, 2=, 3=, 4=, 5=]
	{\tuplecx{\ArnName}{\isttElm}{\WinSet}[#3][#4][#5][#1][#2]}
\newcommandx{\gamstrauxma}[5][1=, 2=, 3=, 4=, 5=]
	{\tuplecx{\ExtName}{\isttElm}{\conElm}[#3][#4][#5][#1][#2]}

\newcommand{\worset}{W}
\newcommandx{\WorSet}[3][1=, 2=, 3=]
	{\mthset{\worset#3}[#1][#2]}

\newcommand{\worsym}{w}
\newcommandx{\worSym}[3][1=, 2=, 3=]
	{\mthsym{\worsym#3}[#1][#2]}

\newcommand{\worelm}{w}
\newcommandx{\worElm}[3][1=, 2=, 3=]
	{\mthelm{\worelm#3}[#1][#2]}

\newcommand{\trnrel}{R}
\newcommandx{\TrnRel}[3][1=, 2=, 3=]
	{\mthrel{\trnrel#3}[#1][#2]}

\newcommand{\trnsym}{r}
\newcommandx{\trnSym}[3][1=, 2=, 3=]
	{\mthsym{\trnsym#3}[#1][#2]}

\newcommand{\trnelm}{r}
\newcommandx{\trnElm}[3][1=, 2=, 3=]
	{\mthelm{\trnelm#3}[#1][#2]}

\newcommand{\labfun}{L}
\newcommandx{\labFun}[4][1=, 2=, 3=, 4=]
	{\mthargfun{\labfun#4}[#1][#2]{#3}}

\newcommand{\krpstr}{KS}
\newcommandx{\KrpStr}[5][1=, 2=, 3=, 4=, 5=]
	{\txtargname{\krpstr#5{\small\argint{$[$}{#1}{$]$}}}[#2][#3]{#4}\xspace}

\newcommand{\krpstrname}{K}
\newcommand{\KrpStrName}
	{\mthname{\krpstrname}}

\newcommand{\krpstrcls}{KS}
\newcommandx{\KrpStrCls}[5][1=, 2=, 3=, 4=, 5=]
	{\mthset[#5]{\krpstrcls#4\text{\small\txtname{\argint{$[$}{#1}{$]$}}}}[#2]%
	[#3]}

\newcommand{\KrpStrStr}
	{\tupleex{\APSet}{\WorSet}{\TrnRel}{\labFun}{\worElm[I]}}

\newcommand{\trkset}{Trk}
\newcommandx{\TrkSet}[3][1=, 2=, 3=]
	{\mthset{\trkset#3}[#1][#2]}

\newcommand{\trksym}{\rho}
\newcommandx{\trkSym}[3][1=, 2=, 3=]
	{\mthsym{\trksym#3}[#1][#2]}

\newcommand{\trkelm}{\rho}
\newcommandx{\trkElm}[3][1=, 2=, 3=]
	{\mthelm{\trkelm#3}[#1][#2]}

\newcommand{\krptree}{KT}
\newcommandx{\KrpTree}[5][1=, 2=, 3=, 4=, 5=]
	{\txtargname{\krptree#5{\small\argint{$[$}{#1}{$]$}}}[#2][#3]{#4}\xspace}

\newcommand{\krptreename}{T}
\newcommand{\KrpTreeName}
	{\mthname{\krptreename}}

\newcommand{\krptreecls}{KT}
\newcommandx{\KrpTreeCls}[5][1=, 2=, 3=, 4=, 5=]
	{\mthset[#5]{\krptreecls#4\text{\small\txtname{\argint{$[$}{#1}{$]$}}}}[#2]%
	[#3]}

\newcommand{\dirset}{Dir}
\newcommandx{\DirSet}[3][1=, 2=, 3=]
	{\mthset{\dirset#3}[#1][#2]}

\newcommand{\dirsym}{d}
\newcommandx{\dirSym}[3][1=, 2=, 3=]
	{\mthsym{\dirsym#3}[#1][#2]}

\newcommand{\direlm}{d}
\newcommandx{\dirElm}[3][1=, 2=, 3=]
	{\mthelm{\direlm#3}[#1][#2]}

\newcommand{\unwfun}{unw}
\newcommandx{\unwFun}[4][1=, 2=, 3=, 4=]
	{\mthargfun{\unwfun#4}[#1][#2]{#3}}

\newcommand{\congamstrkin}{MAC0}

\newcommand{\congamstr}{CGS}
\newcommandx{\ConGamStr}[5][1=, 2=, 3=, 4=, 5=]
	{\txtargname{\congamstr#5{\small\argint{$[$}{#1}{$]$}}}[#2][#3]{#4}\xspace}

\newcommandx{\ConGamStrCls}[5][1=, 2=, 3=, 4=, 5=]
	{\mthset[#5]{\arncls#4\text{\small\txtname{\argint{$[$}{#1}{$]$}}}}[#2][#3]}

\newcommandx{\ConGamStrStr}[1][1=]
	{%
	\IfStrEqCase{\argdef{#1}{\congamstrkin}}
		{%
		{IP}
			{\congamstrstrauxip}%
		{2PT}
			{\congamstrstrauxpt}%
		{MPC0}
			{\congamstrstrauxpca}%
		{MPC1}
			{\congamstrstrauxpcb}%
		{MPC2}
			{\congamstrstrauxpcc}%
		{MPC3}
			{\congamstrstrauxpcd}%
		{IA}
			{\congamstrstrauxia}%
		{2AT}
			{\congamstrstrauxat}%
		{MAC0}
			{\congamstrstrauxaca}%
		{MAC1}
			{\congamstrstrauxacb}%
		{MAC2}
			{\congamstrstrauxacc}%
		{MAC3}
			{\congamstrstrauxacd}%
		}
		[\ensuremath{\clubsuit}]%
	}
\newcommandx{\congamstrstrauxip}[3][1=, 2=, 3=]
	{%
	\def\defini{#1}%
	\def\defsubscr{#2}%
	\def\defsupscr{#3}%
	\congamstrstrauxxip%
	}
\newcommandx{\congamstrstrauxxip}[3][1=, 2=, 3=]
	{%
	\tupledx{\ArnName}{\APSet}{\apFun}{\iposElm}%
		[\defsubscr][\defsupscr][#1][#2][#3][\defini]%
	}
\newcommandx{\congamstrstrauxpt}[3][1=, 2=, 3=]
	{%
	\def\defini{#1}%
	\def\defsubscr{#2}%
	\def\defsupscr{#3}%
	\congamstrstrauxxpt%
	}
\newcommandx{\congamstrstrauxxpt}[5][1=, 2=, 3=, 4=, 5=]
	{%
	\tuplefx{\APSet}{\FPosSet}{\SPosSet}{\MovRel}{\apFun}{\iposElm}%
		[\defsubscr][\defsupscr][#1][#2][#3][#4][#5][\defini]%
	}
\newcommandx{\congamstrstrauxpca}[3][1=, 2=, 3=]
	{%
	\def\defini{#1}%
	\def\defsubscr{#2}%
	\def\defsupscr{#3}%
	\congamstrstrauxxpca%
	}
\newcommandx{\congamstrstrauxxpca}[6][1=, 2=, 3=, 4=, 5=, 6=]
	{%
	\tuplegx{\APSet}{\PlrSet}{\MovSet}{\PosSet}{\trnFun}{\apFun}{\iposElm}%
		[\defsubscr][\defsupscr][#1][#2][#3][#4][#5][#6][\defini]%
	}
\newcommandx{\congamstrstrauxpcb}[3][1=, 2=, 3=]
	{%
	\def\defini{#1}%
	\def\defsubscr{#2}%
	\def\defsupscr{#3}%
	\congamstrstrauxxpcb%
	}
\newcommandx{\congamstrstrauxxpcb}[7][1=, 2=, 3=, 4=, 5=, 6=, 7=]
	{%
	\tuplehx{\APSet}{\PlrSet}{\MovSet}{\PosSet}{\decFun}{\trnFun}{\apFun}%
		{\iposElm}%
		[\defsubscr][\defsupscr][#1][#2][#3][#4][#5][#6][#7][\defini]%
	}
\newcommandx{\congamstrstrauxpcc}[3][1=, 2=, 3=]
	{%
	\def\defini{#1}%
	\def\defsubscr{#2}%
	\def\defsupscr{#3}%
	\congamstrstrauxxpcc%
	}
\newcommandx{\congamstrstrauxxpcc}[8][1=, 2=, 3=, 4=, 5=, 6=, 7=, 8=]
	{%
	\tupleix{\APSet}{\PlrSet}{\MovSet}{\PosSet}{\plrFun}{\movFun}{\trnFun}%
		{\apFun}{\iposElm}%
		[\defsubscr][\defsupscr][#1][#2][#3][#4][#5][#6][#7][#8][\defini]%
	}
\newcommandx{\congamstrstrauxpcd}[3][1=, 2=, 3=]
	{%
	\def\defini{#1}%
	\def\defsubscr{#2}%
	\def\defsupscr{#3}%
	\congamstrstrauxxpcd%
	}
\newcommandx{\congamstrstrauxxpcd}[9][1=, 2=, 3=, 4=, 5=, 6=, 7=, 8=, 9=]
	{%
	\tuplejx{\APSet}{\PlrSet}{\MovSet}{\PosSet}{\plrFun}{\movFun}{\decFun}%
	{\trnFun}{\apFun}{\iposElm}%
		[\defsubscr][\defsupscr][#1][#2][#3][#4][#5][#6][#7][#8][#9][\defini]%
	}
\newcommandx{\congamstrstrauxia}[3][1=, 2=, 3=]
	{%
	\def\defini{#1}%
	\def\defsubscr{#2}%
	\def\defsupscr{#3}%
	\congamstrstrauxxia%
	}
\newcommandx{\congamstrstrauxxia}[3][1=, 2=, 3=]
	{%
	\tupledx{\ArnName}{\APSet}{\apFun}{\isttElm}%
		[\defsubscr][\defsupscr][#1][#2][#3][\defini]%
	}
\newcommandx{\congamstrstrauxat}[3][1=, 2=, 3=]
	{%
	\def\defini{#1}%
	\def\defsubscr{#2}%
	\def\defsupscr{#3}%
	\congamstrstrauxxat%
	}
\newcommandx{\congamstrstrauxxat}[5][1=, 2=, 3=, 4=, 5=]
	{%
	\tuplefx{\APSet}{\FSttSet}{\SSttSet}{\ActRel}{\apFun}{\isttElm}%
		[\defsubscr][\defsupscr][#1][#2][#3][#4][#5][\defini]%
	}
\newcommandx{\congamstrstrauxaca}[3][1=, 2=, 3=]
	{%
	\def\defini{#1}%
	\def\defsubscr{#2}%
	\def\defsupscr{#3}%
	\congamstrstrauxxaca%
	}
\newcommandx{\congamstrstrauxxaca}[6][1=, 2=, 3=, 4=, 5=, 6=]
	{%
	\tuplegx{\APSet}{\AgnSet}{\ActSet}{\SttSet}{\trnFun}{\apFun}{\isttElm}%
		[\defsubscr][\defsupscr][#1][#2][#3][#4][#5][#6][\defini]%
	}
\newcommandx{\congamstrstrauxacb}[3][1=, 2=, 3=]
	{%
	\def\defini{#1}%
	\def\defsubscr{#2}%
	\def\defsupscr{#3}%
	\congamstrstrauxxacb%
	}
\newcommandx{\congamstrstrauxxacb}[7][1=, 2=, 3=, 4=, 5=, 6=, 7=]
	{%
	\tuplehx{\APSet}{\AgnSet}{\ActSet}{\SttSet}{\decFun}{\trnFun}{\apFun}%
		{\isttElm}%
		[\defsubscr][\defsupscr][#1][#2][#3][#4][#5][#6][#7][\defini]%
	}
\newcommandx{\congamstrstrauxacc}[3][1=, 2=, 3=]
	{%
	\def\defini{#1}%
	\def\defsubscr{#2}%
	\def\defsupscr{#3}%
	\congamstrstrauxxacc%
	}
\newcommandx{\congamstrstrauxxacc}[8][1=, 2=, 3=, 4=, 5=, 6=, 7=, 8=]
	{%
	\tupleix{\APSet}{\AgnSet}{\ActSet}{\SttSet}{\agnFun}{\actFun}{\trnFun}%
		{\apFun}{\isttElm}%
		[\defsubscr][\defsupscr][#1][#2][#3][#4][#5][#6][#7][#8][\defini]%
	}
\newcommandx{\congamstrstrauxacd}[3][1=, 2=, 3=]
	{%
	\def\defini{#1}%
	\def\defsubscr{#2}%
	\def\defsupscr{#3}%
	\congamstrstrauxxacd%
	}
\newcommandx{\congamstrstrauxxacd}[9][1=, 2=, 3=, 4=, 5=, 6=, 7=, 8=, 9=]
	{%
	\tuplejx{\APSet}{\AgnSet}{\ActSet}{\SttSet}{\agnFun}{\actFun}{\decFun}%
		{\trnFun}{\apFun}{\isttElm}%
		[\defsubscr][\defsupscr][#1][#2][#3][#4][#5][#6][#7][#8][#9][\defini]%
	}

\newcommand{\trntabkin}{D}

\newcommand{\symset}{Sm}
\newcommandx{\SymSet}[3][1=, 2=, 3=]
	{\mthset{\symset#3}[#1][#2]}

\newcommand{\symsym}{\ell}
\newcommandx{\symSym}[3][1=, 2=, 3=]
	{\mthsym{\symsym#3}[#1][#2]}

\newcommand{\symelm}{\ell}
\newcommandx{\symElm}[3][1=, 2=, 3=]
	{\mthelm{\symelm#3}[#1][#2]}

\newcommand{\DSttSet}[1][]
	{\SttSet[\Delta#1]}

\newcommand{\ESttSet}[1][]
	{\SttSet[\exists#1]}

\newcommand{\ASttSet}[1][]
	{\SttSet[\forall#1]}

\newcommand{\trntab}{tt}
\newcommandx{\TrnTab}[5][1=, 2=, 3=, 4=, 5=]
	{\txtargname{\trntab#5{\small\argint{$[$}{#1}{$]$}}}[#2][#3]{#4}\xspace}

\newcommand{\trntabcls}{TT}
\newcommandx{\TrnTabCls}[5][1=, 2=, 3=, 4=, 5=]
	{\mthset[#5]{\trntabcls#4\text{\txtname{\small\argint{$[$}{#1}{$]$}}}}[#2]%
	[#3]}

\newcommand{\TrnTabStr}[1][]
	{%
	\IfStrEqCase{\argdef{#1}{\trntabkin}}
		{%
		{D}{\tuplecx{\SymSet}{\SttSet}{\trnFun}}%
		{N}{\tupledx{\SymSet}{\DSttSet}{\ESttSet}{\trnFun}}%
		{U}{\tupledx{\SymSet}{\DSttSet}{\ASttSet}{\trnFun}}%
		{A}{\tupleex{\SymSet}{\DSttSet}{\ESttSet}{\ASttSet}{\trnFun}}%
		}
		[\ensuremath{\clubsuit}]%
	}

\newcommandx{\PC}[5][1=, 2=, 3=, 4=, 5=]
	{\txtargname{PC#5{\small\argint{$[$}{#1}{$]$}}}[#2][#3]{#4}\xspace}

\newcommand{\Tt}
	{\mthsym{t}}

\newcommand{\Ff}
	{\mthsym{f}}

\newcommand{\subfun}{sub}
\newcommand{\subFun}
	{\mthargfun{\subfun}}

\newcommand{\bcset}{BC}
\newcommandx{\BCSet}[4][1=, 2=, 3=, 4=]
	{\mthset[3]{\bcset#4}[#1][#2]{#3}}

\newcommand{\bcelm}{\eta}
\newcommandx{\bcElm}[3][1=, 2=, 3=]
	{\mthelm{\bcelm#3}[#1][#2]}

\newcommand{\acset}{AC}
\newcommandx{\ACSet}[4][1=, 2=, 3=, 4=]
	{\mthset[3]{\acset#4}[#1][#2]{#3}}

\newcommand{\acelm}{\eta}
\newcommandx{\acElm}[3][1=, 2=, 3=]
	{\mthelm{\acelm#3}[#1][#2]}

\newcommandx{\QBF}[5][1=, 2=, 3=, 4=, 5=]
	{\txtargname{QBF#5{\small\argint{$[$}{#1}{$]$}}}[#2][#3]{#4}\xspace}

\newcommandx{\FOL}[5][1=, 2=, 3=, 4=, 5=]
	{\txtargname{FOL#5{\small\argint{$[$}{#1}{$]$}}}[#2][#3]{#4}\xspace}

\newcommand{\varset}{Vr}
\newcommandx{\VarSet}[3][1=, 2=, 3=]
	{\mthset{\varset#3}[#1][#2]}

\newcommand{\varsym}{x}
\newcommandx{\varSym}[3][1=, 2=, 3=]
	{\mthsym{\varsym#3}[#1][#2]}

\newcommand{\varelm}{x}
\newcommandx{\varElm}[3][1=, 2=, 3=]
	{\mthelm{\varelm#3}[#1][#2]}

\newcommand{\varfun}{vr}
\newcommandx{\varFun}[4][1=, 2=, 3=, 4=]
	{\mthargfun{\varfun#4}[#1][#2]{#3}}

\newcommand{\qntset}{Qn}
\newcommandx{\QntSet}[3][1=, 2=, 3=]
	{\mthset{\qntset#3}[#1][#2]}

\newcommand{\qntsym}{\wp}
\newcommandx{\qntSym}[3][1=, 2=, 3=]
	{\mthsym{\qntsym#3}[#1][#2]}

\newcommand{\qntelm}{\wp}
\newcommandx{\qntElm}[3][1=, 2=, 3=]
	{\mthelm{\qntelm#3}[#1][#2]}

\newcommand{\qntfun}{qnt}
\newcommandx{\qntFun}[4][1=, 2=, 3=, 4=]
	{\mthargfun{\qntfun#4}[#1][#2]{#3}}

\newcommand{\bndset}{Bn}
\newcommandx{\BndSet}[3][1=, 2=, 3=]
	{\mthset{\bndset#3}[#1][#2]}

\newcommand{\bndsym}{\flat}
\newcommandx{\bndSym}[3][1=, 2=, 3=]
	{\mthsym{\bndsym#3}[#1][#2]}

\newcommand{\bndelm}{\flat}
\newcommandx{\bndElm}[3][1=, 2=, 3=]
	{\mthelm{\bndelm#3}[#1][#2]}

\newcommand{\bndfun}{bnd}
\newcommandx{\bndFun}[4][1=, 2=, 3=, 4=]
	{\mthargfun{\bndfun#4}[#1][#2]{#3}}

\newcommand{\depset}{\Delta}
\newcommandx{\DepSet}[3][1=, 2=, 3=]
	{\mthset{\depset#3}[#1][#2]}

\newcommand{\denfun}{den}
\newcommandx{\denFun}[4][1=, 2=, 3=, 4=]
	{\mthargfun{\denfun#4}[#1][#2]{#3}}

\newcommand{\asgset}{Asg}
\newcommandx{\AsgSet}[3][1=, 2=, 3=]
	{\mthset{\asgset#3}[#1][#2]}

\newcommand{\asgfun}{\chi}
\newcommandx{\asgFun}[4][1=, 2=, 3=, 4=]
	{\mthargfun{\asgfun#4}[#1][#2]{#3}}

\newcommand{\smset}{SM}
\newcommandx{\SMSet}[3][1=, 2=, 3=]
	{\mthset{\smset#3}[#1][#2]}

\newcommand{\smfun}{\delta}
\newcommandx{\smFun}[4][1=, 2=, 3=, 4=]
	{\mthargfun{\smfun#4}[#1][#2]{#3}}

\newcommand{\cmset}{CM}
\newcommandx{\CMSet}[3][1=, 2=, 3=]
	{\mthset{\cmset#3}[#1][#2]}

\newcommand{\cmfun}{\gamma}
\newcommandx{\cmFun}[4][1=, 2=, 3=, 4=]
	{\mthargfun{\cmfun#4}[#1][#2]{#3}}

\newcommand{\schset}{Sch}
\newcommandx{\SchSet}[3][1=, 2=, 3=]
	{\mthset{\schset#3}[#1][#2]}

\newcommand{\schsym}{\sigma}
\newcommandx{\schSym}[3][1=, 2=, 3=]
	{\mthsym{\schsym#3}[#1][#2]}

\newcommand{\schelm}{\sigma}
\newcommandx{\schElm}[3][1=, 2=, 3=]
	{\mthelm{\schelm#3}[#1][#2]}

\newcommand{\entset}{Ent}
\newcommandx{\EntSet}[4][1=, 2=, 3=, 4=]
	{\mthset{\entset#4}[#1][#2]{#3}}

\newcommand{\entfun}{ent}
\newcommandx{\entFun}[4][1=, 2=, 3=, 4=]
	{\mthargfun{\entfun#4}[#1][#2]{#3}}

\newcommandx{\SOL}[5][1=, 2=, 3=, 4=, 5=]
	{\txtargname{SOL#5{\small\argint{$[$}{#1}{$]$}}}[#2][#3]{#4}\xspace}

\newcommandx{\TL}[5][1=, 2=, 3=, 4=, 5=]
	{\txtargname{TL#5{\small\argint{$[$}{#1}{$]$}}}[#2][#3]{#4}\xspace}

\newcommandx{\PL}[5][1=, 2=, 3=, 4=, 5=]
	{\txtargname{PL#5{\small\argint{$[$}{#1}{$]$}}}[#2][#3]{#4}\xspace}

\newcommand{\fvarset}{FVr}
\newcommandx{\FVarSet}[3][1=, 2=, 3=]
	{\mthset{\fvarset#3}[#1][#2]}

\newcommand{\fvarsym}{x}
\newcommandx{\fvarSym}[3][1=, 2=, 3=]
	{\mthsym{\fvarsym#3}[#1][#2]}

\newcommand{\fvarelm}{x}
\newcommandx{\fvarElm}[3][1=, 2=, 3=]
	{\mthelm{\fvarelm#3}[#1][#2]}

\newcommand{\fvarfun}{fvr}
\newcommandx{\fvarFun}[4][1=, 2=, 3=, 4=]
	{\mthargfun{\fvarfun#4}[#1][#2]{#3}}

\newcommand{\svarset}{SVr}
\newcommandx{\SVarSet}[3][1=, 2=, 3=]
	{\mthset{\svarset#3}[#1][#2]}

\newcommand{\svarsym}{X}
\newcommandx{\svarSym}[3][1=, 2=, 3=]
	{\mthsym{\svarsym#3}[#1][#2]}

\newcommand{\svarelm}{X}
\newcommandx{\svarElm}[3][1=, 2=, 3=]
	{\mthelm{\svarelm#3}[#1][#2]}

\newcommand{\svarfun}{svr}
\newcommandx{\svarFun}[4][1=, 2=, 3=, 4=]
	{\mthargfun{\svarfun#4}[#1][#2]{#3}}

\newcommandx{\ML}[5][1=, 2=, 3=, 4=, 5=]
	{\txtargname{ML#5{\small\argint{$[$}{#1}{$]$}}}[#2][#3]{#4}\xspace}

\newcommandx{\MC}[5][1=, 2=, 3=, 4=, 5=]
	{\txtargname{$\mu$Calculus#5{\small\argint{$[$}{#1}{$]$}}}[#2][#3]{#4}\xspace}

\newcommandx{\LTL}[5][1=, 2=, 3=, 4=, 5=]
	{\txtargname{LTL#5{\small\argint{$[$}{#1}{$]$}}}[#2][#3]{#4}\xspace}

\newcommandx{\PTL}[5][1=, 2=, 3=, 4=, 5=]
	{\txtargname{PTL#5{\small\argint{$[$}{#1}{$]$}}}[#2][#3]{#4}\xspace}

\newcommand{\X}
	{\mthsym{X}}

\newcommand{\F}
	{\mthsym{F}}

\newcommand{\G}
	{\mthsym{G}}

\newcommand{\U}
	{\mthsym{U}}

\newcommandx{\CTL}[5][1=, 2=, 3=, 4=, 5=]
	{\txtargname{CTL#5{\small\argint{$[$}{#1}{$]$}}}[#2][#3]{#4}\xspace}

\newcommandx{\CTLP}[5][1=, 2=, 3=, 4=, 5=]
	{\txtargname{CTL$^{+}$#5{\small\argint{$[$}{#1}{$]$}}}[#2][#3]{#4}\xspace}

\newcommandx{\CTLS}[5][1=, 2=, 3=, 4=, 5=]
	{\txtargname{CTL$^{\star}$#5{\small\argint{$[$}{#1}{$]$}}}[#2][#3]{#4}\xspace}

\newcommand{\E}
	{\mthsym{E}}

\newcommand{\A}
	{\mthsym{A}}

\newcommandx{\STL}[5][1=, 2=, 3=, 4=, 5=]
	{\txtargname{STL#5{\small\argint{$[$}{#1}{$]$}}}[#2][#3]{#4}\xspace}

\newcommandx{\STLP}[5][1=, 2=, 3=, 4=, 5=]
	{\txtargname{STL$^{+}$#5{\small\argint{$[$}{#1}{$]$}}}[#2][#3]{#4}\xspace}

\newcommandx{\STLS}[5][1=, 2=, 3=, 4=, 5=]
	{\txtargname{STL$^{\star}$#5{\small\argint{$[$}{#1}{$]$}}}[#2][#3]{#4}\xspace}

\newcommandx{\ATL}[5][1=, 2=, 3=, 4=, 5=]
	{\txtargname{ATL#5{\small\argint{$[$}{#1}{$]$}}}[#2][#3]{#4}\xspace}

\newcommandx{\ATLP}[5][1=, 2=, 3=, 4=, 5=]
	{\txtargname{ATL$^{+}$#5{\small\argint{$[$}{#1}{$]$}}}[#2][#3]{#4}\xspace}

\newcommandx{\ATLS}[5][1=, 2=, 3=, 4=, 5=]
	{\txtargname{ATL$^{\star}$#5{\small\argint{$[$}{#1}{$]$}}}[#2][#3]{#4}\xspace}

\newcommandx{\SL}[5][1=, 2=, 3=, 4=, 5=]
	{\txtargname{SL#5{\small\argint{$[$}{#1}{$]$}}}[#2][#3]{#4}\xspace}

\newcommand{\EExs}[1]
	{\ensuremath{%
	\argint{\mbox{$\langle\!\langle$}}{#1}{\mbox{$\rangle\!\rangle$}}%
	}}

\newcommand{\AAll}[1]
	{\ensuremath{\argint{\mbox{$[\:\!\![$}}{#1}{\mbox{$]\:\!\!]$}}}}

\newcommandx{\EF}[5][1=, 2=, 3=, 4=, 5=]
	{\txtargname{EF#5{\small\argint{$[$}{#1}{$]$}}}[#2][#3]{#4}\xspace}

\newcommandx{\SG}[5][1=, 2=, 3=, 4=, 5=]
	{\txtargname{SG#5{\small\argint{$[$}{#1}{$]$}}}[#2][#3]{#4}\xspace}

\newcommandx{\PG}[5][1=, 2=, 3=, 4=, 5=]
	{\txtargname{PG#5{\small\argint{$[$}{#1}{$]$}}}[#2][#3]{#4}\xspace}

\newcommandx{\LogTime}[4][1=, 2=, 3=, 4=]
	{\txtargname{LogTime#4}[#2][#3]{#1}\xspace}
\newcommandx{\LogTimeH}[4][1=, 2=, 3=, 4=]
	{\LogTime[#1][#2][#3][#4]-\HComp}
\newcommandx{\LogTimeE}[4][1=, 2=, 3=, 4=]
	{\LogTime[#1][#2][#3][#4]-\EComp}
\newcommandx{\LogTimeC}[4][1=, 2=, 3=, 4=]
	{\LogTime[#1][#2][#3][#4]-\CComp}

\newcommand{\NLogTime}
	{\txtname{N}\LogTime}
\newcommandx{\NLogTimeH}[4][1=, 2=, 3=, 4=]
	{\NLogTime[#1][#2][#3][#4]-\HComp}
\newcommandx{\NLogTimeE}[4][1=, 2=, 3=, 4=]
	{\NLogTime[#1][#2][#3][#4]-\EComp}
\newcommandx{\NLogTimeC}[4][1=, 2=, 3=, 4=]
	{\NLogTime[#1][#2][#3][#4]-\CComp}

\newcommand{\CoNLogTime}
	{\txtname{Co}\NLogTime}
\newcommandx{\CoNLogTimeH}[4][1=, 2=, 3=, 4=]
	{\CoNLogTime[#1][#2][#3][#4]-\HComp}
\newcommandx{\CoNLogTimeE}[4][1=, 2=, 3=, 4=]
	{\CoNLogTime[#1][#2][#3][#4]-\EComp}
\newcommandx{\CoNLogTimeC}[4][1=, 2=, 3=, 4=]
	{\CoNLogTime[#1][#2][#3][#4]-\CComp}

\newcommandx{\ALogTime}
	{\txtname{A}\LogTime}
\newcommandx{\ALogTimeH}[4][1=, 2=, 3=, 4=]
	{\ALogTime[#1][#2][#3][#4]-\HComp}
\newcommandx{\ALogTimeE}[4][1=, 2=, 3=, 4=]
	{\ALogTime[#1][#2][#3][#4]-\EComp}
\newcommandx{\ALogTimeC}[4][1=, 2=, 3=, 4=]
	{\ALogTime[#1][#2][#3][#4]-\CComp}

\newcommandx{\LogSpace}[4][1=, 2=, 3=, 4=]
	{\txtargname{LogSpace#4}[#2][#3]{#1}\xspace}
\newcommandx{\LogSpaceH}[4][1=, 2=, 3=, 4=]
	{\LogSpace[#1][#2][#3][#4]-\HComp}
\newcommandx{\LogSpaceE}[4][1=, 2=, 3=, 4=]
	{\LogSpace[#1][#2][#3][#4]-\EComp}
\newcommandx{\LogSpaceC}[4][1=, 2=, 3=, 4=]
	{\LogSpace[#1][#2][#3][#4]-\CComp}

\newcommandx{\NLogSpace}
	{\txtname{N}\LogSpace}
\newcommandx{\NLogSpaceH}[4][1=, 2=, 3=, 4=]
	{\NLogSpace[#1][#2][#3][#4]-\HComp}
\newcommandx{\NLogSpaceE}[4][1=, 2=, 3=, 4=]
	{\NLogSpace[#1][#2][#3][#4]-\EComp}
\newcommandx{\NLogSpaceC}[4][1=, 2=, 3=, 4=]
	{\NLogSpace[#1][#2][#3][#4]-\CComp}

\newcommandx{\CoNLogSpace}
	{\txtname{Co}\NLogSpace}
\newcommandx{\CoNLogSpaceH}[4][1=, 2=, 3=, 4=]
	{\CoNLogSpace[#1][#2][#3][#4]-\HComp}
\newcommandx{\CoNLogSpaceE}[4][1=, 2=, 3=, 4=]
	{\CoNLogSpace[#1][#2][#3][#4]-\EComp}
\newcommandx{\CoNLogSpaceC}[4][1=, 2=, 3=, 4=]
	{\CoNLogSpace[#1][#2][#3][#4]-\CComp}

\newcommandx{\ALogSpace}
	{\txtname{A}\LogSpace}
\newcommandx{\ALogSpaceH}[4][1=, 2=, 3=, 4=]
	{\ALogSpace[#1][#2][#3][#4]-\HComp}
\newcommandx{\ALogSpaceE}[4][1=, 2=, 3=, 4=]
	{\ALogSpace[#1][#2][#3][#4]-\EComp}
\newcommandx{\ALogSpaceC}[4][1=, 2=, 3=, 4=]
	{\ALogSpace[#1][#2][#3][#4]-\CComp}

\newcommandx{\PTime}[4][1=, 2=, 3=, 4=]
	{\txtargname{PTime#4}[#2][#3]{#1}\xspace}
\newcommandx{\PTimeH}[4][1=, 2=, 3=, 4=]
	{\PTime[#1][#2][#3][#4]-\HComp}
\newcommandx{\PTimeE}[4][1=, 2=, 3=, 4=]
	{\PTime[#1][#2][#3][#4]-\EComp}
\newcommandx{\PTimeC}[4][1=, 2=, 3=, 4=]
	{\PTime[#1][#2][#3][#4]-\CComp}

\newcommandx{\UPTime}
	{\txtname{U}\PTime}
\newcommandx{\UPTimeH}[4][1=, 2=, 3=, 4=]
	{\UPTime[#1][#2][#3][#4]-\HComp}
\newcommandx{\UPTimeE}[4][1=, 2=, 3=, 4=]
	{\UPTime[#1][#2][#3][#4]-\EComp}
\newcommandx{\UPTimeC}[4][1=, 2=, 3=, 4=]
	{\UPTime[#1][#2][#3][#4]-\CComp}

\newcommandx{\CoUPTime}
	{\txtname{Co}\UPTime}
\newcommandx{\CoUPTimeH}[4][1=, 2=, 3=, 4=]
	{\CoUPTime[#1][#2][#3][#4]-\HComp}
\newcommandx{\CoUPTimeE}[4][1=, 2=, 3=, 4=]
	{\CoUPTime[#1][#2][#3][#4]-\EComp}
\newcommandx{\CoUPTimeC}[4][1=, 2=, 3=, 4=]
	{\CoUPTime[#1][#2][#3][#4]-\CComp}

\newcommandx{\NPTime}
	{\txtname{N}\PTime}
\newcommandx{\NPTimeH}[4][1=, 2=, 3=, 4=]
	{\NPTime[#1][#2][#3][#4]-\HComp}
\newcommandx{\NPTimeE}[4][1=, 2=, 3=, 4=]
	{\NPTime[#1][#2][#3][#4]-\EComp}
\newcommandx{\NPTimeC}[4][1=, 2=, 3=, 4=]
	{\NPTime[#1][#2][#3][#4]-\CComp}

\newcommandx{\CoNPTime}
	{\txtname{Co}\NPTime}
\newcommandx{\CoNPTimeH}[4][1=, 2=, 3=, 4=]
	{\CoNPTime[#1][#2][#3][#4]-\HComp}
\newcommandx{\CoNPTimeE}[4][1=, 2=, 3=, 4=]
	{\CoNPTime[#1][#2][#3][#4]-\EComp}
\newcommandx{\CoNPTimeC}[4][1=, 2=, 3=, 4=]
	{\CoNPTime[#1][#2][#3][#4]-\CComp}

\newcommandx{\APTime}
	{\txtname{A}\PTime}
\newcommandx{\APTimeH}[4][1=, 2=, 3=, 4=]
	{\APTime[#1][#2][#3][#4]-\HComp}
\newcommandx{\APTimeE}[4][1=, 2=, 3=, 4=]
	{\APTime[#1][#2][#3][#4]-\EComp}
\newcommandx{\APTimeC}[4][1=, 2=, 3=, 4=]
	{\APTime[#1][#2][#3][#4]-\CComp}

\newcommandx{\PSpace}[4][1=, 2=, 3=, 4=]
	{\txtargname{PSpace#4}[#2][#3]{#1}\xspace}
\newcommandx{\PSpaceH}[4][1=, 2=, 3=, 4=]
	{\PSpace[#1][#2][#3][#4]-\HComp}
\newcommandx{\PSpaceE}[4][1=, 2=, 3=, 4=]
	{\PSpace[#1][#2][#3][#4]-\EComp}
\newcommandx{\PSpaceC}[4][1=, 2=, 3=, 4=]
	{\PSpace[#1][#2][#3][#4]-\CComp}

\newcommandx{\NPSpace}
	{\txtname{N}\PSpace}
\newcommandx{\NPSpaceH}[4][1=, 2=, 3=, 4=]
	{\NPSpace[#1][#2][#3][#4]-\HComp}
\newcommandx{\NPSpaceE}[4][1=, 2=, 3=, 4=]
	{\NPSpace[#1][#2][#3][#4]-\EComp}
\newcommandx{\NPSpaceC}[4][1=, 2=, 3=, 4=]
	{\NPSpace[#1][#2][#3][#4]-\CComp}

\newcommandx{\CoNPSpace}
	{\txtname{Co}\NPSpace}
\newcommandx{\CoNPSpaceH}[4][1=, 2=, 3=, 4=]
	{\CoNPSpace[#1][#2][#3][#4]-\HComp}
\newcommandx{\CoNPSpaceE}[4][1=, 2=, 3=, 4=]
	{\CoNPSpace[#1][#2][#3][#4]-\EComp}
\newcommandx{\CoNPSpaceC}[4][1=, 2=, 3=, 4=]
	{\CoNPSpace[#1][#2][#3][#4]-\CComp}

\newcommandx{\APSpace}
	{\txtname{A}\PSpace}
\newcommandx{\APSpaceH}[4][1=, 2=, 3=, 4=]
	{\APSpace[#1][#2][#3][#4]-\HComp}
\newcommandx{\APSpaceE}[4][1=, 2=, 3=, 4=]
	{\APSpace[#1][#2][#3][#4]-\EComp}
\newcommandx{\APSpaceC}[4][1=, 2=, 3=, 4=]
	{\APSpace[#1][#2][#3][#4]-\CComp}

\newcommandx{\ExpTime}[4][1=, 2=, 3=, 4=]
	{\txtargname{ExpTime#4}[#2][#3]{#1}\xspace}
\newcommandx{\ExpTimeH}[4][1=, 2=, 3=, 4=]
	{\ExpTime[#1][#2][#3][#4]-\HComp}
\newcommandx{\ExpTimeE}[4][1=, 2=, 3=, 4=]
	{\ExpTime[#1][#2][#3][#4]-\EComp}
\newcommandx{\ExpTimeC}[4][1=, 2=, 3=, 4=]
	{\ExpTime[#1][#2][#3][#4]-\CComp}

\newcommandx{\NExpTime}
	{\txtname{N}\ExpTime}
\newcommandx{\NExpTimeH}[4][1=, 2=, 3=, 4=]
	{\NExpTime[#1][#2][#3][#4]-\HComp}
\newcommandx{\NExpTimeE}[4][1=, 2=, 3=, 4=]
	{\NExpTime[#1][#2][#3][#4]-\EComp}
\newcommandx{\NExpTimeC}[4][1=, 2=, 3=, 4=]
	{\NExpTime[#1][#2][#3][#4]-\CComp}

\newcommandx{\CoNExpTime}
	{\txtname{Co}\NExpTime}
\newcommandx{\CoNExpTimeH}[4][1=, 2=, 3=, 4=]
	{\CoNExpTime[#1][#2][#3][#4]-\HComp}
\newcommandx{\CoNExpTimeE}[4][1=, 2=, 3=, 4=]
	{\CoNExpTime[#1][#2][#3][#4]-\EComp}
\newcommandx{\CoNExpTimeC}[4][1=, 2=, 3=, 4=]
	{\CoNExpTime[#1][#2][#3][#4]-\CComp}

\newcommandx{\AExpTime}
	{\txtname{A}\ExpTime}
\newcommandx{\AExpTimeH}[4][1=, 2=, 3=, 4=]
	{\AExpTime[#1][#2][#3][#4]-\HComp}
\newcommandx{\AExpTimeE}[4][1=, 2=, 3=, 4=]
	{\AExpTime[#1][#2][#3][#4]-\EComp}
\newcommandx{\AExpTimeC}[4][1=, 2=, 3=, 4=]
	{\AExpTime[#1][#2][#3][#4]-\CComp}

\newcommandx{\ExpSpace}[4][1=, 2=, 3=, 4=]
	{\txtargname{ExpSpace#4}[#2][#3]{#1}\xspace}
\newcommandx{\ExpSpaceH}[4][1=, 2=, 3=, 4=]
	{\ExpSpace[#1][#2][#3][#4]-\HComp}
\newcommandx{\ExpSpaceE}[4][1=, 2=, 3=, 4=]
	{\ExpSpace[#1][#2][#3][#4]-\EComp}
\newcommandx{\ExpSpaceC}[4][1=, 2=, 3=, 4=]
	{\ExpSpace[#1][#2][#3][#4]-\CComp}

\newcommandx{\NExpSpace}
	{\txtname{N}\ExpSpace}
\newcommandx{\NExpSpaceH}[4][1=, 2=, 3=, 4=]
	{\NExpSpace[#1][#2][#3][#4]-\HComp}
\newcommandx{\NExpSpaceE}[4][1=, 2=, 3=, 4=]
	{\NExpSpace[#1][#2][#3][#4]-\EComp}
\newcommandx{\NExpSpaceC}[4][1=, 2=, 3=, 4=]
	{\NExpSpace[#1][#2][#3][#4]-\CComp}

\newcommandx{\CoNExpSpace}
	{\txtname{Co}\NExpSpace}
\newcommandx{\CoNExpSpaceH}[4][1=, 2=, 3=, 4=]
	{\CoNExpSpace[#1][#2][#3][#4]-\HComp}
\newcommandx{\CoNExpSpaceE}[4][1=, 2=, 3=, 4=]
	{\CoNExpSpace[#1][#2][#3][#4]-\EComp}
\newcommandx{\CoNExpSpaceC}[4][1=, 2=, 3=, 4=]
	{\CoNExpSpace[#1][#2][#3][#4]-\CComp}

\newcommandx{\AExpSpace}
	{\txtname{A}\ExpSpace}
\newcommandx{\AExpSpaceH}[4][1=, 2=, 3=, 4=]
	{\AExpSpace[#1][#2][#3][#4]-\HComp}
\newcommandx{\AExpSpaceE}[4][1=, 2=, 3=, 4=]
	{\AExpSpace[#1][#2][#3][#4]-\EComp}
\newcommandx{\AExpSpaceC}[4][1=, 2=, 3=, 4=]
	{\AExpSpace[#1][#2][#3][#4]-\CComp}

\newcommandx{\NonElm}[4][1=, 2=, 3=, 4=]
	{\txtargname{NonElementary#4}[#2][#3]{#1}\xspace}
\newcommandx{\NonElmH}[4][1=, 2=, 3=, 4=]
	{\NonElm[#1][#2][#3][#4]-\HComp}
\newcommandx{\NonElmE}[4][1=, 2=, 3=, 4=]
	{\NonElm[#1][#2][#3][#4]-\EComp}
\newcommandx{\NonElmC}[4][1=, 2=, 3=, 4=]
	{\NonElm[#1][#2][#3][#4]-\CComp}

\newcommandx{\NonElmTime}[4][1=, 2=, 3=, 4=]
	{\txtargname{NonElementaryTime#4}[#2][#3]{#1}\xspace}
\newcommandx{\NonElmTimeH}[4][1=, 2=, 3=, 4=]
	{\NonElmTime[#1][#2][#3][#4]-\HComp}
\newcommandx{\NonElmTimeE}[4][1=, 2=, 3=, 4=]
	{\NonElmTime[#1][#2][#3][#4]-\EComp}
\newcommandx{\NonElmTimeC}[4][1=, 2=, 3=, 4=]
	{\NonElmTime[#1][#2][#3][#4]-\CComp}

\newcommandx{\NonElmSpace}[4][1=, 2=, 3=, 4=]
	{\txtargname{NonElementarySpace#4}[#2][#3]{#1}\xspace}
\newcommandx{\NonElmSpaceH}[4][1=, 2=, 3=, 4=]
	{\NonElmSpace[#1][#2][#3][#4]-\HComp}
\newcommandx{\NonElmSpaceE}[4][1=, 2=, 3=, 4=]
	{\NonElmSpace[#1][#2][#3][#4]-\EComp}
\newcommandx{\NonElmSpaceC}[4][1=, 2=, 3=, 4=]
	{\NonElmSpace[#1][#2][#3][#4]-\CComp}

\newcommandx{\DLHier}[4][2=, 3=, 4=]
	{\mthargset[0]{\Delta#4}[#1][#3]{#2}\xspace}
\newcommandx{\DLHierH}[4][2=, 3=, 4=]
	{\DLHier{#1}[#2][#3][#4]-\HComp}
\newcommandx{\DLHierE}[4][2=, 3=, 4=]
	{\DLHier{#1}[#2][#3][#4]-\EComp}
\newcommandx{\DLHierC}[4][2=, 3=, 4=]
	{\DLHier{#1}[#2][#3][#4]-\CComp}

\newcommandx{\ELHier}[4][2=, 3=, 4=]
	{\mthargset[0]{\Sigma#4}[#1][#3]{#2}\xspace}
\newcommandx{\ELHierH}[4][2=, 3=, 4=]
	{\ELHier{#1}[#2][#3][#4]-\HComp}
\newcommandx{\ELHierE}[4][2=, 3=, 4=]
	{\ELHier{#1}[#2][#3][#4]-\EComp}
\newcommandx{\ELHierC}[4][2=, 3=, 4=]
	{\ELHier{#1}[#2][#3][#4]-\CComp}

\newcommandx{\ULHier}[4][2=, 3=, 4=]
	{\mthargset[0]{\Pi#4}[#1][#3]{#2}\xspace}
\newcommandx{\ULHierH}[4][2=, 3=, 4=]
	{\ULHier{#1}[#2][#3][#4]-\HComp}
\newcommandx{\ULHierE}[4][2=, 3=, 4=]
	{\ULHier{#1}[#2][#3][#4]-\EComp}
\newcommandx{\ULHierC}[4][2=, 3=, 4=]
	{\ULHier{#1}[#2][#3][#4]-\CComp}

\newcommandx{\DBHier}[4][2=, 3=, 4=]
	{\mthargset[3]{\Delta#4}[#1][#3]{#2}\xspace}
\newcommandx{\DBHierH}[4][2=, 3=, 4=]
	{\DBHier{#1}[#2][#3][#4]-\HComp}
\newcommandx{\DBHierE}[4][2=, 3=, 4=]
	{\DBHier{#1}[#2][#3][#4]-\EComp}
\newcommandx{\DBHierC}[4][2=, 3=, 4=]
	{\DBHier{#1}[#2][#3][#4]-\CComp}

\newcommandx{\EBHier}[4][2=, 3=, 4=]
	{\mthargset[3]{\Sigma#4}[#1][#3]{#2}\xspace}
\newcommandx{\EBHierH}[4][2=, 3=, 4=]
	{\EBHier{#1}[#2][#3][#4]-\HComp}
\newcommandx{\EBHierE}[4][2=, 3=, 4=]
	{\EBHier{#1}[#2][#3][#4]-\EComp}
\newcommandx{\EBHierC}[4][2=, 3=, 4=]
	{\EBHier{#1}[#2][#3][#4]-\CComp}

\newcommandx{\UBHier}[4][2=, 3=, 4=]
	{\mthargset[3]{\Pi#4}[#1][#3]{#2}\xspace}
\newcommandx{\UBHierH}[4][2=, 3=, 4=]
	{\UBHier{#1}[#2][#3][#4]-\HComp}
\newcommandx{\UBHierE}[4][2=, 3=, 4=]
	{\UBHier{#1}[#2][#3][#4]-\EComp}
\newcommandx{\UBHierC}[4][2=, 3=, 4=]
	{\UBHier{#1}[#2][#3][#4]-\CComp}

\newcommandx{\DPolHier}[4][2=, 3=, 4=]
	{\DLHier{#1}[#2][\argb{\mathrm{P}}{#3}][#4]}
\newcommandx{\DPolHierH}[4][2=, 3=, 4=]
	{\DPolHier{#1}[#2][#3][#4]-\HComp}
\newcommandx{\DPolHierE}[4][2=, 3=, 4=]
	{\DPolHier{#1}[#2][#3][#4]-\EComp}
\newcommandx{\DPolHierC}[4][2=, 3=, 4=]
	{\DPolHier{#1}[#2][#3][#4]-\CComp}

\newcommandx{\EPolHier}[4][2=, 3=, 4=]
	{\ELHier{#1}[#2][\argb{\mathrm{P}}{#3}][#4]}
\newcommandx{\EPolHierH}[4][2=, 3=, 4=]
	{\EPolHier{#1}[#2][#3][#4]-\HComp}
\newcommandx{\EPolHierE}[4][2=, 3=, 4=]
	{\EPolHier{#1}[#2][#3][#4]-\EComp}
\newcommandx{\EPolHierC}[4][2=, 3=, 4=]
	{\EPolHier{#1}[#2][#3][#4]-\CComp}

\newcommandx{\UPolHier}[4][2=, 3=, 4=]
	{\ULHier{#1}[#2][\argb{\mathrm{P}}{#3}][#4]}
\newcommandx{\UPolHierH}[4][2=, 3=, 4=]
	{\UPolHier{#1}[#2][#3][#4]-\HComp}
\newcommandx{\UPolHierE}[4][2=, 3=, 4=]
	{\UPolHier{#1}[#2][#3][#4]-\EComp}
\newcommandx{\UPolHierC}[4][2=, 3=, 4=]
	{\UPolHier{#1}[#2][#3][#4]-\CComp}

\newcommandx{\DAriHier}[4][2=, 3=, 4=]
	{\DLHier{#1}[#2][\argb{0}{#3}][#4]}
\newcommandx{\DAriHierH}[4][2=, 3=, 4=]
	{\DAriHier{#1}[#2][#3][#4]-\HComp}
\newcommandx{\DAriHierE}[4][2=, 3=, 4=]
	{\DAriHier{#1}[#2][#3][#4]-\EComp}
\newcommandx{\DAriHierC}[4][2=, 3=, 4=]
	{\DAriHier{#1}[#2][#3][#4]-\CComp}

\newcommandx{\EAriHier}[4][2=, 3=, 4=]
	{\ELHier{#1}[#2][\argb{0}{#3}][#4]}
\newcommandx{\EAriHierH}[4][2=, 3=, 4=]
	{\EAriHier{#1}[#2][#3][#4]-\HComp}
\newcommandx{\EAriHierE}[4][2=, 3=, 4=]
	{\EAriHier{#1}[#2][#3][#4]-\EComp}
\newcommandx{\EAriHierC}[4][2=, 3=, 4=]
	{\EAriHier{#1}[#2][#3][#4]-\CComp}

\newcommandx{\UAriHier}[4][2=, 3=, 4=]
	{\ULHier{#1}[#2][\argb{0}{#3}][#4]}
\newcommandx{\UAriHierH}[4][2=, 3=, 4=]
	{\UAriHier{#1}[#2][#3][#4]-\HComp}
\newcommandx{\UAriHierE}[4][2=, 3=, 4=]
	{\UAriHier{#1}[#2][#3][#4]-\EComp}
\newcommandx{\UAriHierC}[4][2=, 3=, 4=]
	{\UAriHier{#1}[#2][#3][#4]-\CComp}

\newcommandx{\DAnaHier}[4][2=, 3=, 4=]
	{\DLHier{#1}[#2][\argb{1}{#3}][#4]}
\newcommandx{\DAnaHierH}[4][2=, 3=, 4=]
	{\DAnaHier{#1}[#2][#3][#4]-\HComp}
\newcommandx{\DAnaHierE}[4][2=, 3=, 4=]
	{\DAnaHier{#1}[#2][#3][#4]-\EComp}
\newcommandx{\DAnaHierC}[4][2=, 3=, 4=]
	{\DAnaHier{#1}[#2][#3][#4]-\CComp}

\newcommandx{\EAnaHier}[4][2=, 3=, 4=]
	{\ELHier{#1}[#2][\argb{1}{#3}][#4]}
\newcommandx{\EAnaHierH}[4][2=, 3=, 4=]
	{\EAnaHier{#1}[#2][#3][#4]-\HComp}
\newcommandx{\EAnaHierE}[4][2=, 3=, 4=]
	{\EAnaHier{#1}[#2][#3][#4]-\EComp}
\newcommandx{\EAnaHierC}[4][2=, 3=, 4=]
	{\EAnaHier{#1}[#2][#3][#4]-\CComp}

\newcommandx{\UAnaHier}[4][2=, 3=, 4=]
	{\ULHier{#1}[#2][\argb{1}{#3}][#4]}
\newcommandx{\UAnaHierH}[4][2=, 3=, 4=]
	{\UAnaHier{#1}[#2][#3][#4]-\HComp}
\newcommandx{\UAnaHierE}[4][2=, 3=, 4=]
	{\UAnaHier{#1}[#2][#3][#4]-\EComp}
\newcommandx{\UAnaHierC}[4][2=, 3=, 4=]
	{\UAnaHier{#1}[#2][#3][#4]-\CComp}

\newcommandx{\DBorHier}[4][2=, 3=, 4=]
	{\DBHier{#1}[#2][\argb{\mathrm{B}}{#3}][#4]}
\newcommandx{\DBorHierH}[4][2=, 3=, 4=]
	{\DBorHier{#1}[#2][#3][#4]-\HComp}
\newcommandx{\DBorHierE}[4][2=, 3=, 4=]
	{\DBorHier{#1}[#2][#3][#4]-\EComp}
\newcommandx{\DBorHierC}[4][2=, 3=, 4=]
	{\DBorHier{#1}[#2][#3][#4]-\CComp}

\newcommandx{\EBorHier}[4][2=, 3=, 4=]
	{\EBHier{#1}[#2][\argb{\mathrm{B}}{#3}][#4]}
\newcommandx{\EBorHierH}[4][2=, 3=, 4=]
	{\EBorHier{#1}[#2][#3][#4]-\HComp}
\newcommandx{\EBorHierE}[4][2=, 3=, 4=]
	{\EBorHier{#1}[#2][#3][#4]-\EComp}
\newcommandx{\EBorHierC}[4][2=, 3=, 4=]
	{\EBorHier{#1}[#2][#3][#4]-\CComp}

\newcommandx{\UBorHier}[4][2=, 3=, 4=]
	{\UBHier{#1}[#2][\argb{\mathrm{B}}{#3}][#4]}
\newcommandx{\UBorHierH}[4][2=, 3=, 4=]
	{\UBorHier{#1}[#2][#3][#4]-\HComp}
\newcommandx{\UBorHierE}[4][2=, 3=, 4=]
	{\UBorHier{#1}[#2][#3][#4]-\EComp}
\newcommandx{\UBorHierC}[4][2=, 3=, 4=]
	{\UBorHier{#1}[#2][#3][#4]-\CComp}

\newcommand{\HComp}
	{\txtname{hard}\xspace}

\newcommand{\EComp}
	{\txtname{easy}\xspace}

\newcommand{\CComp}
	{\txtname{complete}\xspace}

\newtheorem{theorem}{Theorem}

\newtheorem{proposition}{Proposition}

\newtheorem{definition}{Definition}

\newcommandx{\CTLSCD}[5][1=, 2=, 3=, 4=, 5=]
	{\txtargname{CTL$^{\star}_{\circlearrowleft}$#5{\small\argint{$[$}{#1}{$]$}}}
	[#2][#3]{#4}\xspace}

\newcommandx{\GCTLSCD}[5][1=, 2=, 3=, 4=, 5=]
	{\txtargname{GCTL$^{\star}_{\circlearrowleft}$#5{\small\argint{$[$}{#1}{$]$}}}
	[#2][#3]{#4}\xspace}

\newcommandx{\ECTLSCD}[5][1=, 2=, 3=, 4=, 5=]
	{\txtargname{ECTL$^{\star}_{\circlearrowleft}$#5{\small\argint{$[$}{#1}{$]$}}}
	[#2][#3]{#4}\xspace}

\newcommandx{\LTLCD}[5][1=, 2=, 3=, 4=, 5=]
	{\txtargname{LTL$_{\circlearrowleft}$#5{\small\argint{$[$}{#1}{$]$}}}
	[#2][#3]{#4}\xspace}

\newcommand{\CycSet}{\mthset{Cyc}}

\newcommand{\cycle}{\circlearrowleft}

\newcommand{\cycE}{\E^{\cycle}}
\newcommand{\cycA}{\A^{\cycle}}

\newcommand{\new}{\mthsym{new}}
\newcommand{\news}{\mthsym{n}}

\newcommand{\pr}{\mthsym{pr}}
\newcommand{\cy}{\mthsym{cycle}}
\newcommand{\cys}{\mthsym{c}}

\newcommand{\scycE}{\E^{\cycle}_s}
\newcommand{\cytr}[1]{(#1)_s}
\newcommand{\decpnt}{\mthsym{dec}}
\newcommand{\res}{\mthsym{res}}

\usepackage{tikz}
\usetikzlibrary{arrows,shapes,calc}

\usepackage{wrapfig}

\tikzstyle{every node} =
	[draw = none, fill = white, thin]
\tikzstyle{every edge} +=
	[black, thick]

\tikzstyle{noall} =
	[draw = none, fill = none]
\tikzstyle{nodraw} =
	[draw = none, fill = white]
\tikzstyle{nofill} =
	[draw = black, fill = none]

\tikzstyle{cnode} =
	[circle, draw = black]
\tikzstyle{snode} =
	[regular polygon, regular polygon sides = 4, draw = gray!50]
\tikzstyle{lnode} =
	[diamond, draw = gray!75]
\tikzstyle{pnode} =
	[regular polygon, regular polygon sides = 5, draw = gray]

\AfterEndPreamble
	{

	\newcommand{\figexmkrp}
		{
		\begin{wrapfigure}[8]{l}{0.240\textwidth}
			\vspace{-1.50em}
			\centering
			\parbox{0.200\textwidth}
				{%
				\begin{center}
					\footnotesize
					\scalebox{1.00}[1.00]
						{
						\begin{tikzpicture}[node distance = 5em, bend angle = 45]

						\node []
									(I)
									[]
									{$\vElm$};
						\node []
									(X)
									[below of = I]
									{$\wElm$};

						\path[->]
							(I)		edge	[]
													(X)
							(X)		edge	[loop right]
													()
										edge	[bend left]
													(I)
							;

					\end{tikzpicture}
						}
				\end{center}
				\vspace{-1.25em}
				\caption{\label{fig:krpstr0} \small \!\! The Kripke Model $\KName[0]$.}
				}
		\end{wrapfigure}
		}

	\newcommand{\figexmtreeunw}
		{
		\begin{wrapfigure}[10]{r}{0.510\textwidth}
			\vspace{-3.75em}
			\centering

				\begin{center}
					\footnotesize
					\scalebox{0.90}[0.90]
						{
					\begin{tikzpicture}[node distance = 4em, bend angle = 30]

						\node []
									(eps)
									[]
									{$\epsilon$};
						\node []
									(wn)
									[below left of = eps]
									{$(w, n)$};
						\node [node distance = 8em]
									(wnwn)
									[below right of = wn]
									{$(w, n)$};
						\node [node distance = 8em]
									(wnvn)
									[left of = wnwn]
									{$(v, n)$};
						\node [node distance = 12em]
									(wnwc)
									[right of = wnwn]
									{$(w, c)$};
						\node []
									(wnvnwn)
									[below of = wnvn]
									{$(w, n)$};
						\node []
									(wnwnwn)
									[below of = wnwn]
									{$(w, n)$};
						\node [node distance = 4em]
									(wnwnvn)
									[left of = wnwnwn]
									{$(v, n)$};
						\node [node distance = 4em]
									(wnwnvc)
									[right of = wnwnwn]
									{$(v,	 n)$};
						\node []
									(wnwcwn)
									[below of = wnwc]
									{$(w, n)$};
						\node []
									(wnwcvn)
									[left of = wnwcwn]
									{$(v, n)$};
						\node []
									(wnwcwc)
									[right of = wnwcwn]
									{$(w, c)$};
						\node [node distance = 1.75em]
									(dot)
									[below of = wnwnvc]
									{\Huge $\dots$};
						\path[->]
							(eps)		edge	[]
													(wn)
							(wn)		edge	[]
													(wnvn)
										edge	[]
													(wnwn)
										edge	[]
													(wnwc)
							(wnvn)		edge	[]
													(wnvnwn)
							(wnwn)		edge	[]
													(wnwnvn)
										edge	[]
													(wnwnwn)
										edge	[]
													(wnwnvc)
							(wnwc)		edge	[]
													(wnwcvn)
										edge	[]
													(wnwcwn)
										edge	[]
													(wnwcwc)
							(wn)		edge[dotted, bend left]
													(eps)
							(wnvn)		edge[dotted, bend left]
													(wn)
							(wnwn)		edge[dotted, bend left]
													(wn)
							(wnwc)		edge[dotted]
													(eps)
							(wnvnwn)		edge[dotted, bend left]
													(wnvn)
							(wnwnvn)		edge[dotted, bend left]
													(wnwn)
							(wnwnwn)		edge[dotted, bend left]
													(wnwn)
							(wnwnvc)		edge[dotted, bend angle = 20, bend right]
													(wn)
							(wnwcvn)		edge[dotted, bend left]
													(wnwc)
							(wnwcwn)		edge[dotted, bend left]
													(wnwc)
							(wnwcwc)		edge[dotted, bend angle = 45, bend right]
													(eps)
							;

					\end{tikzpicture}
						}
				\end{center}
				\vspace{-1.25em}
				\caption{\label{fig:treelike} \small The tree with back edges
				$\TName(\KName[0])$.}

		\end{wrapfigure}
		}

	}

\begin{document}
	\maketitle



\begin{abstract}

	Temporal logic is a very powerful formalism deeply investigated and used in formal
	system design and verification.
	Its application usually reduces to solving specific decision problems such as model
	checking and satisfiability.
	In these kind of problems, the solution often requires detecting some specific
	properties over cycles.
	For instance, this happens when using classic techniques based on automata, game-theory,
	SCC decomposition, and the like.
	Surprisingly, no temporal logics have been considered so far with the explicit ability of talking about cycles.

	In this paper we introduce Cycle-\CTLS, an extension of the classical
	branching-time temporal logic \CTLS along with cycle quantifications in order
	to predicate over cycles.
	This logic turns out to be very expressive.
	Indeed, we prove that it strictly extends \CTLS and is orthogonal to \MC.
	We also give an evidence of its usefulness by providing few examples involving
	non-regular properties.

	We investigate the model checking problem for Cycle-\CTLS and show that it is
	\PSpaceC as for \CTLS.
	We also study the satisfiability problem for the existential-cycle fragment of
	the logic and show that it is solvable in 2\ExpTime.
	This result makes use of an	automata-theoretic approach along with novel
	\adhoc definitions of bisimulation and tree-like unwinding.

\end{abstract}




\begin{section}{Introduction}

	\emph{Temporal logic} is a suitable framework largely used in formal system
	verification~\cite{Pnu77,CE81,EH85,CGP02}.
	It allows to specify and reasoning in a rigorous manner about the temporal
	evolution of a system, without talking explicitly about the elapsing of time.
	Two fundamental decision problems involving temporal logics have been deeply
	investigated: \emph{model checking} and \emph{satisfiability}.
	The former, given a mathematical model of the system, such as a \emph{Kripke}
	structure, asks whether it satisfies a temporal logic formula specifying its
	desired behavior.
	The latter, instead, checks whether the temporal logic specification is
	consistent and, thus, a corresponding system is feasible~\cite{CGP02}.

	In several situations, reasoning about system correctness and, in particular,
	solving the above decision questions, reduces to detect precise cycle
	properties over the system model.
	For example, in the classical automata-theoretic approach there are settings
	in which the satisfiability question reduces to first build a B\"uchi
	automaton accepting all models of the formula and then to check for its
	non-emptiness~\cite{KVW00}.
	The latter can be solved by looking for a ``lasso'', that is a path from the
	initial state to a final state belonging to a cycle~\cite{KVW00,KMM06}.
	Similarly, if one uses a game-theory approach, solving the model checking or
	the satisfiability questions reduces to first construct a two-player game,
	such as a B\"uchi or a parity game~\cite{EJ88,KVW00,KVW01,alur2004deterministic,GTW02,Zie98}, 
	and then check for the existence of a winning strategy for a designed player.
	The latter can be reduced to check whether it has the ability to confine the
	evolution of the game (a \emph{play}) over some specific cycle over the arena,
	no matter how the other player behaves.

	Depending on the view of the underlying nature of time, two types of temporal
	logics are mainly considered.
	In \emph{linear-time temporal logics}, such as \LTL~\cite{Pnu77}, time is
	treated as if each moment in time has a unique possible future.
	Conversely, in \emph{branching-time temporal logics} such as \CTL~\cite{CE81}
	and \CTLS~\cite{EH86} each moment in time may split into various possible
	futures.
	Then, to express properties along one or all the possible futures we make use
	of existential and universal quantifiers.
	Noticeably, \LTL is suitable to express path properties; \CTL is more
	appropriate to express state-based property; finally, \CTLS has the power to
	express combinations of path and state properties.
	In the years, these logics have been extended in a number of ways in order to
	express very complicated specification properties. 
    Surprisingly, no temporal logic has been introduced so far to reason explicitly about cycles, despite their usefulness. In addition to the technical motivation mentioned above, there are often cases in which it is useful to distinguish between purely infinite behaviors, like those occurring in infinite-state systems, from regular infinite behaviors~\cite{BMP10,KPV02}.
    Moreover, also in finite-state systems there are infinite behaviors that are not regular, like the \emph{prompt}ones~\cite{KPV09,MMS15prompt}, which we can distinguish by using our new concept of cycling path, as we show in an example later in the paper.
  
	In this paper we introduce \emph{Cycle-\CTLS}, an extension of the classical
	logic \CTLS along with the ability to predicate over cycles.
	For a \emph{cycle} we mean a path that passes through its initial state
	infinitely often.
	Syntactically, Cycle-\CTLS is obtained by enriching \CTLS with two novel cycle
	quantifiers, namely the existential one $\cycE$ and the universal one $\cycA$.
	Note that Cycle-\CTLS still uses the classical quantifiers $\E$ and $\A$.
	Hence, we can use it to specify models whose behavior results as an opportune
	combination of standard paths and cycles.
	In particular, Cycle-\CTLS can specify the existence of a lasso within a model.

	We study the expressiveness of Cycle-\CTLS and show that it is strictly more
	expressive than \CTLS but orthogonal to $\mu$-calculus.
	To give an evidence of the power and usefulness of the introduced logic, we
	provide some examples along the paper.
	Precisely, we first show how Cycle-\CTLS can be used to reasoning, in a very
	natural way, about liveness properties restricted to cycles.
	Precisely, we show how to specify that some designed property recurrently
	occurs in the starting state of a cycle.
	As another example, we show the ability of the logic to handle non-regular
	properties such as the ``\emph{prompt-parity condition}''~\cite{MMS15prompt}.
	In temporal logic, we can specify properties that will eventually hold, but
	this gives no bound on the moment they will occur.
	Prompt temporal logics and games have been deeply investigated in order to
	restrict reasoning about properties that only occur in bounded
	time~\cite{CHH09,AH98,KPV12,AMRZ16,MMS15prompt}.

	We investigate both the model checking and the satisfiability questions for
	Cycle-\CTLS and provide some automata-based solutions.
	For the model checking question we provide a \PSpace upper-bound by
	opportunely extending the classical approach that is used for \CTLS~\cite{KVW00}.
	Specifically, we add a machinery consisting of an appropriate B\"uchi
	automaton that checks in parallel whether a path is a cycle and satisfies a
	required formula.
	Concerning the satisfiability question, we introduce instead a novel approach
	that makes use of two-way automata~\cite{Var98}.
	These automata, largely investigate and used in formal
	verification~\cite{BLMV08,FMP08,KPV02}, allow to traverse trees both in forward and
	backward.
	The reason why we cannot use and extend the classical approach provided for
	\CTLS (see~\cite{KVW00}) resides on the fact that such an approach makes
	strongly use of some positive properties that hold for \CTLS, among the
	others the tree- and the finite-model ones.
	Unluckily and unsurprisingly, due to the ability in Cycle-\CTLS to force (and
	even more to forbid) the existence of cycles, we lose in this logic both
	these properties.
	This requires the introduction of novel and \adhoc definitions of bisimulation
	and tree-like unwinding to be used along with the automata-based approach.
	In particular, two-way tree automata are used to collect all tree
	representations of such tree-like unwinding structures.
	By means of this machinery we show that the satisfiability question for the
	full logic is 3\ExpTime.
	We also investigate the satisfiability of the existential-cycle fragment and
	show that it is solvable in 2\ExpTime, thus it is not harder than \CTLS in
	complexity.
	Such fragment is simply obtained by forbidding the use of the universal cycle
	quantifier $\cycA$ and by only allowing negations over atomic propositions.
	Note that this fragment still admits the classical $\E$ and $\A$ quantifiers,
	as well as it strictly subsumes \CTLS.


\end{section}





\begin{section}{Computation-Tree Logic with Cycle Detection}
	\label{sec:ctlcd}

	In this section we introduce and discuss the syntax and semantics of
	Cycle-\CTLS (\CTLSCD, for short) and
discuss some interesting problems that can be expressed in
	our logic.

	\begin{subparagraph}{Models}
		
		We first provide the definition of the underlying model for our Cycle-\CTLS.

	\begin{definition}[Kripke Structure]
		\label{def:krpstr}
		A \emph{Kripke structure} (\KrpStr, for short)~\cite{Kri63} over a finite
		set of \emph{atomic propositions} $\APSet$ is a tuple $\KName \defeq
		\KrpStrStr$, where $\WSet$ is an enumerable non-empty set of \emph{worlds},
		$\wElm[0] \in \WSet$ is a designated \emph{initial world}, $\RRel \subseteq
		\WSet \times \WSet$ is a left-total \emph{transition relation}, and $\labFun
		: \WSet \mapsto \pow{\APSet}$ is a \emph{labeling function} mapping each
		world to the set of atomic propositions true in that world.
	\end{definition}

	A \emph{path} in $\KName$ is an infinite sequence of worlds $\pthElm \in \PthSet \subseteq \WSet^{\omega}$ such that, for all $i \in \SetN$, it holds that $((\pthElm)_{i}, (\pthElm)_{i + 1}) \in \RRel$.
  We denote by $\fst{\pthElm} \defeq \pthElm_{0}$ and $(\pthElm)_{i} \defeq \pthElm_{i}$ the first and $i$-th element of $\pthElm$.
  %
  For a path $\pthElm$, we say that $\pthElm$ is a \emph{cycle} if, for all $i \in \SetN$, there exists $j \in \SetN$, with $j > i$, such that $(\pthElm)_{j} = \fst{\pthElm}$.
  For a given path $\pthElm$, we denote by $\labFun(\pthElm)$ the sequence $\gamma$ in $(\pow{\APSet})^{\omega}$ such that $(\gamma)_{i} =
  \labFun(\pthElm_{i})$ for all $i \in \SetN$.
  Moreover, $(\pthElm)_{\leq i} \defeq \pthElm_{0} \cdots \pthElm_{i}$ and
  $(\pthElm)_{\geq i} \defeq \pthElm_{i} \cdot \pthElm_{i + 1} \cdots$ represent the prefix up to and the suffix from position $i$ of $\pthElm$.
  Prefixes of a path are also called \emph{tracks} and denoted by $\trkElm \in \TrkSet \subseteq \WSet^{+}$.
  We also denote by $\lst{\trkElm}$ the last element occurring in the track $\trkElm$.
  Finally, all the definitions given above for paths naturally apply to tracks.
  
  By $\TrkSet(\worElm)$ and $\PthSet(\worElm)$ we denote the set of tracks
  and paths starting from $\worElm$, respectively.
  By $\CycSet$ and $\CycSet(\worElm)$ we denote the set of cycles and
  the set of cycles starting from $\worElm$, respectively.
	Intuitively, tracks and paths of a \KrpStr $\KName$ are legal sequences, either finite or infinite, of reachable worlds that can be seen as partial or complete descriptions of possible \emph{computations} of the system modelled by $\KName$.
	%


	For a pair $(\worElm[1], \worElm[2]) \in \RRel$, we say that $\worElm[2]$ is
	an $\RRel$-successor of $\worElm[1]$.
	Note that in case $\RRel$ is a function, then each world $\worElm$ has only
	one $\RRel$-successor.
	This implies that, starting from the initial world $\worElm[0]$, there is a
	unique legal path.
	Such structures are called 
	\emph{\LTL models}.

	\end{subparagraph}

	\begin{subparagraph}{Syntax}

		\CTLSCD extends \CTLS~\cite{CES86} by means of two additional path
		operators, $\cycE \psi$ and $\cycA \psi$, which respectively read as
		``there exists a cycle path satisfying $\psi$'' and ``for all cycle paths
		$\psi$ holds''.
		As for \CTLS, the syntax includes path-formulas, expressing properties over
		sequences of words, and state-formulas, expressing properties over a single
		word.
		State and path formulas are defined by mutual induction as follows.


		\begin{definition}[\CTLSCD syntax]
			\label{def:syn}
			\CTLSCD formulas are inductively built from a set of atomic propositions
			$\APSet$, by using the following grammar, where $\pElm \in \APSet$:
			\begin{center}
				$\phi  := \pElm \mid  \neg \phi  \mid  \phi \wedge \phi  \mid  \phi \vee
				\phi \mid \E \psi  \mid  \A \psi \mid \cycE \psi  \mid  \cycA \psi$ \\
				$\psi :=  \phi  \mid   \neg \psi  \mid  \psi \wedge \psi  \mid  \psi
				\vee \psi  \mid  \X \psi  \mid  \psi \U \psi$
			\end{center}
		\end{definition}

		All the formulas generated by a $\phi$-rule are called
		\emph{state-formulas}, while the formulas generated by a $\psi$-rule are
		called \emph{path-formulas}.
		By $\subFun{\varphi}$ we denote the set of all
                subformulas of $\varphi$,
		and by $\subFun[s]{\varphi}$ we denote the set of state subformulas of
		$\varphi$.

	\end{subparagraph}

	\begin{subparagraph}{Semantics}

	The semantics for \CTLSCD is defined \wrt Kripke structures.
	It extends the one for \CTLS, with the addition of two new definitions for
	two cycle path quantifiers.

		\begin{definition}
			\label{def:sem}
			The semantics of \CTLSCD formulas is recursively defined as follows.
			For a Kripke structure $\KrpStrName$, a world $\wElm$, a path $\pthElm$
			and a natural number $i \in \SetN$, we have that:
			\begin{itemize}
				\item
					For all state formulas $\phi$, $\phi_{1}$, and $\phi_{2}$:
					\begin{itemize}
						\item
							$\KrpStrName, \wElm \models \pElm$ if $\pElm \in \LFun(\wElm)$;
						\item
							$\KrpStrName, \wElm \models \neg \phi$ if $\KrpStrName, \wElm
							\not\models \phi$;
						\item
							$\KrpStrName, \wElm \models \phi_{1} \wedge \phi_{2}$ if both
							$\KrpStrName, \wElm \models \phi_{1}$ and $\KrpStrName, \wElm
							\models \phi_{2}$;
						\item
							$\KrpStrName, \wElm \models \phi_{1} \vee \phi_{2}$ if either
							$\KrpStrName, \wElm \models \phi_{1}$ or $\KrpStrName, \wElm
							\models \phi_{2}$;
						\item
							$\KrpStrName, \wElm \models \E \psi$ if there exists a path
							$\pthElm$ in $\PthSet(\wElm)$ such that $\KrpStrName, \pthElm, 0
							\models \psi$;
						\item
							$\KrpStrName, \wElm \models \A \psi$ if, for all paths $\pthElm$
							in $\PthSet(\wElm)$, it holds that $\KrpStrName, \pthElm, 0
							\models \psi$
						\item
							$\KrpStrName, \wElm \models \cycE \psi$ if there exists a path
							$\pthElm$ in $\CycSet(\wElm)$ and $\KrpStrName, \pthElm, 0 \models
							\psi$;
						\item
							$\KrpStrName, \wElm \models \cycA \psi$ if, for all paths
							$\pthElm$ in $\CycSet(\wElm)$, it holds that $\KrpStrName,
							\pthElm, 0 \models \psi$.
					\end{itemize}
				\item
					For path formulas $\phi$, $\psi$, $\psi_{1}$, and $\psi_{2}$:
					\begin{itemize}
						\item
							$\KrpStrName, \pthElm, i \models \phi$ if $\KrpStrName,
							(\pthElm)_{i} \models \phi$;
						\item
							$\KrpStrName, \pthElm, i \models \neg \psi$ if $\KrpStrName,
							\pthElm, i  \not\models \psi$;
						\item
							$\KrpStrName, \pthElm, i \models \psi_{1} \wedge \psi_{2}$ if both
							$\KrpStrName, \pthElm, i \models \psi_{1}$ and $\KrpStrName, \wElm
							\models \psi_{2}$;
						\item
							$\KrpStrName, \pthElm, i \models \psi_{1} \vee \psi_{2}$ if either
							$\KrpStrName, \pthElm, i \models \psi_{1}$ or $\KrpStrName, \wElm
							\models \psi_{2}$;
						\item
							$\KrpStrName, \pthElm, i \models \X \psi$ if $\KrpStrName,
							\pthElm, i + 1  \models \psi$;
						\item
							$\KrpStrName, \pthElm, i \models \psi_{1} \U \psi_{2}$ if there
							exists $k \in \SetN$ such that $\KrpStrName, \pthElm, i + k
							\models \psi_{2}$ and $\KrpStrName, \pthElm, i + j  \models
							\psi_{1}$, for all $j \in \numco{0}{k}$;
					\end{itemize}
			\end{itemize}
			We say that $\pthElm$ satisfies the path formula $\phi$ over
			$\KrpStrName$, and write $\KrpStrName, \pthElm \models \phi$, if
			$\KrpStrName, \pthElm, 0 \models \phi$.
			Also, we say that $\KrpStrName$ satisfies the state formula $\varphi$, and
			write $\KrpStrName \models \varphi$, if $\KrpStrName, \worElm[I] \models
			\varphi$.
		\end{definition}

	\end{subparagraph}

	\begin{subparagraph}{Examples}

		In this section, we provide some properties that are expressible with
		\CTLSCD.

		Assume that there is a system composed by two processes, requesting to
		access a resource, and a scheduler, releasing such resource in a fair way,
		\ie, the resource is never used by the two processes at the same time.
		Every time the scheduler grants the resource to process $i$, such resource
		is exclusively used by process $i$ until the system goes back to the
		decision point, that is, the state in which the scheduler released the
		resource.
		We denote by $\decpnt$ the atomic proposition labeling the states that
		are decision points (that is, the moment where the scheduler makes a
		decision) and by $\res[1]$, $\res[2]$ the atomic propositions representing
		the fact that the resource is released to processes $1$ and $2$,
		respectively.
		The above described situation can be expressed with the \CTLSCD formula
		$\varphi_{i} = \cycE((\decpnt \wedge \neg \res[i] \wedge \G \neg \res[1 -
		i]) \to \F \; \res[i])$, for $i \in \{ 1, 2 \}$.
		Note that in that formula, the use of the cycle operator is crucial as it
		allows us to loop at the decision point.
		As another example, we can also force the system to satisfy the mutual
		exclusion property in each possible decision point by means of the formula
		$\A \G (\decpnt \to \varphi_{1} \wedge \varphi_{2})$.
		Finally, note that, since the system is required to loop on a decision point
		from which it is possible to release the resource for either process $1$ or
		process $2$, this automatically implies the existence of an infinite path
		which is able to satisfy the fairness condition, which is expressible in
		\CTLS by means of the formula $\psi = \E (\G \F \res[1] \wedge \G \F
		\res[2])$.
		In other words, we have that $\varphi_{1} \wedge \varphi_{2} \to \psi$ is a
		valid \CTLSCD formula.

		We now discuss another example involving prompt parity games, introduced
		in~\cite{MMS15prompt}.

		A Parity Game is a tuple of the form $\PName =
		\tuplef{\VSet}{\VSet[0]}{\VSet[1]}{\ERel}{\pFun}{\vElm[0]}$ where $\VSet$ is
		a nonempty finite set of states of the game, partitioned into $\VSet[0]$ and
		$\VSet[1]$, being the set belonging to Player $0$ and Player $1$,
		respectively, $\ERel \subseteq \VSet \times \VSet$ is an edge relation,
		$\pFun: \VSet \to \SetN$ is a priority labeling function, assigning a
		natural number to each state, and $\vElm[0] \in \VSet$ is a designated
		initial state.
		The game is played starting from $\vElm[0]$.
		At each state $\vElm$ of the game, if $\vElm \in \VSet[i]$, then Player $i$
		move to an $\ERel$-successor of $\vElm$.
		Such operation induces an infinite path $\pthElm$ over $\VSet$ called play
		and then, by means of the function $\pFun$, we also consider the infinite
		path $\pFun(\pthElm)$.
		Every occurrence of an odd priority on $\pFun(\pthElm)$ is called request.
		For any request, the successive occurrence of an even and greater priority
		is its response.
		We say that Player $0$ wins the play $\pthElm$ under the parity condition if
		every request occurring infinitely often is responded.
		Moreover, we say that Player $0$ wins the play $\pthElm$ under the prompt
		parity condition if there exists a natural number $n$ such that each request
		occurring infinitely often is responded in less than $n$ steps.
		For both the cases above, we say that Player $1$ wins the game iff Player
		$0$ does not win.
		A strategy for Player $i$ is a function $\fFun[i]: \VSet[][*] \cdot \VSet[i]
		\to \VSet$ assigning an $\ERel$-successor to each partial (finite) path of
		the game.
		Clearly, a pair of strategies $\fFun[0]$ and $\fFun[1]$ determines a unique
		path and therefore, the winner.
		A strategy $\fFun[i]$ is \emph{positional} if, for all partial paths
		$\trkElm$ and $\trkElm'$, with $\lst{\trkElm} = \lst{\trkElm'} \in
		\VSet[i]$, it holds that $\fFun[i](\trkElm) = \fFun[i](\trkElm')$.

		Let $\PName$ be a parity game and $\fFun[0]$ be a positional strategy for
		Player $0$.
		By projecting the strategy on the arena, we obtain a \KrpStr
		$\KName[ {\PName, \fFun[0]} ] = \KrpStrStr$ defined as follows:
		$\APSet = \rng{\pFun} = \numcc{0}{n}$~\footnote{\Wlogx, we can assume that
		the range of a priority function is an initial segment of $\SetN$.}, for
		some $n \in \SetN$, $\WorSet = \VSet$, $\RRel = \fFun[0] \cup \ERel \cap
		(\VSet[1] \times \VSet)$, $\labFun(\worElm) = \{\pFun(\worElm)\}$, for all
		$\worElm \in \WorSet$, and $\worElm[I] = \vElm[0]$.
		We can express that $\fFun[0]$ is winning for Player $0$ by means of the
		formula $\varphi^{par} = \A(\bigvee_{k \equiv_{2}0}(\G \F k \wedge
		\bigwedge_{l \geq k, l \equiv_{2} 1} \F \G \neg l))$.
		Indeed, the formula says that, for all possible paths, there exists an even
		priority $k$ occurring infinitely often such that each odd priority $l$
		greater than $k$ occurs finitely many times.
		Hence, we have that $\fFun[0]$ is winning over $\PName$ iff
		$\KName[ {\PName, \fFun[0]} ], \vElm[0] \models \varphi^{par}$.

		In addition to this, we can express the existence of a path violating the
		prompt condition by means of the formula $\varphi^{npmt} = \bigvee_{n
		\equiv_{2} 0 }\E(\bigvee_{k < n, k \equiv_{2} 1}(\G \F k \wedge \G(k \to
		(\bigwedge_{l \geq k, l \equiv_{2} 0} \G \neg l) \U (\cycE \bigwedge_{l \geq
		k, l \equiv_{2} 0} \allowbreak \G \neg l))))$.
		At this point, the formula $\varphi^{par} \to \varphi^{npmt}$ is able to
		express the existence of a winning strategy for Player $1$ under the prompt
		parity condition.

	\end{subparagraph}

\end{section}




\begin{section}{Model-Theoretic Properties}
	\label{sec:modprp}

	This section consists of two parts.
	First, we present invariance properties of \CTLSCD.
	As trees do not contain any cycle and bisimulation do not preserve cycles, it
	does not come at a surprise that \CTLSCD is not invariant under bisimulation
	and does not have a tree-model property.
	Therefore, we introduce a new notion of bisimulation namely
	\emph{cycle-bisimulation}, which takes cycles into account.
	We prove that \CTLSCD is invariant under cycle-bisimulation.
	Using that property, we show that \CTLSCD has a tree-like model property.

	In the second part of the section, we investigate the expressive power of
	\CTLSCD.
	We show that \CTLSCD strictly extends \CTLS and is orthogonal to the \MC.

	\begin{subparagraph}{Invariance Properties}

		We start by establishing that \CTLSCD is not invariant under
		bisimulation and does not have a tree-model or
		finite-model property.

		\begin{theorem}[\CTLSCD Negative Model Properties]
			\label{thm:ctlscd:negmodprp}
			\CTLSCD  has neither the finite-model property, nor the tree-model
			property.
			It is also not invariant under bisimulation.
		\end{theorem}

		\begin{proof}
			Consider the formula $\varphi_1 = \A \G \neg \cycE \top$ stating that all paths starting from the initial state, do not contain any cycle.
			This formula is satisfiable.
			However, since the transition relation is such that each state has a
			successor, $\varphi_1$ can only be true in an infinite model.

			Consider now the formula $\varphi_{2} = \cycE \top$.
			It is true in a model iff its initial state is the first point of a cycle.
			So $\varphi_{2}$ is satisfiable but is never true at the root of a tree.
			Hence, \CTLSCD does not have the tree-model
                        property and thus, 
is not invariant under bisimulation.
%
		\end{proof}


		\begin{definition}[Bisimulation]
			\label{def:bis}
			Let $\KName[1] = \KrpStrStr[1]$ and $\KName[2] = \KrpStrStr[2]$ be two
			Kripke structures.
			Then, a relation $\BRel \subseteq \WorSet[1] \times \WorSet[2]$
			is a \emph{cycle-bisimulation relation} if the following hold:
			\begin{enumerate}
				\item
					\label{def:bis:init}
					$(\worElm[I][1] ,\worElm[I][2])$ belongs to $\BRel$;
				\item
					\label{def:bis:bis}
					for all $\wElm[1] \in \WorSet[1]$ and $\wElm[2] \in \WorSet[2]$, if
					$(\wElm[1],\wElm[2])$ belongs to $ \BRel$, then:
					\begin{enumerate}
						\item
						\label{def:bis:bis:ap}
							$\labFun[1 ](\wElm[1]) = \labFun[2](\wElm[2])$;
						\item
						\label{def:bis:bis:forth}
							for all $\vElm[1] \in \WorSet[1]$ such that $(\wElm[1],\vElm[1])
							\in \TrnRel[1]$, there is $\vElm[2] \in \WorSet[2]$ such that
							$(\wElm[2],\vElm[2]) \in \TrnRel[2]$ and $(\vElm[1], \vElm[2]) \in
							\BRel$;
						\item
						\label{def:bis:bis:back}
							for all $\vElm[2] \in \WorSet[2]$ such that $(\wElm[2], \vElm[2])
							\in \TrnRel[2]$, there is $\vElm[1] \in \WorSet[1]$ such that
							$(\wElm[1],\vElm[1]) \in \TrnRel[1]$ and $(\vElm[1], \vElm[2]) \in
							\BRel$;

						\item
						\label{def:bis:bis:cycforth}
							for all cycles $\pthElm[1]$ with beginning state $\wElm[1]$, there
							is a cycle $\pthElm[2]$ with beginning state $\wElm[2]$ such that
							for all $i \in \SetN$, the pair $((\pthElm[1])_i,(\pthElm[2])_i)$
							belongs to $\BRel$,

						\item
						\label{def:bis:bis:cycback}
							for all cycles $\pthElm[2]$ with beginning state $\wElm[2]$
							there is a cycle $\pthElm[1]$ with beginning state $\wElm[1]$ such
							that for all $i \in \SetN$, the pair
							$((\pthElm[1])_i,(\pthElm[2])_i)$ belongs to $\BRel$.
					\end{enumerate}
			\end{enumerate}

			We say that $\KName[1]$ and $\KName[2]$ are \emph{cycle-bisimilar} 
			\wrt a relation $\BRel \subseteq \WorSet[1] \times \WorSet[2]$ if $\BRel$
			is a cycle bisimulation.
			Moreover, two paths $\pthElm[1]$ and $\pthElm[2]$ are \emph{bisimilar}
			\wrt a cycle-bisimulation 
			$\BRel$ if for all $i \in \SetN$, the pair $((\pthElm[1])_i,
			(\pthElm[2])_i)$ belongs to $\BRel$.
		\end{definition}

		The notion of cycle-bisimulation is quite intuitive.
		While the usual definition of a bisimulation allows us to ``mimic'' the
		transition relation from one model to the other, a cycle-bisimulation also
		ensures that we can ``mimic'' cycles from one model to the other.
    
    As a remark, the cycle-bisimulation notion is interesting by itself, as it gives rise to a new notion of equivalence among structures, that might lead to model-reduction characterization of the logic.
    We plan to investigate this aspect in a future work.

		\begin{theorem}[Invariance under bisimulation]
			\label{thm:ctlscd:posmodprp}
					\CTLSCD is invariant under cycle-bisimulation.
		\end{theorem}

		Using the invariance under cycle-bisimulation, we establish a tree-like
		model property for \CTLSCD.
		Intuitively, the tree-model property for \CTLSCD fails as trees do not
		admit any cycle.
		Hence, the idea is to consider structures obtained by adding some
		restricted form of cycles over trees.
		We call those structures \emph{trees with back edges} and they are defined
		as follows.

		\begin{definition}
			\label{def:treelike}
			A Kripke model $\KName = \KrpStrStr$  is a \emph{tree with back edges} if
			there are a Kripke model $\KrpTreeName[0] = (\APSet, \WSet, \RRel[0],
			\labFun, \wElm[I])$ and a partial map $\fFun: \WSet \to \WSet$ such that
			\begin{itemize}

				\item[(i)]
					$(\WSet,\RRel[0])$ is a tree with root $\wElm[I]$ over the
					alphabet~\footnote{The relation $\RRel[0]$ is the child relation of
			    the tree.} $\APSet$,

				\item[(ii)]
					$\RRel$ is equal to $\RRel[0] \cup \{ (\wElm,\fFun(\wElm)) : \wElm
					\text{ belongs to the domain of } \fFun \}$,

				\item[(iii)]
					for all $\wElm \in \WSet$, $\fFun (\wElm)$ is an ancestor of $\wElm$,

				\item[(iv)]
					\mbox{for all $\wElm[1], \wElm[2] \in \WSet$, if
					$\fFun(\wElm[1])$ is defined, $(\fFun(\wElm[2]), \wElm[1]),
					(\wElm[1],\wElm[2]) \in \RRel[0]^+$~\footnote{As usual, $\RRel[0][+]$
					is the transitive closure of $\RRel[0]$ and is the ancestor relation
					of the tree $(\WSet, \RRel[0])$.},
					then $\fFun(\wElm[1]) = \fFun(\wElm[2])$.}
			\end{itemize}

			We say that $(\TName[0],\fFun)$ is a \emph{tree decomposition} of
			$\KName$, where $\KrpTreeName[0]$ is the \emph{associated tree} and
			$\fFun$ is the \emph{back-edge map}.
			If a pair $(\wElm,\vElm)$ belongs to $\RRel[0]$, we say that $(\wElm,
			\vElm)$ is associated with a \emph{forward edge}, while if $\vElm =
			f(\wElm)$, the pair $(\wElm,\vElm)$ is associated with a
			\emph{back edge}.
		\end{definition}

		Note that if for every pair $(\wElm,\vElm)$ in $\RRel$ we know whether
		$(\wElm, \vElm)$ is associated with a forward or back edge, then this
		uniquely defines a tree decomposition.

		Intuitively, a tree with back edges is a structure obtained from a
		tree by adding edges (called back edges) from some nodes to their
		ancestors.
		More precisely, we add a back edge from each node $\wElm$ in the domain of
		$\fFun$ to its image $\fFun(\wElm)$.
		Such back edges need to satisfy two conditions.
		First, each node must admit at most one outgoing back edge.
		The second condition (condition~(iv)) is a bit less intuitive.
		It requires that the partial map $\fFun$ preserves the ancestor relation,
		and, in addition, that the back edges cannot ``superpose'', that is, in a
		tree back edges never cross each other.

		We prove now the tree-like model property and show that each satisfiable
		formula of \CTLSCD is satisfiable in a tree with back edges.
		More specifically, given a Kripke model $\KName$, we show how to define a
		tree with back edges $\UName[\KName]$ such that $\KName$ and
		$\UName[\KName]$ are cycle-bisimilar.
		Together with Theorem~\ref{thm:ctlscd:posmodprp}, this implies that each
		satisfiable formula of \CTLSCD, is satisfiable in a tree with back edges.
		Before defining $\UName[\KName]$, we need to introduce
		two preliminaries notions: the projection map and the initial cycle
		state.

		Let $\KName = \KrpStrStr$ be a Kripke model and consider two constants
		$\new$ and $\cy$.
		We define the \emph{projection map} $\pr: (\WSet \times \{ \new,\cy\})^* \to
		\WSet$ as the unique surjective map such that:

				$\pr(\epsilon)=\wElm[I]$ 
and
				for all $\wElm[][\bullet] \neq \epsilon$, we have
				$\pr(\wElm[][\bullet]) = \wElm$, where $\lst{\wElm[][\bullet]} =
				(w,\alpha)$ and $\alpha \in \{\new,\cy \}$.

		Given a state $\wElm[][\bullet] \in (\WSet \times \{ \new,\cy\})^*$, we say
		that $\wElm[][\bullet]$ admits a sequence $\vElm[][\bullet]$ as an
		\emph{initial cycle state} if there is a sequence $\vElm[1] \dots \vElm[k]$
		(where $k \geq 2$) such that $\wElm[][\bullet]$ is equal to
		$\vElm[][\bullet] \; (\vElm[1],\new) (\vElm[2],\cy) \dots (\vElm[k],\cy)$.
		Given a sequence $\wElm[][\bullet] \in (\WSet \times \{ \new,\cy\})^*$ such
		that $\lst{\wElm[][\bullet] } = (w, \alpha)$, we say that $\wElm[][\bullet]
		$ is labeled by $w$ and $\alpha$.
		Intuitively, the initial cycle state of a given state $\wElm[][\bullet]$ is
		simply the parent of the closest ancestor of $\wElm[][\bullet]$ that is
		labeled by $\new$.
		Note that a state admits at most one initial cycle state.
		We are now ready to define $\UName[\KName]$.

		\begin{definition} \label{def:treelikeunw}
			Given a Kripke model $\KName = \KrpStrStr$, we define the \emph{tree-like
			unwinding} $\UName[\KName] = \KrpStrStr[][\bullet]$ of $\KName$ in the
			following way:

			\begin{itemize}
				\item
					$\WSet[][\bullet] = (\WSet \times \{ \new,\cy\})^*$;

				\item
					$\wElm[I][\bullet] = \epsilon$;

				\item
					for all $\wElm[][\bullet] \in \WSet[][\bullet]$, we have
					$\LFun[][\bullet](\wElm[][\bullet])=\LFun(\pr(\wElm[][\bullet]))$;

				\item
					for all $\wElm[][\bullet] \in \WSet[][\bullet]$ and for all
					$(\pr(\wElm[][\bullet]), v) \in \RRel$:
					\begin{itemize}
						\item
							the pair $(\wElm[][\bullet], \wElm[][\bullet](v,\new))$ belongs
							to $\RRel[][\bullet]$ and is associated with a forward edge;

						\item
							if $\wElm[][\bullet]$ admits an initial cycle state
							$\uElm[][\bullet]$ such that $\pr(\uElm[][\bullet])\neq v$, then
							the pair $(\wElm[][\bullet], \wElm[][\bullet](v, \cy))$ belongs
						  to $\RRel[][\bullet]$ and is associated with a forward edge;

						\item
							if $\wElm[][\bullet]$  admits an initial cycle state
							$\uElm[][\bullet]$ such that $\pr(\uElm[][\bullet])= v$, then the
						  pair $(\wElm[][\bullet],\uElm[][\bullet])$ belongs to
						  $\RRel[][\bullet]$ and is associated with a back edge.
					\end{itemize}
			\end{itemize}

			As mentioned earlier, knowing which edges are forward edges or back edges,
			uniquely  determines a tree decomposition.
			We denote by $(\TName[0](\KName),\fFun(\KName))$ the tree decomposition
			associated with the above definition.
		\end{definition}

		Note that $\epsilon$ is the only state of $\UName[\KName]$ that does
		not admit any initial cycle state.
		It follows from the definition of $\RRel[][\bullet]$ that all the successors
		of $\epsilon$ in $\UName[\KName]$ are of the form $(\wElm, \new)$ (where
		$\wElm$ is a successor in $\KName$ of the initial state of $\KName$).
		Intuitively, the tree with back edges $\UName[\KName]$ is defined as
		follows.
		We consider the usual unwinding construction~\footnote{That is, the Kripke
		model with domain $\{ \trkElm : \trkElm \text{ is a track in } \KName \}$,
		initial state $\epsilon$, transition relation $\{ (\trkElm, \trkElm\wElm):
		\trkElm \text{ and } \trkElm \cdot \wElm \text{ are tracks in } \KName \}$
		and a labeling function mapping each track $\trkElm$ to the set
		$\LFun{\lst{\trkElm}}$.} of a Kripke model and we modify it in two steps.
		First, in the unwinding construction, given a track $\trkElm$ with
		$\lst{\trkElm} = \wElm$ and given a pair $(\wElm, \vElm)$ in the transition
		relation $\RRel$, we construct \emph{one successor} of $\trkElm$ of the
		form $\trkElm \cdot \vElm$.
		Here, we make two ``copies'' of the successor  $\trkElm \cdot \vElm$, one
		labeled by $\new$ and the other one labeled by $\cy$.

		\figexmkrp

		The second modification is as follows: we delete certain edges and replace
		them with back edges (and finally, delete all the states that are reachable
		from $\epsilon$).
		An edge from track  $\trkElm[1]$ to $\trkElm[2]$ is deleted iff $\trkElm[2]$
		is labeled by $\cy$ and $\trkElm[2]$ and the initial cycle state of
		$\trkElm[1]$ are labeled by the same state of $\KName$.


		In order to illustrate the construction $\UName[\KName]$, we provide
		an example in Figure~\ref{fig:krpstr0} and Figure~\ref{fig:treelike}.
		To make notation easier in the figure, we abbreviate $\new$ by $\news$ and
		$\cy$ by $\cys$. Also, instead of writing $\trkElm$ for a state, we only write the pair
		of labels $\lst{\trkElm}$.
		The back edges are those that are not straight lines.


		\figexmtreeunw

\vspace{0.1cm}

		\begin{theorem}
			\CTLSCD has a tree-like model property.
			Every satisfiable formula of \CTLSCD is satisfiable in a tree with back
			edges.
		\end{theorem}

		This follows immediately from the following proposition.

		\begin{proposition} \label{prop:cyclebisim}
			Let $\KName = \KrpStrStr$ be a Kripke model, let $\UName[\KName] =
			\KrpStrStr[][\bullet]$ be its tree-like unwinding.
			Then the relation $\{ (\wElm[][\bullet], \pr(\wElm[][\bullet])):
			\wElm[][\bullet] \in \WorSet[][\bullet] \}$ is a cycle-bisimulation.
			Hence, $\KName$ and $\UName[\KName]$ satisfy exactly the same formulas in
			\CTLSCD.
		\end{proposition}

		Before finishing the section on model properties, we state one more
		property concerning the tree-like unwinding of a model.
		It states that if a formula is true in a tree-like unwinding, then we may
		assume the ``witness'' cycles (for the subformulas of the form $\scycE
		\psi$) to be simple cycles (defined below).
		The property is not that interesting in itself, but it will play an
		important role in the next section for obtaining a 2\ExpTime upper-bound
		for the satisfiability problem of the existential fragment of \CTLSCD.

		\begin{definition}
			A cycle $\pthElm$ is a \emph{simple cycle} if there is a sequence
			$(n_i)_{i \in \SetN}$ such that
			\begin{itemize}
				\item
					$n_i < n_{i+1}$ and $\pthElm[n_i] = \pthElm[0]$, for all $i \in
					\SetN$;
				\item
					for all $i \in \SetN$ and for all $n_i < j < k < n_{i+1}$, we have
					$\pthElm[j] \neq \pthElm[k]$.
			\end{itemize}

			A (state or path) formula $\varphi$ is in \emph{normal form} if for all
			subformulas $\neg \psi$ in $\subFun{\varphi}$, the formula $\psi$ is a
			variable.
			Given a formula $\varphi$ in normal form, we define its \emph{simple cycle
			translation} as the formula obtained by replacing each symbol $\cycE$ in
			the formula $\varphi$, by the symbol $\scycE$.
			The simple cycle translation of $\varphi$ is denoted by $\cytr{\varphi}$.

			The semantics of the formulas of the form  $\cytr{\varphi}$ is defined
			by induction on $\varphi$.
			The basic and induction cases are defined as in Definition~\ref{def:sem},
			with the additional induction step: $\KName, \wElm \models \scycE \psi$ if
			there is a {\bf simple} cycle $\pi$ with beginning state $\wElm$, such
			that $\KName, \pi \models \psi$.
		\end{definition}

		\begin{proposition}
			\label{prop:simplecycle}
			Let $\varphi$ be a formula in \CTLSCD in normal form and let $\KName$ be
			a Kripke model.
			Then $\KName \models \varphi$ iff $\UName[\KName] \models
			\cytr{\varphi}$, where $\UName[\KName]$ is the tree with back edges as
			in Definition~\ref{def:treelike}.
		\end{proposition}


	\end{subparagraph}

	\begin{subparagraph}{Expressiveness}

		We now investigate the expressive power of \CTLSCD 
		\wrt the usual temporal logics.
		All the results are collected in the following theorem.

		\begin{theorem}[Expressiveness comparison]
			\label{thm:ctlscd:exprcmp}
			\CTLSCD is strictly more expressive than 	\CTLS and is incomparable with
			the \MC.
		\end{theorem}

		\begin{proof}
			We observed in the proof of Theorem~\ref{thm:ctlscd:negmodprp} that
			$\varphi_{1} = \A \G \neg \cycE \top$ is satisfiable but does not admit
			any finite model.
			Since \CTLS and the \MC have the finite-model property, this implies that
			$\varphi_1$ 
			is not equivalent to any formula in \CTLS or in the \MC.

			By using the result that there is no \LTL formula expressing that a
			proposition $p$ is true in every even state~\cite{Wol83}, we can show that
			the \MC formula $\psi = \nu x. p \wedge \Box \Box x$ is not equivalent
			to any formula in \CTLSCD.
			Note that $\psi$ is true in a model if for all paths $ \pthElm$ starting
			from the initial state, $p$ is true in every even state $(\pthElm)_{2i}$
			of the path $\pthElm$.
		\end{proof}

	\end{subparagraph}


%
%
%
%
%
%

\end{section}




\begin{section}{Decision Problems}
	\label{sec:decprb}

	In this section, we deal with the solution of the model-checking and
	satisfiability problems for \CTLSCD.
	Regarding the former, we show that we retain the same complexity as for \CTLS,
	that is \PSpace.
	Concerning satisfiability, we also retain the same complexity of \CTLS if we restrict to the existential-cycle fragment of the logic, that is 2\ExpTime.
	Conversely, we show that it is 3\ExpTime for the whole logic.

	\begin{subparagraph}{Model Checking}

		For the solution of the model-checking problem of \CTLSCD, we employ a
		standard bottom-up procedure on the nesting of the path quantifiers of the
		specification under exam, which extends the one originally proposed for
		\CTLS~\cite{CGP02}.
		With more details, starting from the innermost state formulas $\varphi$ of
		the kind $\E \psi$, $\A \psi$, $\cycE \psi$, and $\cycA \psi$, we determine
		their truth value over a \KrpStr $\KName$ at a world $\worElm \in \WSet$ by
		checking the emptiness of a suitable nondeterministic B\"uchi word automaton
		$\NName[\KName, \worElm][\varphi]$.
		In case of a positive result, we enrich the labeling of the world $\worElm$
		with a fresh proposition $\varphi$ representing the formula $\varphi$
		itself.
		Obviously, the path formula $\psi$ is just seen as a classic \LTL formula,
		where all its subformulas of the kind described above are interpreted as
		atomic propositions whose truth values on the worlds of $\KName$ are already
		computed in some previous step of the algorithm.
		It is important to observe that the difference between the automata for $\E
		\psi$ or $\A \psi$ and those for $\cycE \psi$ or $\cycA \psi$ resides in the
		fact that, for the latter, we have to further verify that the initial state
		of the path is seen infinitely often.
		This can be done by means of the standard B\"uchi acceptance condition.
		Hence, we directly obtain that the model checking for \CTLSCD is not more
		complex than the same problem for \CTLS.

		\begin{theorem}
			\label{thm:modchk}
			The model-checking problem for \CTLSCD is \PSpaceC \wrt the formula
			complexity and \NLogSpaceC \wrt the data complexity.
		\end{theorem}

	\end{subparagraph}

	\begin{subparagraph}{Satisfiability}

		Differently from the model checking, the two introduced looping quantifiers
		$\cycE \psi$ and $\cycA \psi$ heavily affect the satisfiability of \CTLSCD.
		In particular, since this logic lacks of the standard tree-model property,
		we cannot use, for the \CTLS part of \CTLSCD, the automata approach as
		proposed in~\cite{KVW00}.
		Instead, we use symmetric two-way alternating tree automata~\cite{BLMV08},
    a simplified version of two-way graded alternating parity tree automata~\cite{BLMV08}, which simply extend classic two-way alternating
    automata over ranked trees~\cite{Var98} to ``unranked trees'', \ie, trees with possibly unbounded width.
		These are automata that allow to traverse a tree in forward and backward.
		We use these automata here to search for tree representation of the
		tree-like unwinding of a structure, as described in the previous section.
		With more details, for every \CTLSCD state formula $\varphi$, we build an
		alternating parity two-way tree automaton $\AName[\varphi]$ such that a
		\KrpStr $\KName = \KrpStrStr$ is a model of $\varphi$ iff $\AName[\varphi]$
		accepts a tree $\TName[\KName] = \tuplee{\APSet \cup \{ \new, \uparrow \}}
		{\WorSet[][\bullet]} {\TrnRel[][\star]} {\labFun[][\star]}
		{\worElm[I][\bullet]}$ associated with the tree-like unwinding
		$\UName[\KName] = \tuplee{\APSet} {\WorSet[][\bullet]} {\TrnRel[][\bullet]}
		{\labFun[][\bullet]} {\worElm[I][\bullet]}$ of $\KName$ via the following
		properties: \emph{(i)} $\RRel[][\star] = \set{ (\worElm[][\bullet],
		\vElm[][\bullet]) \in \RRel[][\bullet] }{ \card{\worElm[][\bullet]} <
		\card{\vElm[][\bullet]}}$, \emph{(ii)} $\LFun[][\star](\worElm[][\bullet])
		\cap \APSet = \LFun[][\bullet](\worElm[][\bullet])$, \emph{(iii)} $\new \in
		\LFun[][\star](\worElm[][\bullet])$ iff $\lst{\worElm[][\bullet]} =
		(\worElm, \new)$, for some $\worElm \in \WSet$, and \emph{(iv)} $\uparrow\:
		\in \LFun[][\star](\worElm[][\bullet])$ iff there exists
		$(\worElm[][\bullet], \vElm[][\bullet]) \in \RRel[][\bullet]$ with
		$\card{\vElm[][\bullet]} < \card{\worElm[][\bullet]}$.
		Intuitively, $\TName[\KName]$ is built from $\UName[\KName]$ by deleting all
		back edges (property \emph{(i)}) and enriching the original labeling of
		every world $\worElm[][\bullet]$ (property \emph{(ii)}) with $\new$,
		if the last letter of $\worElm[][\bullet]$ contains the flag with the same
		name (property \emph{(iii)}), and with $\uparrow$, if $\worElm[][\bullet]$
		is the origin of a back edge (property \emph{(iv)}).
		It is not hard to see that, for every unwinding $\UName[\KName]$ of a
		\KrpStr $\KName$, there exists one and only one tree $\TName[\KName]$
		satisfying the previous four properties.
		Therefore, instead of looking for a model $\KName$ of $\varphi$ or its
		tree-like unwinding $\UName[\KName]$, we just look for its tree
		representation $\TName[\KName]$.
		This idea is at the basis for the automata-theoretic approach described in
		the proofs of the following theorems.

		\begin{theorem}
			\label{thm:ctlscdsat}
			The satisfiability problem for \CTLSCD can be solved in 3\ExpTime and is
			2\ExpTimeH.
		\end{theorem}
		\begin{proof}
			The 2\ExpTime lower bound for \CTLSCD immediately follows from the one of
			\CTLS.
			For the 3\ExpTime upper bound, given a \CTLSCD state formula $\varphi$, we
			reduce the associated satisfiability question to the emptiness problem of
			an alternating parity two-way tree automaton $\AName[\varphi]$, whose size
			and index are, respectively, doubly and single exponential in
			$\card{\varphi}$. For a detailed definition of symmetric alternating
			parity two-way tree automata and the related concepts of size and
			index, we refer to~\cite{BLMV08}.
			Since the emptiness of $\AName[\varphi]$ can be checked in time exponential
			\wrt both its states and index~\cite{BLMV08}, we obtain
			the desired result~\footnote{In particular, Theorem 6.7 in~\cite{BLMV08} can be used for the translation. 
      Observe that, since we do not make use of any graded modalities (our box and diamond symbols stand for $\AAll{0}$ and $\EExs{0}$ in their syntax) the resulting automaton is simply a symmetric non-deterministic tree automaton.}.
      
			As mentioned above, $\AName[\varphi]$ needs to recognize all and only the
			tree representations $\TName[\KName]$ of the tree-like unwindings
			$\UName[\KName]$ of \KrpStr models $\KName$ of $\varphi$.
			As it is usually done for \CTLS, we slightly weaken this property by
			allowing $\AName[\varphi]$ to run on trees that also contain, as labeling
			of its worlds, the subformulas of $\varphi$ of the form $\E \psi$, $\A
			\psi$, $\cycE \psi$, and $\cycA \psi$, which are interpreted as fresh
			atomic propositions. We denote by
                        $\subFun[Q]{\varphi}$ the set of subformulas
                        of $\varphi$ of the form $\E \psi$, $\A
			\psi$, $\cycE \psi$, and $\cycA \psi$. We also 
                        let $\subFun[Q][\neg]{\varphi}$ be the the closure under negation of the
			set $\subFun[Q]{\varphi}$, \ie, for every $\E \psi$ (resp., $\A \psi$, $\cycE
			\psi$, $\cycA \psi$) in $\subFun{\varphi}$, we have $\A \neg\psi$ (resp.,
			$\E \neg\psi$, $\cycA \neg\psi$, $\cycE \neg\psi$) in
			$\subFun[][\neg]{\varphi}$. 
So, instead of considering a model $\KName = \KrpStrStr$ of
			$\varphi$, we work on the enriched \KrpStr $\KName[][\star] =
			\KrpStrStr[][][ {\APSet[][\star]} ][][][ {\LFun[][\star]} ]$ such that
			\emph{(i)} $\APSet[][\star] = \APSet \cup \subFun[][\neg]{\varphi}$,
			\emph{(ii)} $\LFun[][\star](\worElm) \cap \APSet = \LFun(\worElm)$, and
			\emph{(iii)} $\eta \in \LFun[][\star](\worElm)$ iff $\KName, \worElm
			\models \eta$, for all $\eta \in \subFun[][\neg]{\varphi}$.
			set $\subFun{\varphi}$, \ie, for every $\E \psi$ (resp., $\A \psi$, $\cycE
			\psi$, $\cycA \psi$) in $\subFun{\varphi}$, we have $\A \neg\psi$ (resp.,
			$\E \neg\psi$, $\cycA \neg\psi$, $\cycE \neg\psi$) in
			$\subFun[][\neg]{\varphi}$.

			The automaton $\AName[\varphi]$ is built as the conjunction of an
			automaton $\AName[\eta]$, for every subformula $\eta \in
			\subFun[][\neg]{\varphi}$, and a deterministic safety (\ie, without
			acceptance condition) automaton $\DName[\varphi]$ used to verify that
			$\varphi$ is satisfied at the root of the input tree $\TName[\KName]$,
			when $\varphi$ is interpreted as a Boolean formula on $\APSet[][\star]$.
			In addition, $\AName[\varphi]$ needs to check that, if a world is not
			labeled by a state formula $\eta \in \subFun[][\neg]{\varphi}$, it is
			necessarily labeled by a formula $\dual{\eta} \in
			\subFun[][\neg]{\varphi}$ equivalent to its negation, \ie, $\dual{\eta}
			\equiv \neg \eta$.
			The automaton $\AName[\eta]$ is committed to check that a world
			labeled by $\eta \in \subFun[][\neg]{\varphi}$ really satisfies this
			formula.
			Formally, we have $\AName[\varphi] \defeq \DName[\varphi] \wedge
			\bigwedge_{\eta \in \subFun[][\neg]{\varphi}} \AName[\eta]$.
			So, its size is the sum of the sizes of the components.
			The construction of $\DName[\varphi]$ is trivial.
			Moreover, the automata for $\A \psi$ and $\cycA \psi$ can be directly
			derived from the automaton for $\E \neg \psi$ and $\cycE \neg \psi$ by replacing
			$\vee$ and $\Diamond$ with $\wedge$ and $\Box$ in their definitions.
			Hence, we just focus on the constructions for the latter.

			We start with the construction of $\AName[\E \psi]$ for $\E \psi$.
			Consider the nondeterministic B\"uchi word automaton $\NName[\psi] =
			\tuplee{\pow{\APSet[][\star]}\!} {\QSet} {\delta} {\QSet[I]} {\FSet}$
			obtained by applying the Vardi-Wolper
                        construction to $\psi$ which is read as an
			\LTL formula over $\APSet[][\star]$~\cite{VW86b}.
			We set as a two-way B\"uchi tree automaton $\AName[\E \psi] \defeq
			\tuplee{\Sigma^{\star}} {\QSet[][\star]} {\delta^{\star}}
			{\qElm[I][\star]} {\FSet[][\star]}$, where the alphabet $\Sigma^{\star}
			\defeq \pow{\APSet[][\star] \cup \{ \new, \uparrow \}}$ augments the set
			of extended atomic propositions $\APSet[][\star]$ with the symbols $\new$
			and $\uparrow$, as required by the definition of the tree representations
			$\TName[\KName]$.
			The set of states $\QSet[][\star] \defeq \{ \qElm[I][\star] \} \cup \QSet
			\times \{ \downarrow, \uparrow \}$ contains the initial state
			$\qElm[I][\star]$ plus two copies of the states of $\NName[\psi]$, one
			for each direction of navigation over the tree $\TName[\KName]$.
			For the B\"uchi acceptance condition we consider the set $\FSet[][\star]
			\defeq \{ \qElm[I][\star] \} \cup \FSet \times \{ \downarrow \}$.
			The definition of the transition function $\delta^{\star}$ follows.
			For the sake of readability, we divide it in three parts, depending on
			whether it predicates on $\qElm[I][\star]$, a state $\qElm$ flagged with
			$\downarrow$ or a state $\qElm$ flagged with $\uparrow$.

			\begin{itemize}
			  \item
					The initial state $\qElm[I][\star]$ is used to start the evaluation of
					the formula $\E \psi$ on every world of the input tree labeled by $\qElm[I][\star]$ .
					This is done by starting the simulation of $\NName[\psi]$.
					Formally, we have that $\delta^{\star}(\qElm[I][\star], \sigma) \defeq
					(\Box, \qElm[I][\star]) \wedge \bigvee_{\qElm \in \QSet[I]} (\epsilon,
					(\qElm, \downarrow))$, if $\E \psi \in \sigma$, and
					$\delta^{\star}(\qElm[I][\star], \sigma) \defeq (\Box,
					\qElm[I][\star])$, otherwise.
				\item
					Every copy of a state $\qElm \in \QSet$ flagged with $\downarrow$ is
					used to effectively verify the existence of an infinite path in
					$\UName[\KName]$ satisfying $\psi$.
					This is done by guessing an extension of the finite path built up to
					now and sending, to the corresponding direction, a successor $\pElm$
					of $\qElm$ that complies with the transition function $\delta$ of
					$\NName[\psi]$, when the labeling $\sigma$ of the world under exam is
					read.
					As the input tree $\TName[\KName]$ is a representation of the tree
					with back edges $\UName[\KName]$, we have also to take them into
					account when we guess the extension of the path from a world labeled
					with $\uparrow$.
					This is done by sending up along the tree the copy of the state
					$\pSym$ flagged with $\uparrow$, which is used to simulate a jump to
					the world destination of the back edge.
					Formally, we have $\delta^{\star}((\qElm, \downarrow), \sigma) \defeq
					\bigvee_{\pElm \in \delta(\qElm, \sigma \cap \APSet[][\star])}\:
					(\Diamond, (\pElm, \downarrow)) \vee {\uparrow\!(\pElm)}$, where
					${\uparrow\!(\pElm)}$ is set to $(\epsilon, (\pElm, \uparrow))$ if
					$\uparrow\: \in \sigma$, and to $\Ff$, otherwise.
				\item
					Finally, for every copy of a state $\qElm \in \QSet$ flagged with
					$\uparrow$, we only have to
                                        modify the state and the
                                        direction of the automaton 
          when we are approaching to the
					destination of the back edge that gave rise to the evaluation of
					$(\qElm, \uparrow)$.
					Fortunately, due to the structure of the tree-like unwinding
					$\UName[\KName]$ and, consequently, of its tree representation
					$\TName[\KName]$, when we reach a world labeled by $\new$, we are
					sure that the immediate ancestor of this world is the destination of
					the back edge.
					Thus, we can immediately change the flag of the state $\qElm$ to
					$\downarrow$ in order to resume the verification of the path formula
					$\psi$.
					Formally, $\delta^{\star}((\qElm, \uparrow), \sigma) \defeq (\uparrow,
					(\qElm, \downarrow))$, if $\new \in \sigma$, and
					$\delta^{\star}((\qElm, \uparrow), \sigma) \defeq \allowbreak
					(\uparrow, (\qElm, \uparrow))$, otherwise.
			\end{itemize}
			Now, by construction, it is not hard to prove that $\AName[\E \psi]$
			correctly verifies that every world of $\TName[\KName]$ labeled by $\E
			\psi$ satisfies $\E
			\psi$  in $\UName[\KName]$.
			Also, by the Vardi-Wolper procedure, it follows that $\card{\QSet} =
			\AOmicron{2^{\card{\psi}}}$.
			Consequently, the size of $\AName[\E \psi]$ is exponential in the length
			of $\E \psi$.

			The construction of $\AName[\cycE \psi]$ is quite more complex than the
			one previously described, as it also requires a projection operation that
			is the reason behind the exponential gap between the upper and lower
			bounds.
			Differently from the automata for classic path quantifiers, we cannot
			evaluate the correctness of the labeling $\cycE \psi$ on all worlds of the
			tree in one shot.
			This is because of the possible interactions among the cycles starting in
			different worlds, which does not allow us to determine which is the origin
			of the path we are interested in.
			Consequently, we have to focus on one world labeled by $\cycE \psi$ at a
			time and check the existence of a path passing infinitely often through
			that world, which also satisfies the property $\psi$.
			This unique world is identified by a fresh symbol $\#$.
			Then, an universal projection operation over such a symbol will take care
			of the fact that this check has to be done for every possible world
			labeled by $\cycE \psi$.
			Formally, $\AName[\cycE \psi]$ is built as follows:
			$\Pi_{\#}^{\forall}(\NName[\#] \vee \AName[\cycE \psi][\#])$.
			Intuitively, we make a universal projection over $\#$ of a disjunction
			between the automaton $\NName[\#]$, accepting all trees where the labeling
			$\#$ is incorrect (\ie, there are more than one occurrences of $\#$ or
			this symbol is on a world that is not labeled by $\cycE \psi$), and the
			automaton $\AName[\cycE \psi][\#]$, verifying the existence of a path
			satisfying $\psi$ that starts and passes infinitely often through the
			world labeled by $\#$.
			The construction of $\NName[\#]$ is trivial.
			For the computation of the projection, we use the equality
			$\Pi_{\#}^{\forall} \AName = \neg \Pi_{\#}^{\exists} \neg \AName$.
			Note however that there is no known projection operation
			that can act directly on a two-way automaton.
			Instead, we have first to translate it into a nondeterministic one-way
			automaton~\cite{BLMV08} and then apply the standard projection.
			Due to the nondeterminization procedure, $\Pi_{\#}^{\forall}
			\AName$ has exponential size \wrt that of $\AName$. So, $\AName[\cycE
			\psi]$ is exponential in the size of $\AName[\cycE \psi][\#]$.

			It remains to define the latter automaton.
			As above, let $\NName[\psi] = \tuplee{\pow{\APSet[][\star]}\!} {\QSet}
			{\delta} {\QSet[I]} {\FSet}$ be the nondeterministic B\"uchi word
			automaton obtained by applying the Vardi-Wolper construction to $\psi$.
			Then, we set $\AName[\cycE \psi][\#] \defeq \tuplee{\Sigma^{\star}}
			{\QSet[][\star]} {\delta^{\star}} {\qElm[I][\star]} {\FSet[][\star]}$ as a
			two-way B\"uchi tree automaton having alphabet $\Sigma^{\star} \defeq
			\pow{\APSet[][\star] \cup \{ \new, \uparrow, \# \}}$.
			The set of states $\QSet[][\star] \defeq \{ \qElm[I][\star] \} \cup \QSet
			\times \{ \Ff, \Tt \} \times \{ \#, \downarrow, \uparrow \}$ contains the
			initial state $\qElm[I][\star]$ plus six copies of the states of
			$\NName[\psi]$.
			Each of them is flagged with a Boolean value keeping track of the original
			acceptance condition derived from $\NName[\psi]$ and a symbol indicating
			the direction of navigation over the tree.
			Differently from the previous case, we have also $\#$ as a flag in order
			to indicate the passage though the state labeled by the flag itself.
			For the B\"uchi acceptance condition we consider the set $\FSet[][\star]
			\defeq \{ \qElm[I][\star] \} \cup \QSet \times \{ \Tt \} \times \{ \# \}$.
			Intuitively, apart from the initial state, we assume as final those states
			that certify both the passage through the origin of the path indicated by
			$\#$ and the possibly previous occurrence of an accepting state.
			It remains to define the transition function $\delta^{\star}$.
			Here we use $\alpha(\qElm, \alpha)$ to denote the Boolean value $\Tt$, if
			$\qElm \in \FSet$, and $\alpha$, otherwise.

			\begin{itemize}
			  \item
					The initial state $\qElm[I][\star]$ is used to start evaluating 
					the formula $\cycE \psi$ on the unique world of the input tree labeled
					by $\#$.
					Formally, we have $\delta^{\star}(\qElm[I][\star], \sigma) \defeq
					\bigvee_{\qElm \in \QSet[I]} (\epsilon, (\qElm, \alpha(\qElm, \Ff),
					\downarrow))$, if $\# \in \sigma$, and
					$\delta^{\star}(\qElm[I][\star], \sigma) \defeq (\Box,
					\qElm[I][\star])$, otherwise.
					Note that, since we are just starting with the simulation of
					$\NName[\psi]$, the flag $\alpha(\qElm, \Ff)$ concerning the memory on
					the acceptance condition only depends on the state $\qElm$, as the
					second argument is fixed to $\Ff$.
				\item
					Since a state $(\qElm, \alpha, \#)$ is simply used to verify the
					passage through the starting point of the path satisfying $\psi$, the
					automaton has to reset the memory on the acceptance condition and
					continue with the simulation of $\NName[\psi]$.
					Formally, $\delta^{\star}((\qElm, \alpha, \#), \sigma) \defeq
					(\epsilon, (\qElm, \Ff, \downarrow))$.
				\item
					The automaton $\AName[\cycE \psi][\#]$ on the state $(\qElm, \alpha,
					\downarrow)$ behaves similar to $\AName[\E \psi]$ on $(\qElm, \downarrow)$.
					One difference resides in the update $\alpha(\pElm, \alpha)$ of the
					memory on the acceptance condition, which takes into account both the
					previous memory $\alpha$ and the membership of $\pElm$ in $\FSet$.
					The other difference is that, if $\sigma$ contains the symbol $\#$, we
					have to record this fact in the state, by swapping the flag from
					$\downarrow$ to $\#$.
					Formally, we have $\delta^{\star}((\qElm, \alpha, \downarrow), \sigma)
					\defeq \bigvee_{\pElm \in \delta(\qElm, \sigma \cap
					\APSet[][\star])}\: (\Diamond, (\pElm, \alpha(\pElm, \alpha), \beta))
					\vee {\uparrow\!(\pElm)}$, where $\beta = \#$, if $\# \in \sigma$,
					and $\beta = \:\downarrow$, otherwise; moreover, ${\uparrow\!(\pElm)}$
					is set to $(\epsilon, (\pElm, \alpha(\pElm, \alpha), \uparrow))$ if
					$\uparrow\: \in \sigma$, and to $\Ff$, otherwise.
				\item
					Finally, as for $\AName[\E \psi]$, a state of the form $(\qElm,
					\alpha, \uparrow)$ identifies the destination of a back edge.
					Thus, we have $\delta^{\star}((\qElm, \alpha, \uparrow),
					\sigma) \defeq (\uparrow, (\qElm, \alpha, \downarrow))$, if
					$\new \in \sigma$, and $\delta^{\star}((\qElm, \alpha, \uparrow),
					\sigma) \defeq (\uparrow, (\qElm, \alpha, \uparrow))$, otherwise.
			\end{itemize}
			Finally, the size of $\AName[\cycE \psi][\#]$ is exponential in the length
			of $\cycE \psi$, which implies that $\AName[\cycE \psi]$ is doubly
			exponential \wrt the same length.
		\end{proof}

		In case we want to restrict our attention to the satisfiability of the
		\CTLSCD fragment having only existential looping quantifiers, we can improve
		the previous proof, obtaining a tight 2\ExpTime procedure, by providing a
		single exponential construction for the automaton $\AName[\cycE \psi]$.
		Indeed, thanks to the simple cycle property of the verification of the
		formula $\cycE \psi$ on the tree-like unwinding $\UName[\KName]$, we can
		just focus on cycle paths of $\UName[\KName]$ going through the successors
		of their origin labeled by $\new$.
		In this way, there are no interactions among the paths that start at
		different worlds labeled by $\cycE \psi$, since two paths passing through
		the same world necessarily use different successors.
		Consequently, we can always uniquely identify the origin of a path on
		which we have to pass infinitely often.

		Unfortunately, the same idea cannot be exploited for the verification of the
		universal looping quantifiers $\cycA \psi$, as we have to check the property
		$\psi$ on all cycle paths and not only on those that are simple.
		At the moment, it is left open whether a 2\ExpTime satisfiability procedure
		for the whole \CTLSCD logic exists.

		\begin{theorem}
			\label{thm:ctlcdelsat}
			The satisfiability problem for the existential-cycle fragment of \CTLSCD
			is 2\ExpTimeC.
		\end{theorem}

	\end{subparagraph}

\end{section}


	
	
	
	



\begin{section}{Discussion}
	\label{sec:discussion}

	To conclude, we give a concise overview of the main properties of the cycle-logic
	extension we have introduced.
	We also explain why this extension is natural and why, given our results, we have
	decided to focus our presentation on the cycle-logic \CTLSCD.

	\CTLSCD allows us to quantify only over cycles, that is, paths such that the
	initial state occurs infinitely often.
	Hence, it has been natural to consider also a more general logic (denoted by
	\ECTLSCD) allowing us to test whether \emph{any arbitrary} state of a given
	path occurs infinitely often in the path.
	More formally, \ECTLSCD is the extension of \CTLS obtained by adding the
	symbol $\cycle$.
	This symbol is treated as an atomic path formula and is true at a
	state in a path iff the state occurs infinitely often in the path.
	It is easy to see that \ECTLSCD is an extension of \CTLSCD.

	We have studied several properties about \ECTLSCD and, among the others, we have shown that this logic does not preserve the cycle-bisimulation property.
	Clearly, one can use a stronger notion of bisimulation under which
	\ECTLSCD can still retain the invariance. However, we came up with notions that are not very intuitive (as the notion of cycle-bisimulation) and not useful to prove any kind of tree-like model property. Given these negative results, we decided to not present extensively this part.

	We would like to mention that we also considered the extension of
	\LTL with the symbol $\cycle$.
	It is not hard to show that it is a proper extension of \LTL and is
	orthogonal to $\omega$-regular expressions.
	We can also prove that the finite satisfiability problem for that logic is
	decidable (using an adaption of the proof for \LTL~\cite{VW86b}).
	We did not present these results by lack of space.
	
	Finally, as future work we would like to investigate the use of the introduced cycle construct in the realm of logics for multi-agent systems such as \ATLS~\cite{AHK02} and Strategy Logic\cite{MMPV14}. These logics have been proved to be useful to reasoning about strategic abilities in a number of complicated settings. In particular, the latter is able to express sophisticated solution concepts such as Nash Equilibria and Subgame Perfect Equilibria, as well as they it has been used to express iterative extensive game forms such as the iterated prisoner dilemma. In all these contexts, talking explicitly about cycles could play a central role in solving the related game questions.

\end{section}


  
\section*{Acknoledgments}
 Aniello Murano and Loredana Sorrentino are partially supported by the GNCS 2016 project:  Logica, Automi e Giochi per Sistemi Auto-adattivi.
 Giuseppe Perelli thanks the support of the ERC Advanced Grant 291528 (``Race'') at Oxford.



\bibliographystyle{eptcs}
\bibliography{References}

	\newpage
	\normalsize
	\appendix

\end{document}